\documentclass{aa}  

\usepackage{graphicx}
\usepackage{txfonts}
\usepackage{placeins}

\begin{document} 

   \title{
   CO line observations of OH/IR stars in the inner Galactic Bulge: Characteristics of stars at the tip of the AGB
   }

   \titlerunning{OH/IR stars in the inner Galactic Bulge}

   \author{H.~Olofsson  \inst{1}
          \and
          T.~Khouri    \inst{1}
          \and
          B.A. Sargent  \inst{2,3}
          \and
          A. Winnberg \inst{1}
          \and
          J.A.D.L. Blommaert \inst{4}
          \and
          M.A.T. Groenewegen \inst{5}
          \and
          S. Muller \inst{1}
          \and
          J.H. Kastner \inst{6}
          \and
          M. Meixner \inst{7}
          \and
          M. Otsuka \inst{8}
          \and
          N. Patel \inst{9}
          \and
          N. Ryde \inst{10}
          \and
          S. Srinivasan \inst{11}
          }

\institute{Dept of Space, Earth and Environment, Chalmers Univ. of Technology,
           Onsala Space Observatory, SE-43992 Onsala, Sweden\\
           \email{hans.olofsson@chalmers.se}
         \and
         Space Telescope Science Institute, 3700 San Martin Drive, Baltimore, MD 21218, USA
         \and
         Center for Astrophysical Sciences, The William H. Miller III Department of Physics and Astronomy, Johns Hopkins University, Baltimore, MD 21218, USA
         \and
         Astronomy and Astrophysics Research Group, Department of Physics and Astrophysics, Vrije Universiteit Brussel, Pleinlaan 2, B-1050 Brussels, Belgium
         \and
         Koninklijke Sterrenwacht van Belgi\"e, Ringlaan 3, B--1180 Brussels, Belgium
         \and
         Chester F. Carlson Center for Imaging Science, School of Physics \& Astronomy, and Laboratory for Multiwavelength Astrophysics, Rochester Institute of Technology, Rochester, NY 14623, USA
         \and
         Stratospheric Observatory for Infrared Astronomy/USRA, NASA Ames Research Center, MS 232-11, Moffett Field, 94035 CA, USA 
         \and
         Okayama Observatory, Kyoto University Honjo, Kamogata, Asakuchi, Okayama, 719-0232, Japan
         \and
         Center for Astrophysics | Harvard \& Smithsonian, 60 Garden Street, MS 78, Cambridge, MA 02138, USA
         \and
         Lund Observatory, Department of Astronomy and Theoretical Physics, Lund University, Box 43, SE-22100 Lund, Sweden
         \and
         Instituto de Radioastronom\'ia y Astrof\'isica, Universidad Nacional Aut\'onoma de M\'exico, Morelia, Michoac\'an, 58089 M\'exico              
}

   \date{Received 19 May 2022 / Accepted 10 July 2022}

 \abstract{}{}{}{}{} 
 
  \abstract
   {OH/IR stars are examples of late stellar evolution on the asymptotic giant branch (AGB), and they are, as such, important objects to study. They are also excellent probes of stellar populations, in particular in regions of high interstellar extinction such as the central regions of our Galaxy.}
   {Our goal is to characterise the stellar and circumstellar properties of high-mass-loss-rate OH/IR stars in the inner Galactic Bulge using the Atacama Large Millimeter/submillimeter Interferometer (ALMA).}
   {Rotational lines of $^{12}$CO and $^{13}$CO, as well as a millimetre-wave continuum, have been observed for a sample of 22 OH/IR stars in directions within 2$^\circ$ of the Galactic Centre. Photometry data ($\approx$\,1$-$30\,$\mu$m) have been gathered from the literature to construct spectral energy distributions (SEDs) and to determine pulsational variability. Radiative transfer models have been used to interpret the line and photometry data. }
  {All stars in the sample were detected in at least one CO line, and eight objects were detected in 324\,GHz continuum. Based on luminosity criteria, the sample is divided into 17 objects that most likely lie within the inner Galactic Bulge, and five objects that are most likely foreground objects. The median luminosity of the inner-Galactic-Bulge sub-sample, 5600\,$L_\odot$, corresponds to an initial mass in the range 1.2\,$-$\,1.6\,$M_\odot$, indicating that these inner-Galactic-Bulge OH/IR stars descend from solar-type stars. The objects in this sub-sample are further divided into two classes based on their SED characteristics: Eleven objects have SEDs that are well matched by models invoking dust envelopes extending from a few stellar radii and outwards, while six objects are better modelled as having detached dust envelopes with inner radii in the range 200\,$-$\,600\,au and warmer central stars. The former objects have periodic variability, while the latter objects are predominantly non-periodic. The median gas-mass-loss rate, gas terminal expansion velocity, gas-to-dust mass ratio, and circumstellar $^{12}$CO/$^{13}$CO abundance ratio have been estimated to be 2$\times$10$^{-5}$\,$M_\odot$\,yr$^{-1}$, 18\,km\,s$^{-1}$, 200 (excluding the sources with detached dust envelopes, which show markedly lower gas-to-dust ratios), and 5, respectively, for the inner-Galactic-Bulge sub-sample. All line brightness distributions are resolved at an angular scale of $\approx$\,0\farcs15, but only two objects show a structure in their circumstellar envelopes at our resolution and sensitivity. In both cases, this structure takes the form of a cavity and a bipolar morphology.  }
   {The inner-Galactic-Bulge sub-sample consists of high mass-loss-rate stars that descend from solar-type progenitors and that lie near the tip of the AGB. Some of the sample stars may have recently ceased mass loss and, hence, have begun to evolve beyond the AGB, as evidenced by a change in circumstellar characteristics and indications of warmer central stars. The inferred very low stellar $^{12}$C/$^{13}$C isotope ratios are indicative of CNO-cycle nuclear processing, and they are most likely established at the surfaces of the stars during the first dredge-up on the red giant branch since these stars are not expected to experience hot-bottom burning. The inner-Galactic-Bulge OH/IR stars studied here constitute an excellent sample of equidistant objects for the purpose of understanding the evolution of the mass-loss-rate characteristics at the tip of the AGB.}


   \maketitle
%
%
%
\section{Introduction}

In the early 1970s point-like objects with two main observational characteristics were found: unusually strong `satellite' 1612\,MHz maser line emission from hydroxyl (OH), compared to that of the `main lines' at 1665 and 1667\,MHz, and a double-peaked line profile with sharp outer edges and inner edges decreasing more slowly towards the systemic velocity \citep[e.g.][]{wilsbarr72,winnetal73,caswhayn75,bowe78}. As data on these objects were gathered, the suspicion that they are highly evolved stars of solar type strengthened \citep{johaetal77b,baudetal81a}. The association with highly reddened sources provided conclusive evidence that these objects are stars that lie at the upper end of the asymptotic giant branch (AGB) \citep{baudhabi83,habi96}. Optically thick circumstellar envelopes (CSEs), produced by intense mass loss, led to high extinction at optical wavelengths. These objects were given the name OH/IR stars. The presence of OH means that they must be of the O-type, that is O is more abundant than C in number (in the CSE, and by inference also in the stellar atmosphere). In fact, mass-losing red supergiants also exhibit these characteristic OH 1612\,MHz line profiles \citep{coheetal87}.

\citet{goldscov76} presented the first physical-chemical model of an AGB CSE in which the OH molecules are formed through dissociation of circumstellar H$_2$O by interstellar UV radiation. The OH molecules subsequently become excited by infrared radiation from dust at 35 and 53\,$\mu$m to produce strong maser emission in the satellite line at 1612\,MHz \citep{elitetal76}. Additional work on circumstellar OH chemistry and excitation has been done by \citet{huggglas82}, \citet{degu82}, and \citet{netzknap87}, for instance.

The fact that the OH maser emission propagation is unaffected by interstellar extinction has made the OH/IR stars important probes of both stellar evolution and galactic structure. Large surveys have been performed over the years \citep[e.g.][]{ederetal88,telietal91,lindetal92a,lesqetal92,engelewi96,seveetal97a,sjouetal98} [see \citet{chenetal01} for a summary of OH observations of OH/IR stars]. Until today, more than a thousand OH/IR stars have been detected in the Milky Way. In addition, about a dozen have been detected in the Large Magellanic Cloud \citep{goldetal17}. 

The penetrating ability of the long-wavelength emission has been particularly important for studies of stars in the Galactic Plane, the Galactic Bulge [GB; defined here as in \citet{barbetal18} as the region within $|\ell|$\,$\la$\,10$^\circ$ and $|b|$\,$\la$\,10$^\circ$ of the Galactic Centre (GC)], and the GC. \citet{seveetal97a} detected more than 250 objects with OH/IR star characteristics in the GB, with $\approx$\,80 of them lying within the inner GB [within $\la$\,2$^\circ$ of the GC; sometimes called the nuclear bulge \citep{mezgetal96} or a disky pseudobulge \citep{kormbend19}], and at higher sensitivity \citet{lindetal92a} and \citet{sjouetal98} detected $\approx$\,190 OH/IR objects within 0.3$^\circ$ of the GC. \citet{vandvehabi90} concluded that the OH/IR stars in the GB are dominated by $\approx$\,1.0--1.4\,$M_\odot$ progenitor stars with an age $>$\,7\,Gyr. \citet{woodetal98} used photometry for a sample of OH/IR stars within 0.7$^\circ$ of the GC to conclude that closer to the GC there are also younger AGB stars belonging to an intermediate-age population. This is a conclusion  supported by the study of Mira variables in the inner Galaxy by \citet{groeblom05}. They found an inner-GB population of age \mbox{1--3}\,Gyr, and an even younger population closer to the GC, $\approx$\,1\,Gyr \citep[see also][]{vanloetal03}. In general, it appears that the OH/IR stars cover the full mass range of AGB stars, from $\approx$\,1 to  $\approx$\,8\,$M_\odot$ \citep{habi96}. Apart from this, these OH/IR stars are not particularly well characterised, although near-IR spectroscopic work is now making good progress \citep{schuetal20}.

The OH maser emission provides interesting insights for studying OH/IR stars, but only brings (at best) order-of-magnitude estimates of the mass-loss rate, and no information on elemental isotopic ratios, such as $^{12}$C/$^{13}$C, which are key signposts of stellar evolution. Observations of rotational lines from carbon monoxide (CO) have proven to be essential in this context \citep[e.g.][]{schoolof01,debeetal10,ramsolof14}. The first major CO-based study of OH/IR stars were done by \citet{hesketal90} on 13 objects. The major conclusions were that the estimated mass-loss rates from the CO lines (using a semi-emperical formula) were much lower than those estimated from dust continuum, dust features, and OH maser emission, and that the $J$\,=\,2--1/$J$\,=\,1--0 line intensity ratios were unexpectedly high. A possible explanation for both results would be a recent increase in mass-loss rate, the start of a superwind. Such a conclusion was further strengthened by the work of \citet{justtiel92} and \citet{justetal94,justetal96a}. However, the mass-loss-rate-estimate methods used in these papers had substantial uncertainties. In a recent paper based on ALMA CO observations of two OH/IR stars, \citet{decietal19} concluded that a fair fraction of the dust is assembled into an equatorial density enhancement due to the binary nature of the objects, and that this would explain  the discrepant mass-loss rates found in previous papers. This explanation was extrapolated to all OH/IR stars, concluding that they reach maximum mass-loss rates of the order a few times 10$^{-5}$\,$M_\odot$\,yr$^{-1}$. In a study of a sample of five OH/IR stars, \citet{delfetal97} found that the circumstellar $^{12}$CO/$^{13}$CO abundance ratios were very low, about 3.5 [compared to 89 in the solar photosphere \citep{claynitt04} and 68 in the ISM \citep{milaetal05}]. Arguments were given that these reflect the stellar $^{12}$C/$^{13}$C ratios and that the low values, close to the equilibrium value of the CNO cycle, were due to the hot-bottom-burning process at the end of the AGB \citep{karaluga16}. The consequence being that the stars in their sample are more massive than about 4$M_\odot$.

A detailed study of the mass-loss characteristics of OH/IR stars based on observations of CO rotational lines is therefore warranted. In particular, a sample of equidistant objects can be formed by selecting OH/IR stars in the inner Galaxy.  However, the presence of ubiquitous interstellar CO along the lines-of-sight complicates such observations. The first OH/IR stars detected in CO line emission towards the GB and GC were therefore restricted to high-velocity objects with emission well outside the velocity range of interstellar CO line emission \citep[2 objects;][]{winnetal91}, or objects well outside the regions of highest extinction \citep[7 objects;][]{blometal18}. The ability of interferometers to filter out extended emission was used to increase the number of objects detected in CO line emission in higher-extinction regions \citep{winnetal09,sargetal13}, bet led to only four new detections. In this paper, we present observations of both $^{12}$CO and $^{13}$CO lines performed with the Atacama Large Millimeter/submillimeter Array (ALMA) that substantially increase the number of OH/IR stars detected in circumstellar CO line emission in the inner GB. In total, 22 stars were observed, all of them were detected in at least one CO line.

%
%
%
%
\section{The sample}

\begin{figure}
  \includegraphics[width=8.8cm]{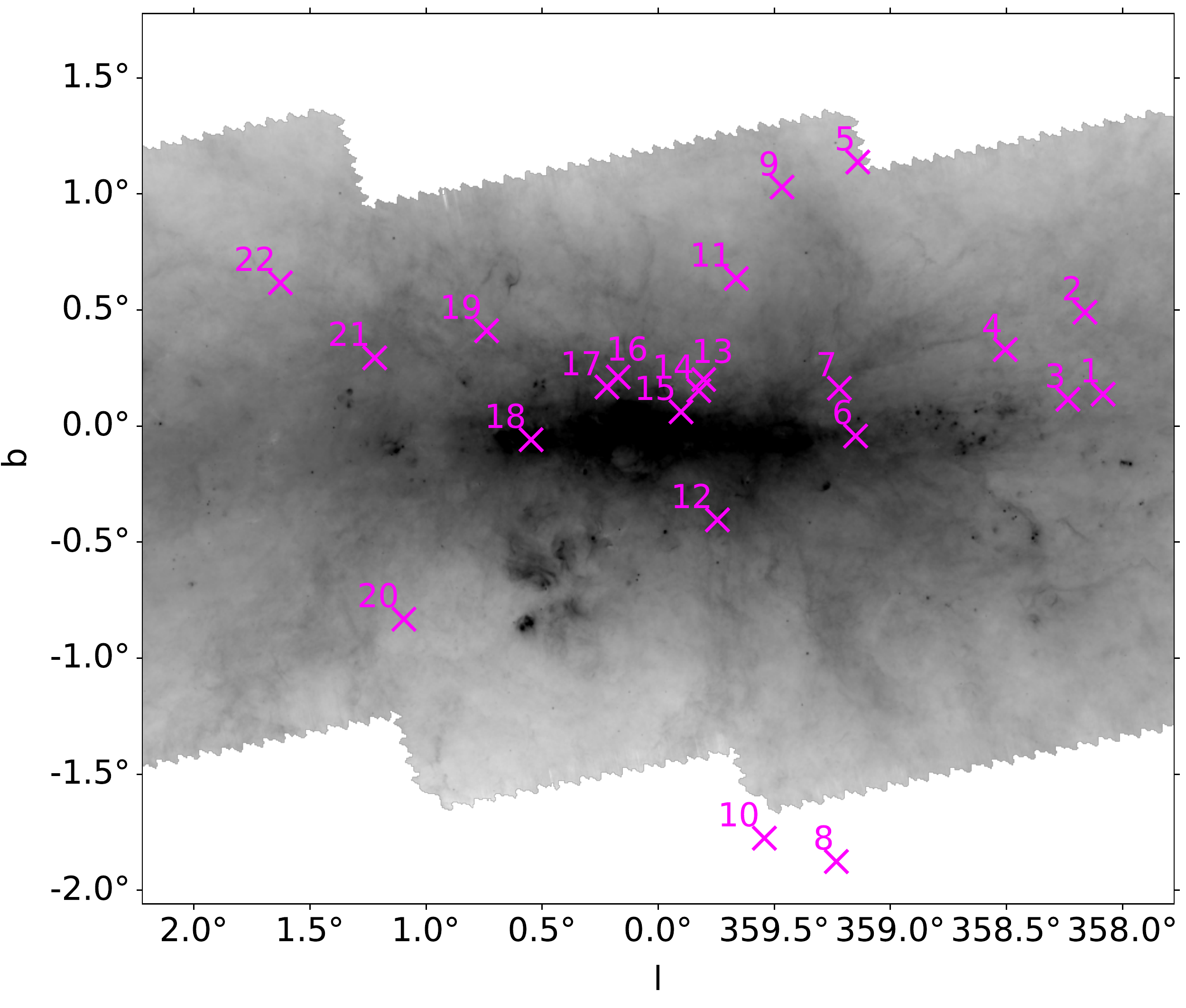}
    \caption{Positions of the sources marked as crosses on a 70\,$\mu$m image obtained with the PACS instrument on the Herschel Space Observatory. The numbers refer to the source identifications in Table~\ref{t:sample}. 
    }
   \label{f:sample_70mu}
\end{figure}

The objects were selected among the OH/IR stars detected in OH 1612\,MHz emission by \citet{lindetal92a}, \citet{seveetal97a}, and \citet{sjouetal98} within about 2$^\circ$ of the GC (corresponding to $\la$\,300\,pc), that is the inner GB. Stars in directions where the CO($J$\,=\,2--1) line emission obtained by \citet{sawaetal01} exceeds 2.2\,K in intensity were omitted in order to minimise the effect of interstellar CO line emission along the line of sight. For the rest, GRAMS models of M-type AGB stars \citep{sargetal11} were fitted to extinction-corrected spectral energy distributions (SEDs; the SEDs and the method of extinction correction were different than the ones eventually adopted in this paper)\footnote{Photometric data for the SEDs were taken from 2MASS ($J$, $H$, and $K_{\rm s}$), Spitzer-IRAC (3.6, 4.5, 5.8, and 8.0\,$\mu$m), Spitzer-MIPS (24\,$\mu$m), WISE bands W1 (3.4\,$\mu$m), W2 (4.6\,$\mu$m), W3 (12\,$\mu$m), and W4 (22\,$\mu$m), and Akari (9 and 18\,$\mu$m). Extinction-correction was performed in the following way: all targets were assumed to have an intrinsic $H-K$ colour of 0.25, as assumed for red giants in the fields observed by \citet{woodetal98}. If a given target had $H$ and $K$ photometry (from 2MASS; in a few cases DENIS $K_{\rm s}$ data were used), $E(H-K)$ could be computed directly; if not, the average of the values where $E(H-K)$ could be computed directly was adopted. The extinction in $K$ was estimated from $A_{\rm K}$\,=\,1.5$E(H-K)$ \citep{woodetal98}. Finally, the extinction law of \citet{mccl09} was used to determine the extinction at all the other photometric bands for objects where $A_{\rm K}$\,$>$\,1.}. The 22 objects with the largest dust production rates were eventually selected to form the sample. The objects are listed with their OH/IR identifications in Table~\ref{t:sample}, and they are shown overlaid on a 70\,$\mu$m image in Fig.~\ref{f:sample_70mu} (the numbering of the sources introduced here is used in the tables throughout this paper). In this way, the sample is biased towards the highest-mass-loss-rate OH/IR objects in the inner GB that are not severely affected by interstellar extinction (as a consequence only five of our objects lie within 0.3$^\circ$ of the GC, corresponding to $\la$\,40\,pc). Our sources must be regarded as the `tip of the iceberg' since \citet{seveetal97a} found about 80 OH/IR objects within 2$^\circ$ of the GC, and \citet{lindetal92a} and \citet{sjouetal98} in much deeper surveys found about 190 such objects within 0.3$^\circ$ of the GC.

We have adopted a distance to the GC of 8.2\,kpc, based on the recent estimate of the distance to the black hole in the GC by the Gravity Collaboration \citep{gravcoll19}. This measurement has a very small formal error (8178\,$\pm$\,13$_{\rm stat}$\,$\pm$\,22$_{\rm sys}$\,pc), but we note that our sources will be spread out along the line of sight even if they are located in the inner GB. None of our sources have a measured reliable parallax in Gaia EDR3 \citep{gaiaetal20}.

\begin{table}[t]
\caption{Sample sources.}
\begin{tabular}{l c c c }
\hline \hline
Source\,$^a$             &  $\alpha$(J2000)\,$^b$  & $\delta$(J2000)\,$^b$ &  \\
                               & [h:m:s]                 & [$^\circ$:\arcmin:\arcsec] \\
\hline 
1\,-\,\phantom{0}\object{OH358.083+0.137}      &  17:40:26.49            & --30:29:38.9 &  \\
2\,-\,\phantom{0}\object{OH358.162+0.490}      &  17:39:14.93            & --30:14:24.3 \\
3\,-\,\phantom{0}\object{OH358.235+0.115}      &  17:40:54.14            & --30:22:38.1 \\
4\,-\,\phantom{0}\object{OH358.505+0.330}      &  17:40:43.37            & --30:02:04.7\\
5\,-\,\phantom{0}\object{OH359.140+1.137}      &  17:39:07.70            & --29:04:03.0\\
6\,-\,\phantom{0}\object{OH359.149$-$0.043}    &  17:43:44.96            & --29:41:00.9\\
7\,-\,\phantom{0}\object{OH359.220+0.163}      &  17:43:06.74            & --29:30:56.9\\
8\,-\,\phantom{0}\object{OH359.233$-$1.876}    &  17:51:12.13            & --30:33:40.5 \\
9\,-\,\phantom{0}\object{OH359.467+1.029}      &  17:40:20.32            & --28:50:55.3 \\
10\,-\,\object{OH359.543$-$1.775}              &  17:51:32.08            & --30:14:36.3  \\
11\,-\,\object{OH359.664+0.636}                &  17:42:20.47            & --28:53:21.4 \\
12\,-\,\object{OH359.745$-$0.404}              &  17:46:35.66            & --29:21:51.5 \\
13\,-\,\object{OH359.805+0.200}                &  17:44:22.27            & --28:59:54.5 \\
14\,-\,\object{OH359.826+0.153}                &  17:44:36.46            & --29:00:19.8 \\
15\,-\,\object{OH359.902+0.061}                &  17:45:08.81            & --28:59:16.7 \\
16\,-\,\object{OH0.173+0.211}                  &  17:45:12.43            & --28:40:44.4  \\
17\,-\,\object{OH0.221+0.168}                  &  17:45:29.41            & --28:39:35.6 \\
18\,-\,\object{OH0.548$-$0.059}                &  17:47:08.98            & --28:29:56.3 \\
19\,-\,\object{OH0.739+0.411}                  &  17:45:46.66            & --28:05:29.3 \\
20\,-\,\object{OH1.095$-$0.832}                &  17:51:26.77            & --28:25:37.1 \\
21\,-\,\object{OH1.221+0.294}                  &  17:47:21.74            & --27:44:23.9 \\
22\,-\,\object{OH1.628+0.617}                  &  17:47:03.91            & --27:13:30.9 \\                        
\hline
\end{tabular}
\label{t:sample}
\tablefoot{$^{(a)}$ Source names are taken from \citet{lindetal92a}, \citet{seveetal97a}, or \citet{sjouetal98}. The SIMBAD identifiers for \object{OH359.805+0.200} and \object{OH359.826+0.153} are \object{OH359.80+0.20} and \object{OH359.800+0.165}, respectively. $^{(b)}$ Coordinates determined from the $^{13}$CO($J$\,=\,\mbox{3--2}) data presented in this paper, except for \object{OH359.149$-$0.043} and \object{OH0.173+0.211} where the  near-IR and $^{12}$CO($J$\,=\,\mbox{2--1}) positions are given, respectively.}
\end{table}

%
%
%
%
\section{Observational data}

\subsection{ALMA data}
\label{s:alma_obs}

The ALMA dataset consists of three different tunings to cover the $^{12}$CO $J$\,=\,\mbox{2--1} line in Band~6 ($\approx$\,230\,GHz) and the $^{12}$CO and $^{13}$CO $J$\,=\,\mbox{3--2} lines in Band~7 ($\approx$\,345 and 330\,GHz, respectively). The $^{12}$CO tunings were observed in one execution each in October 2016, with a fast snapshot observation of about 12\,s on each of the 22 targets. The $^{13}$CO tuning was observed three times, twice in October 2016 and once in July 2017, with a longer integration time of about 50\,s per target for each execution. The total amount of on-source observing time was only about an hour for the whole project. A journal of the observations is given in Table~\ref{t:journal}.

\begin{table*}
\caption{Journal of the observations.}
\label{t:journal}
\begin{tabular}{lcccccc}
 \hline \hline
Band & Line & Date of      & $N_{\rm ant}$\,$^{(a)}$   & $B_{\rm min}$ / $B_{\rm max}$\,$^{(b)}$ & PWV\,$^{(c)}$ & Flux calibrator\,$^{(d)}$   \\
     &      & observation  &                           & [m / km]                           & [mm]        & [Jy] \\
    \hline 

B6    & $^{12}$CO($J$\,=\,2--1) & 04 Oct 2016 & 41 & 18 / 3.1 & 0.8--1.3 & J1924$-$2914 ($3.45 \pm 0.06$ ) \\ 
B7(a) & $^{12}$CO($J$\,=\,3--2) & 11 Oct 2016 & 41 & 18 / 3.1 & 1.2--1.3 & J1733$-$1304 ($1.0\phantom{0} \pm 0.2\phantom{0}$ ) \\ 
B7(b) & $^{13}$CO($J$\,=\,3--2) & 13 Oct 2016 & 41 & 18 / 3.1 & 0.6--0.8 & J1924$-$2914 ($2.68 \pm 0.07$ ) \\ 
      &               & 15 Oct 2016 & 42 & 18 / 1.8 & 0.7--0.9 & J1924$-$2914 ($2.99 \pm 0.58$ ) \\ 
      &               & 04 Jul 2017 & 45 & 21 / 2.6 & 0.5--0.6 & J1733$-$1304 ($1.24 \pm 0.05$ ) \\ 
\hline
\end{tabular}
\tablefoot{$^{(a)}$ Number of 12\,m antennas in the array. $^{(b)}$ Minimum and maximum projected baselines. $^{(c)}$ Amount of precipitable water vapour in the atmosphere. $^{(d)}$ Flux density at the corresponding CO line frequency as retrieved from the ALMA flux monitoring database.} 
\end{table*}

The observations were performed in array configurations with baselines ranging between $\approx$\,20\,m to 2\,$-$\,3\,km. The achieved synthesised beams ($\theta_{\rm b}$, the full width at half maximum, FWHM) vary, but are typically about 0\farcs16\,$-$\,0\farcs18 in Band~6 and 0\farcs11\,$-$\,0\farcs17 in Band~7. The maximum recoverable angular scales (MRSs) are $\approx$\,2\farcs5 (the 230\,GHz data), $\approx$\,1\farcs5 (the 330\,GHz data), and $\approx$\,1\arcsec\ (the 345\,GHz data).

For all tunings, the correlator was set up with one 1.875\,GHz spectral window centred on the relevant CO line, with 1920 channels separated by 0.977\,MHz, and three additional 2\,GHz spectral windows with 128 channels and a coarser spacing of 15.625\,MHz for continuum measurements (these spectral windows have no strong spectral lines that will affect the flux density estimates). For the continuum this means three measurements with 6\,GHz bandwidth centred at the effective frequencies of 222, 324, and 339 GHz for each source. After Hann smoothing\footnote{A smoothing procedure introduced by the Austrian meteorologist Julius von Hann; often referred to as hanning smoothing in astronomical literature.}, the native velocity resolution for the CO spectral windows is between 0.8 and 1.3\,km\,s$^{-1}$. For the continuum windows, the velocity resolution is between 27 and 40\,km\,s$^{-1}$. For analysis, the final CO line velocity resolution was set to 5\,km\,s$^{-1}$.

The calibration was done following the ALMA standard procedures. The bandpass response of the antennas was calibrated on the quasar J1924$-$2914. The flux calibration was done using the quasars J1924$-$2914 or J1733$-$1304, which are regularly monitored by ALMA. The flux accuracy is expected to be within the 5\,$-$\,20\,\% range (see Table~\ref{t:journal}). The gain (phase and amplitude) calibration  was always performed on the quasar J1744$-$3116.

The data were reduced using different versions of the Common Astronomy Software Applications package [CASA; \citet{mcmuetal07}]. After corrections for the time and frequency dependence of the system temperatures, and rapid atmospheric variations at each antenna using water vapour radiometer data, bandpass and gain calibration were done. Imaging was done using the CASA {\tt tclean} algorithm. The final line images were created using Briggs weighting with a robust parameter of 0.5. This resulted in beam sizes of 0\farcs19 and 0\farcs14 for the Bands~6 and 7 data, respectively. The $^{13}$CO $J$\,=\,\mbox{3--2} data have the highest sensitivity, the rms noise per channel in the images is $\approx$\,4\,mJy\,beam$^{-1}$ at 5\,km\,s$^{-1}$ resolution. The corresponding values for the $^{12}$CO $J$\,=\,\mbox{2--1} and \mbox{3--2} data are $\approx$\,9 and 24\,mJy\,beam$^{-1}$, respectively. The continuum rms values are $\approx$\,0.23\,mJy\,beam$^{-1}$, 0.13\,mJy\,beam$^{-1}$, and 0.65\,mJy\,beam$^{-1}$ at 222, 324, and 339\,GHz, respectively. 

The presence of interstellar CO line emission along the line of sight occasionally makes it difficult to isolate the circumstellar emission. In such cases, we rejected visibility data from the shortest baselines, up to 50\,m. Eliminating the shortest baselines, $\le$\,50\,m (and those are in the far minority of all the baselines) will effectively suppress any extended emission on scales larger than 5$-$10\arcsec . Excluding data from baselines longer than this leads to a loss of flux density and poorer S/N. The effect of interstellar emission contamination is less for the $^{13}$CO($J$\,=\,\mbox{3--2}) line due to the fact that the circumstellar $^{13}$CO/$^{12}$CO abundance ratios of O-type AGB stars are in general (much) higher than the interstellar values \citep{milaetal05,ramsolof14}. In the end, omission of short-baseline data were used for four objects in this line. In the case of the $^{12}$CO lines this method was used for nine sources, but here the isolation of the circumstellar emission was guided by the $^{13}$CO results.

The determinations of source positions and sizes of the line brightness distributions are done based on images integrated over the velocity range of the circumstellar line. For the majority of the sources, the line brightness distribution is centrally peaked and eventually fades out into the noise. Hence, the sizes of the molecular envelopes cannot be determined from our data, and we report the FWHM (deconvolved with the synthesised beam) of a two-dimensional Gaussian fitted to the velocity-integrated brightness distribution. The location of the peak of the Gaussian fit gives the source position. We estimate the astrometric accuracy to be better than 0\farcs1.

The line profiles were extracted over an aperture, centred on the peak of the brightness distribution, larger than the estimated FWHM of the brightness distribution, normally 2--3 times larger. The size is set to recover as much as possible of the line flux density, while maintaining an as high as possible S/N and suppressing the influence from the interstellar CO line emission. The uncertainties in the flux densities are affected by a combination of the S/N, the presence of interstellar CO line emission, and the size of the aperture used, and thus difficult to determine in detail. We estimate for the $^{13}$CO($J$\,=\,\mbox{3--2}) emission that for lines stronger than 0.2\,Jy the uncertainty is about 10\,\%, while for lines weaker than 0.1\,Jy the uncertainty increases to about 50\,\% for the weakest lines. To this should be added the possible amount of resolved-out flux as discussed in Sect.~\ref{s:massloss}. For the $^{12}$CO lines the uncertainty in the flux density estimate is about 10\,\% for lines stronger than 1.0 and 2.0\,Jy for the $J$\,=\,\mbox{2--1} and \mbox{3--2} lines, respectively, and below 0.4 and 1.0\,Jy the uncertainties increase to about 50\,\% for the weakest $J$\,=\,\mbox{2--1} and \mbox{3--2} lines, respectively. Flux densities, centre velocities (in the local standard of rest frame), and expansion velocities were determined through fitting the line shape function
\begin{equation}
S(\varv) = S(\varv_{\rm c})\, \left[ 1 - \left( \frac{\varv - \varv_{\rm c}}{\varv_\infty} \right)^2  \right]^\beta
\end{equation}
to the data, so that $\beta$\,=\,1 means a parabolic, $\beta$\,=\,0 a flat-topped, and $\beta$\,$<$\,0 a double-peaked line shape. The uncertainties in the expansion and centre velocities depend on the S/N and the presence of interstellar CO line contamination, and are estimated to be about $\pm$1.5 and $\pm$1\,km\,s$^{-1}$, respectively for the $^{13}$CO($J$\,=\,\mbox{3--2}) data. The corresponding values for the $^{12}$CO($J$\,=\,\mbox{2--1}) and $^{12}$CO($J$\,=\,\mbox{3--2}) data are about $\pm$2 and $\pm$1.5\,km\,s$^{-1}$ and about $\pm$3 and $\pm$2\,km\,s$^{-1}$, respectively.

The $^{12}$CO($J$\,=\,\mbox{3--2}) setting also allowed detection of the H$^{13}$CN($J$\,=\,\mbox{4--3}) line at 345.3398\,GHz. Three sources were detected, and the line profile results were extracted in the same way as for the CO lines. The uncertainties in the line characteristics are the same as those of the $^{12}$CO($J$\,=\,\mbox{3--2}) line. 

The continuum brightness distributions are sharply peaked and we determine their sizes, positions, and flux densities (and their associated uncertainties) by fitting  two-dimensional Gaussians to the data in the image plane. The results are summarised in Table~\ref{t:continuum}. There is likely to exist also low-surface-brightness, extended dust emission, but it is effectively resolved out by the interferometer due to the small MRSs of our data, about 1\,$-$\,2.5\arcsec . Since the brightness distributions for the detected sources are comparable to the size of the synthesised beam, we use the rms flux density per beam in the image to estimate the upper limit for a non-detection.

%
 
\subsection{Photometry and variability}
\label{s:photo}

We gathered photometric data from archives through the VizieR Photometry viewer and the NASA/IPAC infrared science archive. The specific catalogues used were the ISOGAL Point Source Catalogue \citep{omonetal03}, the Midcourse Space Experiment Point Source Catalog Version 2.3 \citep{eganetal03}, OH-selected AGB and post-AGB objects \citep{seve02}, the UKIDSS-DR6 Galactic Plane Survey \citep{lucaetal08}, the MIPSGAL 24\,$\mu$m point source catalogue \citep{guteheye15}, AllWISE \citep{cutretal13}, the Pan-STARRS release 1 (PS1) Survey - DR1 \citep{chametal16}, the Hi-GAL inner Milky Way: +68\,$\ge$\,$\ell$\,$\le$\,+70 \citep{molietal16}, the Hi-GAL compact source catalogue \citep{eliaetal17}, the IRAS PSC/FSC Combined Catalogue \citep{abraetal15}, Catalog of 24\,$\mu$m sources toward Galactic Center \citep{hinzetal09}, NOMAD Catalog \citep{zachetal05}, the Galactic Legacy Infrared Midplane Survey Extraordinaire \citep{spitscie09}, AKARI/IRC mid-IR all-sky Survey \citep{ishietal10}, and the Herschel/PACS Point Source Catalogs \citep{hpsc20}. The photometry data used are available at the CDS as tables for each source containing the wavelength, the flux density, its uncertainty, and the reference.

Since the stars in our sample are in the inner GB, interstellar extinction can be very high. We have retrieved the reddening $E(J-K_{\rm s})$ towards each source using the results of \citet{gonzetal18}. The reddening is converted into extinction in the $K_{\rm s}$ band, $A_{K_{\rm s}}$ given in Table~\ref{t:sed_summary}, and extrapolated up to 8\,$\mu$m using the relationships provided by \citet{nishetal09}. In the range 8 to 30\,$\mu$m we use the results of \citet{wangetal15} and \citet{xueetal16}. No corrections were applied for data at wavelengths longer than 30\,$\mu$m because $A_{\lambda}/A_{K_{\rm s}}$ is expected to be lower than 0.2, which would imply relatively small corrections even for the higher-extinction sources. 
 
Pulsation periods have been estimated using data from the VISTA Variables in the V\'{i}a L\'{a}ctea (VVV) ESO Public Survey \citep{minnetal10}, and data from the combined Wide-field Infrared Survey Explorer (WISE) \citep{wrigetal10} and the Near-Earth Object WISE (NEOWISE \citep{mainetal11,mainetal14}. Using preliminary coordinates for the sources the NIR counterpart was identified in the VVV images. In all cases the coordinates of the NIR counterpart coincided with the ALMA counterpart at the level of 0\farcs1. For the counterpart, the available data in the $Z$ (0.88\,$\mu$m), $Y$ (1.02\,$\mu$m), $J$ (1.25\,$\mu$m), $H$ (1.64\,$\mu$m), and $K_{\rm s}$ (2.15\,$\mu$m) filters were retrieved. In $ZYJH$ these are typically four or less observations, while dozens of data points are available in $K_{\rm s}$. The $K_{\rm s}$-band light curves have been analysed using the codes and methodology outlined in \citet{groeetal20} assuming a single period. For the WISE data a very similar approach was adopted to analyse the light curves in the W1 (3.4\,$\mu$m) and W2 (4.6\,$\mu$m) bands, as outlined in \citet{groe22}. The results, including the adopted period are collected in Table~\ref{t:summary_periods} that also cites periods found in the literature.

Pulsation periods could be determined for 13 objects. In one case, the period was taken from the literature. One object shows non-Mira multi-periodic variability. In a few stars saturation in the $K$ or WISE bands are an issue, but only in one case did this mean that a period could not be determined. The remaining six stars show scatter beyond what is expected based on the error in the individual measurements, but none of them show a periodic behaviour, that is they are classified as non-periodic.

\begin{sidewaystable*}
\caption{Pulsation periods of the sources.}
\begin{tabular}{lrrrrrrllrl} 
\hline \hline 
Source                       &  $P_{\rm K_s}$  & $\sigma_{\rm K_s}$  &  $P_{\rm W1}$  &  $\sigma_{\rm W1}$  & $P_{\rm W2}$  &  $\sigma_{\rm W2}$  &  $P_{\rm literature}$  & Ref. &  $P_{\rm adopted}$  &  Comment  \\
                             &  [days]       &  [days]           &  [days]    &  [days]             &  [days]   &  [days]             &  [days]                &      &  [days]             &  \\
  \hline
1\,-\,\phantom{0}\object{OH358.083+0.137}              & $\ldots$      &     $\ldots$      &   $\ldots$ &  $\ldots$           &  $\ldots$ &    $\ldots$         &   $\ldots$       &            &      $\ldots$    & scatter. not periodic      \\
2\,-\,\phantom{0}\object{OH358.162+0.490}              & $\ldots$      &     $\ldots$      &     1228   &    29\phantom{.0}   &     1272  &    60\phantom{.0}   & 1150, 1120       &   2, 3     &        1230       & scatter in $K$ \\  
3\,-\,\phantom{0}\object{OH358.235+0.115}              & $\ldots$      &     $\ldots$      &   $\ldots$ &  $\ldots$           &  $\ldots$ &    $\ldots$         &     874          &   4        &         874       & saturation in $K$ and WISE \\ 
4\,-\,\phantom{0}\object{OH358.505+0.330}              & $\ldots$      &     $\ldots$      &   $\ldots$ &  $\ldots$           &  $\ldots$ &    $\ldots$         &   $\ldots$       &            &       $\ldots$     & scatter. not periodic   \\
5\,-\,\phantom{0}\object{OH359.140+1.137}              & $\ldots$      &     $\ldots$      &     1219   &   104\phantom{.0}   &      989  &    19\phantom{.0}   &   $\ldots$       &            &        1000       & very low-amplitude variability      \\
6\,-\,\phantom{0}\object{OH359.149$-$0.043}            & $\ldots$      &     $\ldots$      &   $\ldots$ &  $\ldots$           &  $\ldots$ &    $\ldots$         &   $\ldots$       &            &       $\ldots$    & scatter. not periodic \\
7\,-\,\phantom{0}\object{OH359.220+0.163}              & $\ldots$      &     $\ldots$      &   $\ldots$ &  $\ldots$           &  $\ldots$ &    $\ldots$         &   $\ldots$       &            &       $\ldots$    & scatter in $K$. too few datapoints in WISE \\
8\,-\,\phantom{0}\object{OH359.233$-$1.876}            & $\ldots$      &     $\ldots$      &   $\ldots$ &  $\ldots$           &  $\ldots$ &    $\ldots$         &   $\ldots$       &            &       $\ldots$    & scatter in $K$. too few datapoints in WISE \\
9\,-\,\phantom{0}\object{OH359.467+1.029}              &    604        &        3          &      604   &     1.3             &      605  &     2.4             &   $\ldots$       &            &         605       &  \\
10\,-\,\object{OH359.543$-$1.775}            &   $\ldots$    &    $\ldots$       &   $\ldots$ &    $\ldots$         &  $\ldots$ &     $\ldots$        &     790          &   5        &      $\ldots$     & multi-periodic, non-Mira \\
11\,-\,\object{OH359.664+0.636}              &   $\ldots$    &    $\ldots$       &      538   &     3.3             &      581  &     4.3             &   $\ldots$       &            &         550       & scatter in $K$ \\
12\,-\,\object{OH359.745$-$0.404}            &    733        &        4          &      714   &     4.4             &      732  &     5.0             &  806, 766        &   1, 6     &         730       & \\
13\,-\,\object{OH359.805+0.200}              &    533        &       10          &      488   &     3.8             &      501  &     6.5             &   $\ldots$       &            &         500       &  \\
14\,-\,\object{OH359.826+0.153}              &    501        &        3          &      483   &     2.1             &      489  &     4.3             &  493, 510        &   1, 6     &         500       & \\
15\,-\,\object{OH359.902+0.061}              &    586        &       13          &      611   &     3.4             &      588  &     5.1             &  558, 572, 550, 580  & 1, 7, 7, 8 &     590       &  \\ 
16\,-\,\object{OH0.173+0.211}                &    512        &        4          &      510   &     5.5             &      516  &     6.8             &   $\ldots$       &            &         512       & \\
17\,-\,\object{OH0.221+0.168}                &    664        &        4          &      663   &     4.7             &      662  &     3.6             &      697         &     1      &         663       & \\
18\,-\,\object{OH0.548$-$0.059}            &   $\ldots$      &    $\ldots$       &   $\ldots$ &    $\ldots$         &  $\ldots$ &     $\ldots$        &   $\ldots$       &            &      $\ldots$     & scatter. not periodic \\
19\,-\,\object{OH0.739+0.411}                &    577        &       10          &      557   &     2.8             &      602  &     2.8             &   $\ldots$       &            &         570    &  \\
20\,-\,\object{OH1.095$-$0.832}            &   $\ldots$    &      $\ldots$     &     $\ldots$ &    $\ldots$         &  $\ldots$ &     $\ldots$        &   $\ldots$       &            &      $\ldots$     & saturated          \\
21\,-\,\object{OH1.221+0.294}                &    699        &        3          &      715   &     2.0             &      715  &     6.8             &   $\ldots$       &            &         715       &  \\
22\,-\,\object{OH1.628+0.617}              &    921        &       43          &      827   &     6.5             &      839  &     8.0             &     $\ldots$     &            &         830    &  \\
\hline
\end{tabular} 
\tablefoot{
  Period with error in the $K_{\rm s}$ (2.15\,$\mu$m) (columns 2, 3), $W1$ (3.4\,$\mu$m) (columns 4, 5), and  $W2$ (4.6\,$\mu$m) (columns 6, 7) bands.
  Column 9: references for periods in the literature quoted in column 8:
  (1)= \citet{woodetal98}, (2)= \citet{lebe93}, (3)= \citet{olivetal01},
  (4)= present work, data from the Bochum Galactic Disk Survey \citep{hacketal15} for source GDS\_J1740541-302238,
  (5)= present work, data from OGLE \citep{soszetal13} for source OGLE-BLG-LPV-074942 classified as a semi-regular variable with principal period of 78.2 days,
  (6)= \citet{bragetal19} (also based on VVV data, but fewer data points than in the present study),
  (7)= \cite{glasetal01} (two periods from data in overlapping fields), (8)= \citet{matsunetal09}.
}
\label{t:summary_periods}
\end{sidewaystable*}
%
%
%
%
\section{Observational results}

\begin{figure*}
   \includegraphics[width=16cm]{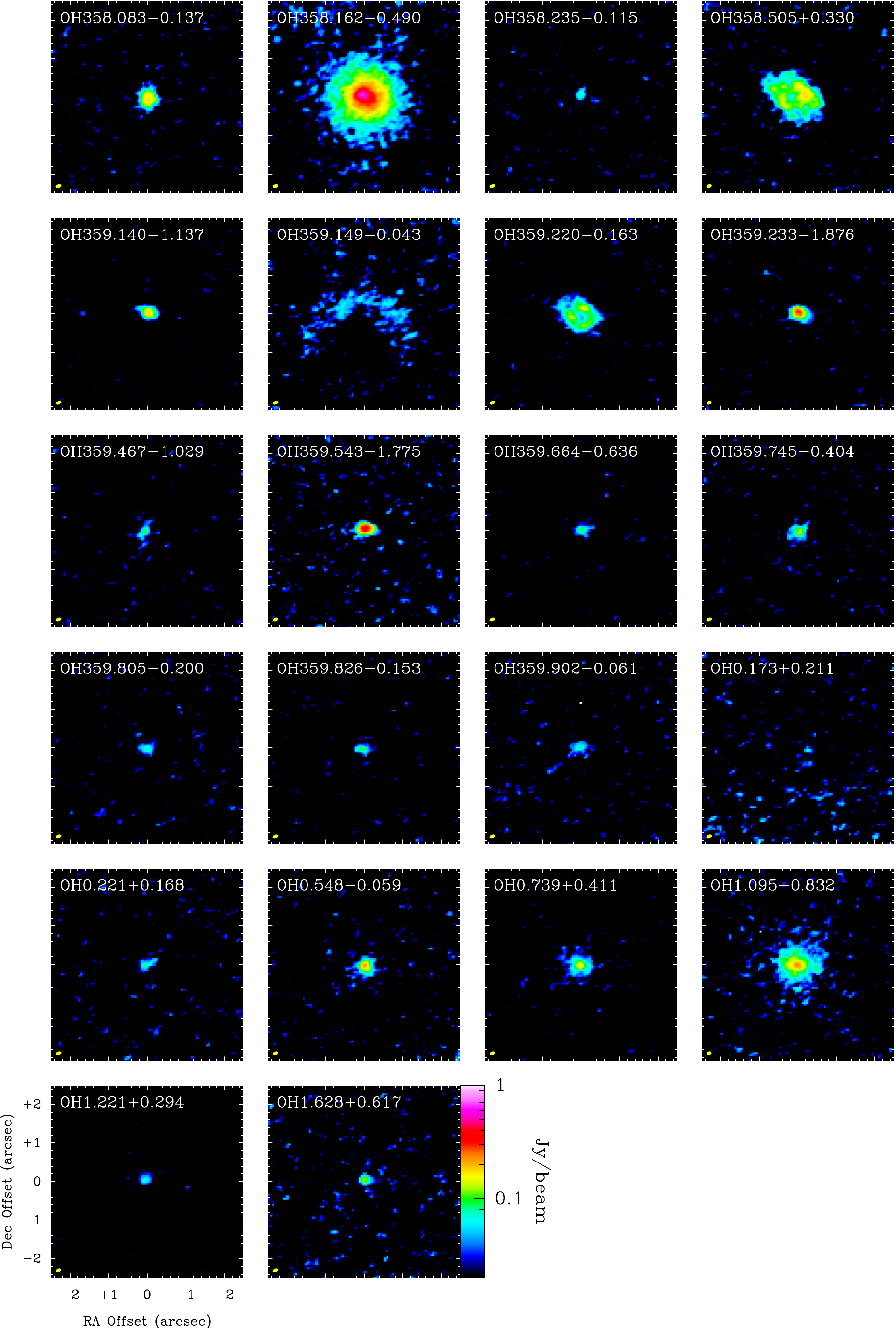}
    \caption{$^{13}$CO($J$\,=\,\mbox{3--2}) brightness distribution integrated over the line for all sources in the sample. A logarithmic scale is used in order to have the same scale for all objects. The synthesised beams are shown in the lower left corner of each panel.
    }
   \label{f:13co_images}
\end{figure*}   

\begin{figure*}
\includegraphics[width=18cm]{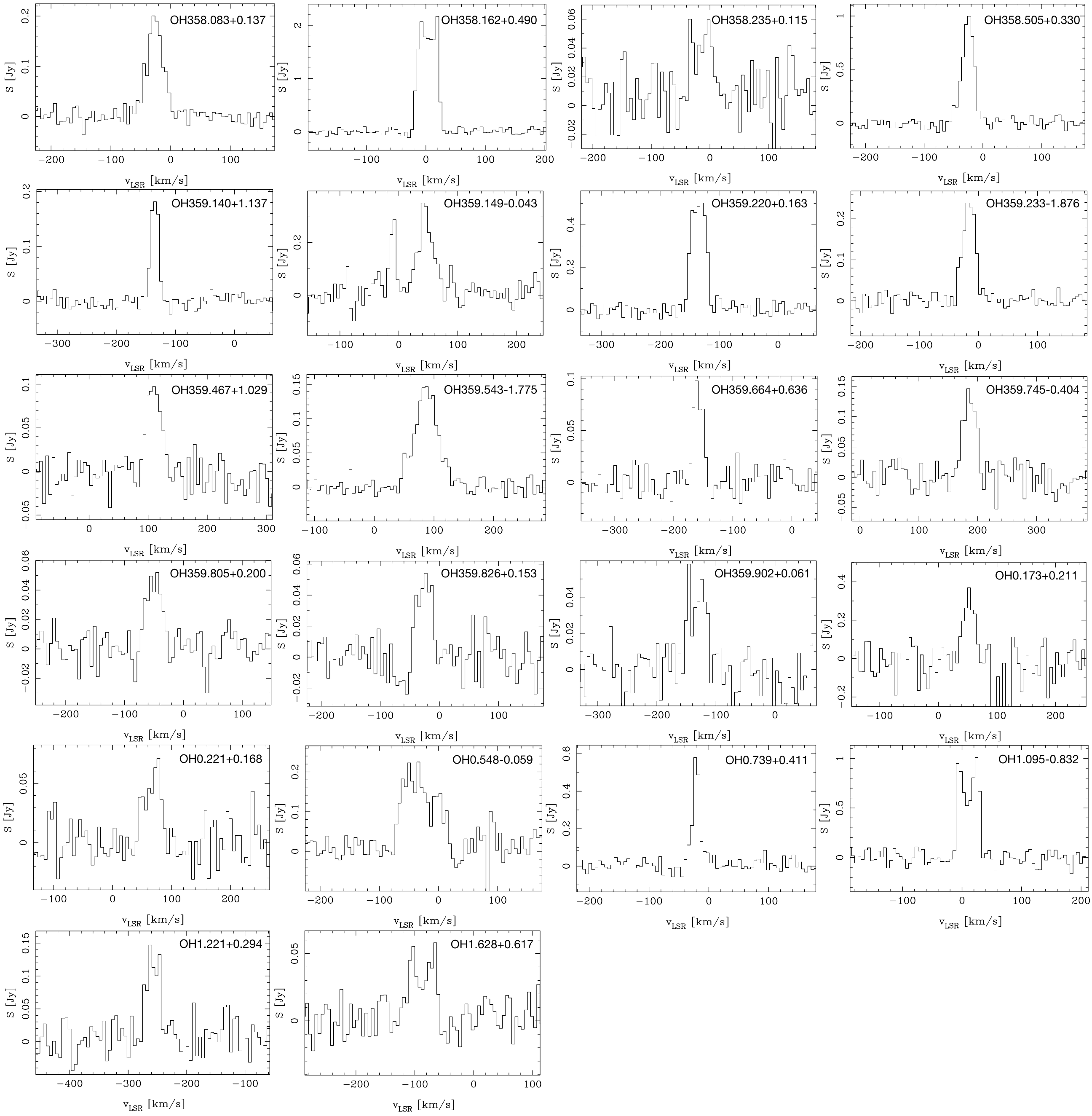}
    \caption{$^{13}$CO($J$\,=\,\mbox{3--2}) spectra at a velocity resolution of 5\,km\,s$^{-1}$, except for \object{OH0.173+0.211} where the $^{12}$CO($J$\,=\,\mbox{2--1}) spectrum is shown. 
    }
   \label{f:13co_spectra}
\end{figure*}   

%
\subsection{The $^{12}$CO and $^{13}$CO lines}

The $^{13}$CO($J$\,=\,\mbox{3--2}) data have the highest S/N and are less affected by contamination from interstellar CO line radiation (as explained above). Consequently, we start the discussion with this line in which we detect 21 (out of 22) sources. Figure~\ref{f:13co_images} shows the $^{13}$CO($J$\,=\,\mbox{3--2}) images (obtained by integrating the brightness distribution over the velocity range of the line) and Fig.~\ref{f:13co_spectra} shows the spectra. For the majority of the sources (19), the brightness distributions are centrally peaked and no source structure is discernible. One source, OH359.149$-$0.043, shows a very strange brightness distribution (which is repeated also in the two $^{12}$CO lines). Two sources with resolved structure are discussed below. The observational results are summarised in Appendix~\ref{a:tables}.

The $^{12}$CO $J$\,=\,\mbox{2--1} and \mbox{3--2} lines are detected towards 22 and 20 sources, respectively. In quite a number of these cases, the detections would have been characterised as tentative if it were not for the detection of the $^{13}$CO($J$\,=\,3--2) line. \object{OH0.173+0.211} was detected in only the $^{12}$CO($J$\,=\,2--1) line\footnote{The position used for the ALMA observations, the one given by \citet{sjouetal98}, is about 6\arcsec\ north of the position of the OH source \citep{lindetal92a}, but still within the half power width of the primary beam.}. Its brightness distribution is centrally peaked and no source structure is discernible. Tables with the observational results are presented in Appendix~\ref{a:tables}. The images and spectra for these lines are shown in Appendix~\ref{a:images} and \ref{a:spectra}, respectively.

The $^{13}$CO positions are given in Table~\ref{t:sample}, and their uncertainties (including the astrometric uncertainty, Sect.~\ref{s:alma_obs}) are estimated to be 0\farcs1 (3$\sigma$) for the highest S/N data, increasing to 0\farcs2 (3$\sigma$) for the poorest S/N data. All brightness distributions peak in the vicinity of a near-IR source in the VVV data. The median and maximum offsets between the $^{13}$CO($J$\,=\,\mbox{3--2}) and near-IR positions are 0\farcs11 and 0\farcs23, respectively. The only offset larger than the combined uncertainties is obtained for \object{OH358.505+0.330}, the source with the largest resolved structure in the CO lines. The median and maximum offsets between the $^{13}$CO($J$\,=\,\mbox{3--2}) and the $^{12}$CO($J$\,=\,\mbox{2--1} and \mbox{3--2}) positions are 0\farcs09 and 0\farcs26 and 0\farcs07 and 0\farcs20, respectively. Half of the $^{13}$CO($J$\,=\,\mbox{3--2}) positions are within 0\farcs5 of the OH position, all but two are within 1\farcs0 of the OH position, and the largest offset between the two is 2\farcs2. The uncertainties in the OH positions are significantly larger than those of our CO line data.

The estimated sizes of the $^{13}$CO($J$\,=\,3--2) brightness distributions ($\theta_{\rm s}$) lie in the range 0.2--1\arcsec, that is they are all resolved. The $^{12}$CO $J$\,=\,\mbox{2--1} and \mbox{3--2} brightness distributions are only marginally larger. For the sources located at the distance of the GC, 1\arcsec\ corresponds to $\approx$\,1.2$\times$10$^{17}$\,cm, or $\approx$\,8200\,au. The implications of these sizes are further discussed in Sect.~\ref{s:phdiss}.

Two sources, \object{OH358.505+0.330} and \object{OH359.220+0.163}, show evidence of resolved structure in the $^{13}$CO($J$\,=\,3--2) brightness distribution. In both cases there appears to be a cavity and a bipolar structure surrounding the star, see Fig~\ref{f:13co_images}. For both sources the line with the, by far, highest S/N, the $^{13}$CO($J$\,=\,3--2) line, shows a line profile expected from circumstellar thermal line emission, see Fig~\ref{f:13co_spectra}. \citet{zijletal89} classified \object{OH358.505+0.330} as a planetary nebula (PN) based on a radio continuum detection at 6\,cm using the Very Large Array (VLA), although they stated that the detection should be repeated due to a low S/N. More recently, \citet{uscaetal12} argued that the claimed radio continuum detection is wrong (by looking at VLA archival data), and hence there is presently no evidence that this object is a PN. For \object{OH359.220+0.163} there exists only OH maser observations in the literature. We note here, that for neither of these two objects do we detect 324\,GHz continuum emission.

\subsection{The H$^{13}$CN($J$\,=\,\mbox{4--3}) line}

The H$^{13}$CN($J$\,=\,\mbox{4--3}) line was detected in three of our sample objects. The results and spectra are presented in Appendix~\ref{a:tables} and \ref{a:spectra}, respectively.

%
\subsection{ALMA continuum measurements}

In total eight sources were detected in 324\,GHz continuum, five were detected at 222\,GHz, and four detected at 339\,GHz, see Table~\ref{t:continuum}. This is fully explained by the lower S/N in the 222 and 339\,GHz data. When detections are made in all three bands, the flux densities are consistent with a spectral index $\alpha$ of about 1.5\,$-$\,3 ($S_\nu$\,$\propto$\,$\nu^\alpha$). Only the upper limits to the 339\,GHz flux densities of \object{OH359.233$-$1.876} and \object{OH0.739+0.411} are surprising, since they are about two times below the flux densities at 324\,GHz. We have no explanation for this. There is nothing obviously wrong with the observations of these particular sources. All objects were observed consecutively at 339\,GHz, in snap-shot mode, using the same source as phase reference.  The images for all sources and frequencies are presented in Appendix~\ref{a:images}. The continuum brightness distributions are significantly smaller than those of the CO lines. Deconvolved FWHMs are comparable to, or smaller than, the beams, $\la$\,0\farcs15. The median and maximum offsets between the $^{13}$CO($J$\,=\,\mbox{3--2}) and 324\,GHz continuum positions are 0\farcs02 and 0\farcs12, respectively (\object{OH359.149$-$0.043} is excluded in the comparison because of its strange line brightness distribution). The largest offset is found for \object{OH358.162+0.490}, a result due to the fact that this source has the apparently largest $^{13}$CO($J$\,=\,\mbox{3--2}) brightness distribution among the sources detected in 324\,GHz continuum. The other offsets are within the combined uncertainties.

A simple calculation shows that the expected continuum flux at 324\,GHz from a 2500\,K blackbody of luminosity 5000\,$L_\odot$ (typical for our sources as shown below) at the distance of the GC would be about 0.04\,mJy. This is well below our detection limit, and it means that any detected continuum emission from sources at that distance must be due to circumstellar dust emission. However, this means that there is an issue with resolved-out continuum flux. The reason is that the larger fraction of the total millimetre/submillimetre flux density comes from extended emission from cold dust, that is from extended low-surface brightness emission which is not seen by the interferometer. The size of the emitting region is likely much larger than the radius of the MRS, $\approx$\,1\arcsec\ at 324\,GHz, corresponding to about 8200\,au if placed at the GC (a distance covered by an 18\,km\,s$^{-1}$ outflow in about 2200\,yr). The high-mass-loss-rate phase may have lasted longer than this, and consequently the dust-CSE may be larger than this. Using the dust radiative transfer modelling (described in Sect.~\ref{s:dustradtrans}) we estimate that the measured ALMA flux densities are in general, at least, a factor of two below the true flux densities, that is the reported ALMA flux density is a conservative lower limit to the total flux density for the detected sources.

%
%
%
\section{Radiative transfer}

%
\subsection{Dust continuum radiation}
\label{s:dustradtrans}

To reproduce the SED of each source, we calculated radiative transfer models using the code MCMax \citep{minetal09} assuming spherical symmetry of the dust CSEs. We consider astronomical silicate grains \citep{osseetal92} with a number size distribution given by a power law with index $-$3.5 and minimum and maximum grains sizes equal to 0.001\,$\mu$m and 1.0\,$\mu$m, respectively. The spherical dust CSEs were assumed to extend from an inner radius $R_{\rm in}$ to 10000\,au. The former is either determined by the radius at which the dust temperature is 1000\,K (as calculated by the code), or as the required inner radius to give a good fit to the data. The latter corresponds to roughly half the ALMA MRS for our observations at the distance of the GC. In this way the SED models should, in principle, be able to reproduce the observed ALMA fluxes. The particular choice of the outer radius does not affect the estimated emission at wavelengths shorter than $\approx$\,70\,$\mu$m, only that at longer wavelengths. The density distribution is proportional to $r^{-2}$, where $r$ is the radial distance from the central star (corresponding to constant mass-loss rate and expansion velocity). The stellar temperature, $T_\ast$, was set to 2500\,K initially. The free parameters in the fit are $R_{\rm in}$ (where required), $T_\ast$ (where required), and the dust optical depth at 9.7\,$\mu$m ($\tau_{9.7}$).

In the fitting procedure, we only include photometric points for wavelengths larger than 1\,$\mu$m because of the large extinction towards some of the sources and the relatively small contribution to the total luminosity of the emission at optical wavelengths. The SED was fitted by eye by varying mainly $\tau_{9.7}$. When an acceptable fit could not be obtained, we varied $R_{\rm in}$. If necessary, we also increased the stellar temperature to decrease the stellar emission in the near-infrared. 

The dust-mass-loss rate is obtained assuming that the expansion velocity of the dust equals that estimated from the observed CO lines. This is a reasonable assumption, since the drift between the dust and the gas is believed to be limited to at most a few km\,s$^{-1}$ for high-mass-loss-rate objects \citep{kwok75}. The stellar luminosity is obtained by integrating the fit to the SED over wavelength and using the distance to the given source.

%
\subsection{CO line radiation}
\label{s:coradtrans}

For the line emission analysis, we used a well-tested, non-local-thermodynamic-equilibrium radiative transfer code based on the Monte Carlo method \citep{bern79,schoolof01}. We assume a CSE expanding with constant velocity (the terminal gas expansion velocity $\varv_\infty$), which is formed by a constant and isotropic gas-mass-loss rate ($\dot{M}_{\rm g}$). 

We have observational data for only two rotational lines of $^{12}$CO, and this is not enough to put constraints on the radial distribution of the kinetic temperature for each individual source. We have therefore opted for using the same circumstellar model for all sources, the one obtained for the high-mass-loss-rate (10$^{-5}$\,$M_\odot$\,yr$^{-1}$) M-type AGB star \object{IK~Tau} by \citet{ramsolof14}. In this model the kinetic temperature law is given by
\begin{equation}
\label{e:tkin}
T_{\rm k} = 940\,\left[ \frac{4\times10^{14}}{r} \right]^{0.9}\,\,[{\rm K}]\, ,
\end{equation}
where $r$ is given in cm, and with a lower limit of $T_{\rm min}$\,=\,7.5\,K at large radii. In addition to thermal motion of the molecules, we assume a micro-turbulent velocity field of 1\,km\,s$^{-1}$. Likewise, \citet{ramsolof14} derived the following dust temperature law for IK~Tau,
\begin{equation}
T_{\rm d} = 580\,\left[ \frac{4\times10^{14}}{r} \right]^{0.4}\,\,[{\rm K}]\, ,
\end{equation}
where $r$ is given in cm, and the dust optical depth is 0.5 at 10\,$\mu$m. The dust optical properties of \citet{justtiel92} are used. The optical depth is scaled with the mass-loss rate for each star. The particular choices of the dust temperature law and optical depth play no role in our analysis, since our stars have relatively high mass-loss rates and the CO lines are dominantly excited through collisions. For example, changing the optical depth by a factor of 10 up or down leads to non-significant changes ($\la$\,1\,\%) in the CO line intensities for a $\dot{M}_{\rm g}$\,=\,2$\times$10$^{-5}$\,$M_\odot$\,yr$^{-1}$ and $\varv_\infty$\,=\,18\,km\,s$^{-1}$ model (our median values, see below). The inner radius of the CO envelope is chosen to be 4$\times$10$^{14}$\,cm. The resulting line intensities are not sensitive to this choice at the high mass-loss rates relevant for our sample.

The radiation fields considered in the excitation of the CO molecules are those of the star (assumed to be a blackbody at 2500\,K and with a luminosity obtained from the SED fit), the dust (as calculated from the above dust temperature and optical depth), and the cosmic microwave background at 2.7\,K. For the high mass-loss rates of our sample stars, collisional excitation dominates completely and these particular choices play no role.

We assume an initial fractional abundance of $^{12}$CO/H$_2$, $f_{\rm i}$, of 4$\times$10$^{-4}$, which corresponds to about 80\,\% complete formation of CO assuming solar abundances of C and O \citep[see e.g.][]{agunetal20}. The abundance distribution of CO is eventually determined by photodissociation, in lines in the case of this species. We have used the recent results of \citet{sabeetal19} on circumstellar CO photodissociation to calculate the size of the CO envelope. The abundance distribution follows the law
\begin{equation}
f(r) = f_{\rm i}\, \exp[-(\ln 2) (r/R_{1/2})^\alpha],
\end{equation}
where $R_{1/2}$ and $\alpha$ are given by
\begin{equation}
\label{e:rph}
R_{1/2} = 1.68\times10^{17}\,  \left[ \frac{\dot{M}_{\rm g}}{10^{-5}} \right]^{0.57}\, \left[ \frac{\varv_\infty}{15} \right]^{-0.35}\, \left[ \frac{f_{\rm i}}{4\times10^{-4}}  \right]^{0.32}\,\,[{\rm cm}],
\end{equation}
and
\begin{equation}
\alpha = 3.09\, \left[ \frac{\dot{M}_{\rm g}}{10^{-5}} \right]^{0.09}\, \left[ \frac{\varv_\infty}{15} \right]^{-0.13}\, \left[ \frac{f_{\rm i}}{4\times10^{-4}}  \right]^{-0.04}\,,
\end{equation}
respectively, where $\varv_\infty$ is given in km\,s$^{-1}$, and $\dot{M}_{\rm g}$ in $M_\odot$\,yr$^{-1}$. These fits to the \citet{sabeetal19} results are appropriate for the mass-loss-rate, expansion velocity, and fractional abundances ranges relevant for this study. The same formulae have been used for $^{13}$CO with the appropriate abundance. In general, the $J$\,=\,\mbox{2--1} and, in particular, the $J$\,=\,\mbox{3--2} model line emissions are relatively insensitive  to the choice of $R_{1/2}$, discussed further in Sect.~\ref{s:phdiss}. The outer radius of the CSE is set to 3\,$\times$\,$R_{1/2}$.

The CO excitation analysis includes 82 rotational energy levels, up to $J$\,=\,40 within the ground  ($\nu$\,=\,0) and first vibrationally excited ($\nu$\,=\,1) states. Radiative transitions within and between the vibrational states and collisional transitions within the vibrational states are taken into account. The collisional rate coefficients between CO and para- and ortho-H$_2$ are taken from \citet{yangetal10}. An ortho-to-para-H$_2$ ratio of 3 was adopted for weighting the ortho-to-para-H$_2$ coefficients together. The rates cover 25 temperatures between 2 to 3000\,K. The same ranges of transitions and rates are used for both CO isotopologues.

The best-fit circumstellar model for the $^{12}$CO line emission is obtained by minimising the difference between the model and observational results in terms of the velocity-integrated flux densities by adjusting the only free parameters, the gas expansion velocity and the mass-loss rate. The estimate of the $^{13}$CO fractional abundance is based on this model and the requirement that the model and the observational $^{12}$CO/$^{13}$CO $J$\,=\,3--2 velocity-integrated flux density ratios are the same. The $^{13}$CO fractional abundance is the only free parameter in this case. 

The availability of only two $^{12}$CO lines (and their generally low S/N) and the use of the same circumstellar model for all stars make our mass-loss rate estimates particularly uncertain, at least by a factor of five. To this should be added the uncertainties due to the $^{12}$CO abundance and the source distance. Likewise, the estimated $^{12}$CO/$^{13}$CO abundance ratios are uncertain since they are based on only one $^{13}$CO line. However, since this is a ratio estimate it has the advantage of being much less sensitive to the adopted circumstellar model than those of the mass-loss-rate. The ratio is also relatively insensitive to the adopted $^{12}$CO abundance. A decrease/increase of the latter leads to an increase/decrease of the mass-loss rate and therefore a decrease/increase in also the estimated $^{13}$CO abundance. In total, the uncertainty of the $^{12}$CO/$^{13}$CO abundance ratio is about a factor of four.

In fact, there are indications that for most of our objects the adopted circumstellar model is not appropriate. Alternatively, the $^{12}$CO line data have too low S/N. The signs of this are found in the $^{12}$CO $J$\,=\,3--2/2--1 line intensity ratios. In the models, this ratio depends sensitively on the mass-loss rate and to some extent on the kinetic temperature. Adopting our circumstellar model and $\varv_\infty$\,=18\,km\,s$^{-1}$ we find that the $J$\,=\,\mbox{3--2}/\mbox{2--1} line intensity ratio goes from being $>$\,1 to becoming $<$\,1 as the mass-loss rate increases above 3$\times$10$^{-5}$\,$M_\odot$\,yr$^{-1}$. At 10$^{-6}$ and 10$^{-4}$\,$M_\odot$\,yr$^{-1}$ the ratio is 1.5 and 0.6, respectively. A change of the kinetic temperature by $-20$\,\% and $+20$\,\%,with respect to that given in Eq.(\ref{e:tkin}), leads to a mass-loss rate of 2$\times$10$^{-5}$ and 4$\times$10$^{-5}$\,$M_\odot$\,yr$^{-1}$, respectively, at the crossing point, that is a relatively modest change. In the same way, a larger/smaller exponent (in absolute terms) in the kinetic temperature law leads to a cooler/warmer CSE and the crossing point moves to lower/higher mass-loss rates, but the effect is limited for reasonable changes. A comparison with the observational data is discussed in Sect.~\ref{s:massloss}.
%
%
%
%
\section{Results and discussion}

For reasons to be explained in Sect.~\ref{s:lum} we have divided our 22 objects into two groups: 17 objects are likely to be located within the inner GB, while five objects are likely to be lying in the foreground with unknown distances. The latter are discussed separately in Sect.~\ref{s:foreground}. 

%
%
\subsection{Spectral energy distributions}

\begin{figure*}
  \includegraphics[width=18cm]{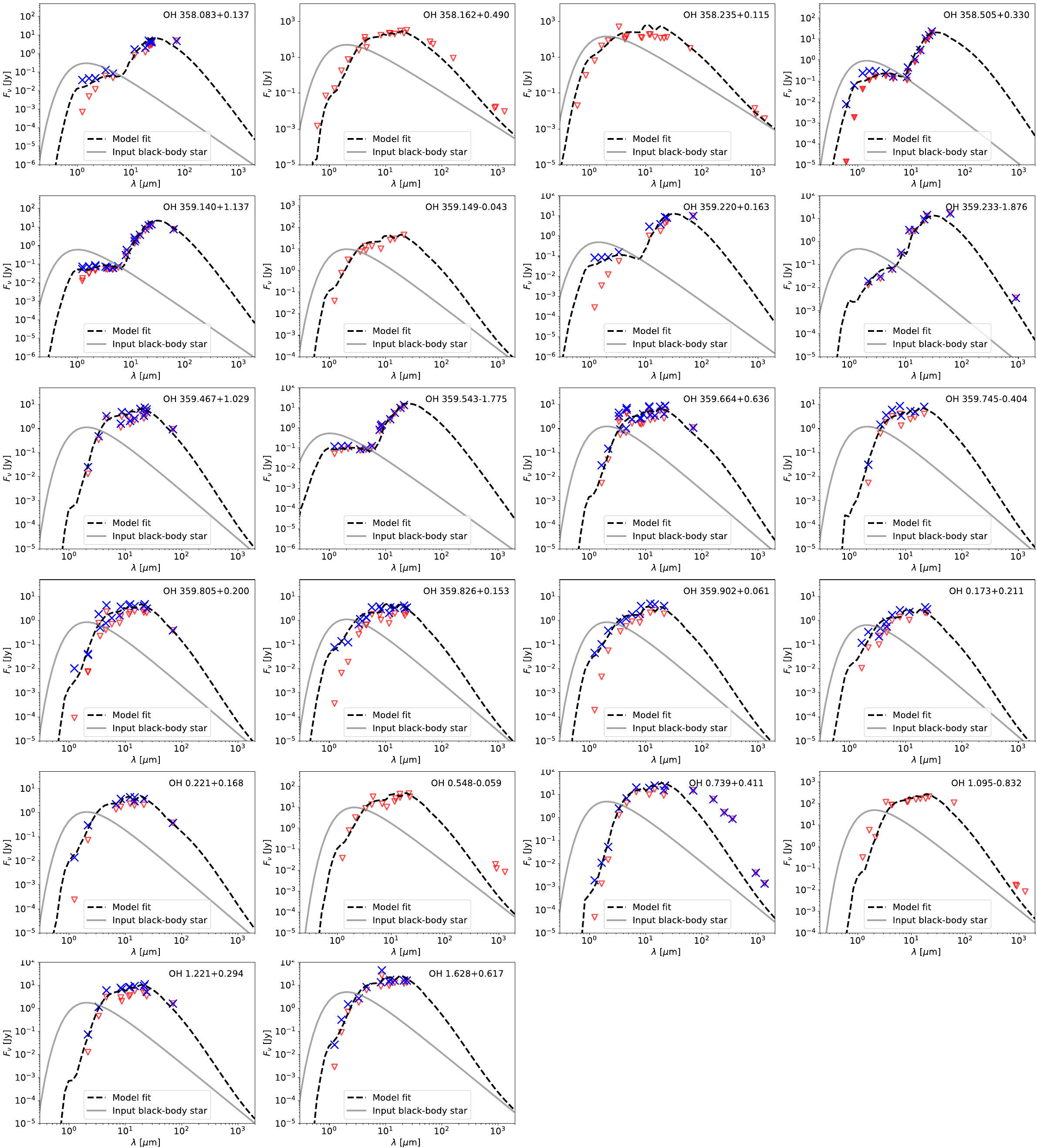} 
    \caption{SEDs composed of photometry data corrected (blue crosses) and not corrected (red triangles) for interstellar extinction. For the objects deemed to be foreground sources only the non-extinction-corrected data are shown. The best-fit model SED obtained from the radiative transfer calculations, input parameters given in Table~\ref{t:sed_summary}, is given as a dashed, black line. The input stellar spectrum, approximated by a blackbody, is shown as the grey, solid line.
    }
   \label{f:all_seds}
\end{figure*}

A model SED has been fitted for all objects assuming isotropic dust density distributions following an $r^{-2}$-law, Sect.~\ref{s:dustradtrans}. The parameters of the best-fit models are given in Table~\ref{t:sed_summary}, and all SEDs are shown in Fig~\ref{f:all_seds}. It is apparent from the SEDs and the model fits that they can be divided into two distinct classes in terms of their appearance. For 16 sources the SED can be fitted using a standard CSE of dust (hereafter SE), that is it is formed by a mass-loss rate constant with time. For the remaining six sources a detached dust CSE (hereafter DE), with an inner radius in the range 200\,$-$\,600\,au, is required to fit the SED. The two objects with resolved source structure in the CO line data, \object{OH358.505+0.330} and \object{OH359.220+0.163}, lie in this category. The classification of each object is given in Tables~\ref{t:summary_inner_GB} and \ref{t:summary_foreground}. Some aspects of the DE objects are discussed in more detail in Sect.~\ref{s:de_objects}.

It is noteworthy that the SEDs of the SE-type objects can all be well-fitted using a stellar (blackbody) temperature of 2500 K, a reasonable stellar effective temperature for a high-mass-loss-rate AGB star. However, the data put no strong constraints on this parameter. On the contrary, the SEDs of the DE-type objects are only well-fitted when a higher stellar temperature is used (this results in a better fit in the range 1 to 7\,$\mu$m).

It is apparent that in the 13 objects where data at wavelengths longer than about 70\,$\mu$m are available, the SED fits are often not very good. In eight objects (5 are SE objects, 3 are DE objects), the longer wavelength points are  underestimated by a factor of a few, while they are severely underestimated in two objects (both are SE objects), \object{OH0.548$-$0.059} and \object{OH0.739+0.411}. Increasing the outer radius by a factor of ten (i.e. 10$^5$\,au, corresponding to a high-mass-loss-rate epoch of about 30\,000\,yr) leads to significantly improved, but not perfect, fits. Alternative ways of improving the fits are a mass-loss rate varying with time, a different dust CSE geometry, and/or different grain properties. However, the aim of this paper is not to provide the best possible fits to the long-wavelength regime of the SEDs by individually selecting the dust envelope geometry or dust composition for each source, but rather to put our CO line data in a broader context that is not dependent on the long-wavelength behaviour of the SED.

\begin{table}
\caption{Results from the SED fits.}
\begin{tabular}{l c c c c c}
\hline \hline
Source                         & $A_{K_{\rm s}}$ & $T_\star$              & $R_{\rm in}$   & $T_{\rm d,in}$   & $\tau_{9.7}$ \\
                               & [mag]          & [K]                     &  [au]          &  [K]            \\
\hline
{\it Inner-GB objects:}\\
1\,-\,\phantom{0}\object{OH358.083+0.137}                & 1.51           & 3500                    & 200            & 210              & 1.5 \\ 
4\,-\,\phantom{0}\object{OH358.505+0.330}                & 0.98           & 4000                    & 500            & 170              & 0.5 \\ 
5\,-\,\phantom{0}\object{OH359.140+1.137}                & 0.56           & 5000                    & 600            & 180              & 1.2 \\
7\,-\,\phantom{0}\object{OH359.220+0.163}                & 2.16           & 3500                    & 500            & 160              & 1.2 \\
8\,-\,\phantom{0}\object{OH359.233$-$1.876}              & 0.42           & 3500                    & 300            & 200              & 2.2 \\
9\,-\,\phantom{0}\object{OH359.467+1.029}                & 0.71           & 2500                    & \phantom{0}15  & $\ldots$         & 3.0 \\ 
10\,-\,\object{OH359.543$-$1.775}                        & 0.32           & 5000                    & 350            & 190              & 0.9 \\
11\,-\,\object{OH359.664+0.636}                          & 1.10           & 2500                    & \phantom{0}15  & $\ldots$         & 2.5 \\
12\,-\,\object{OH359.745$-$0.404}                        & 1.89           & 2500                    & \phantom{0}16  & $\ldots$         & 2.9 \\  
13\,-\,\object{OH359.805+0.200}                          & 1.80           & 2500                    & \phantom{0}13  & $\ldots$         & 2.4 \\ 
14\,-\,\object{OH359.826+0.153}                          & 2.04           & 2500                    & \phantom{0}13  & $\ldots$         & 1.3 \\
15\,-\,\object{OH359.902+0.061}                          & 2.08           & 2500                    & \phantom{0}11  & $\ldots$         & 1.4 \\ 
16\,-\,\object{OH0.173+0.211}                            & 1.62           & 2500                    & \phantom{0}10  & $\ldots$         & 1.5 \\ 
17\,-\,\object{OH0.221+0.168}                            & 1.54           & 2500                    & \phantom{0}12  & $\ldots$         & 1.2 \\ 
19\,-\,\object{OH0.739+0.411}                            & 1.40           & 2500                    & \phantom{0}35  & $\ldots$         & 2.9 \\  
21\,-\,\object{OH1.221+0.294}                            & 1.92           & 2500                    & \phantom{0}19  & $\ldots$         & 2.3 \\  
22\,-\,\object{OH1.628+0.617}                            & 0.84           & 2500                    & \phantom{0}30  & $\ldots$         & 2.0 \\
\hline
{\it Foreground objects:}\\
2\,-\,\phantom{0}\object{OH358.162+0.490}                & $\ldots$       & 2500                    & \phantom{0}14  &  $\ldots$         & 2.5 \\
3\,-\,\phantom{0}\object{OH358.235+0.115}                & $\ldots$       & 2500                    & \phantom{0}12  &  $\ldots$        & 1.1 \\
6\,-\,\phantom{0}\object{OH359.149$-$0.043}              & $\ldots$       & 2500                    & \phantom{0}13  &  $\ldots$        & 1.7 \\
18\,-\,\object{OH0.548$-$0.059}                          & $\ldots$       & 2500                    & \phantom{0}14  &  $\ldots$        & 1.7 \\
20\,-\,\object{OH1.095$-$0.832}                          & $\ldots$       & 2500                    & \phantom{0}14  &  $\ldots$        & 2.6 \\ 
\hline
\end{tabular}
\label{t:sed_summary}
\tablefoot{The far-IR part and the ALMA points are not well fitted for many objects. See text for details on this.}
\end{table}

%
%
\subsection{Stellar variability}

Pulsation periods have been determined for 14 of our sample stars, Table~\ref{t:summary_periods}. All but one, are long-period (in the range 500--1200$^{\rm d}$), large-amplitude (a maximum to minimum ratio in the range \mbox{0.8\,--\,2.2} magnitudes in the WISE W1 and W2 bands) pulsators, that is they have Mira-like variability. This agrees well with the results of \citet{woodetal98} for OH/IR stars in the GC, except the tail towards periods shorter than 500$^{\rm d}$ seen in the \citet{woodetal98} data. \object{OH359.140+1.137} shows periodic variability, but with a very low amplitude (0.05$^{\rm m}$ and 0.1$^{\rm m}$ in the W1 and W2 bands, respectively), and hence cannot be classified as having Mira variability. The rest of the objects have either non-periodic (six objects; for these objects there are enough data to determine a periodicity if present) or non-Mira multi-periodic variability (one object), or the object is so bright that the fluxes are saturated (one object).

There is a clear difference between the SE and DE sources in terms of their variability. Among the 16 SE sources, 13 show Mira-like variability, two are non-periodic, and for one the fluxes are saturated. On the contrary, for the six DE sources four are non-periodic, one shows only low-amplitude variability, and one shows multi-period, non-Mira-like variability.

%
%
\subsection{Luminosities}
\label{s:lum}

We have found that 17 objects have luminosities, obtained by integrating the fit to the extinction-corrected SED for each source, consistent with them being AGB stars if placed at our adopted distance of the GC, 8.2\,kpc. Eleven of these objects have systemic velocities (in the local standard of rest frame) in the range $|\varv_{\rm sys}|$\,$>$\,50\,km\,s$^{-1}$, further strengthening their association with the inner GB, where radial systemic velocities can be high also at a line-of-sight direction close to $\ell$\,=0$^\circ$. When looked at separately in terms of their luminosities and mass-loss-rate characteristics, there is nothing that distinguishes the remaining six stars (with $|\varv_{\rm sys}|$\,$<$\,50\,km\,s$^{-1}$) from the `higher-velocity' stars. Therefore, these 17 stars form our sub-sample of inner-GB OH/IR stars, see Table~\ref{t:summary_inner_GB}. About two thirds of them are SE sources and one third DE sources. Only three of the inner-GB objects are detected in 324\,GHz continuum. 

Two of the inner-GB objects, \object{OH0.739+0.411} and \object{OH1.628+0.617}, have markedly higher luminosities than the others, $\approx$\,25000\,$L_\odot$, and they are among the three objects in this sub-sample that are detected in 324\,GHz continuum. Except for this, they seem not to be different from the inner-GB objects. Therefore, we retain them in this sub-sample, but note that there is a strong possibility that they are instead foreground objects.

This leaves five objects out of the original 22 sources, all of them SE sources, for which the estimated luminosities become higher than expected for an AGB star if placed at the distance of the GC, for three of them even higher than expected for a red supergiant. Most likely these are foreground AGB stars, and they will be discussed separately in Sect.~\ref{s:foreground}. All of them are detected in 324\,GHz continuum.

\begin{figure}
  \includegraphics[width=7.5cm]{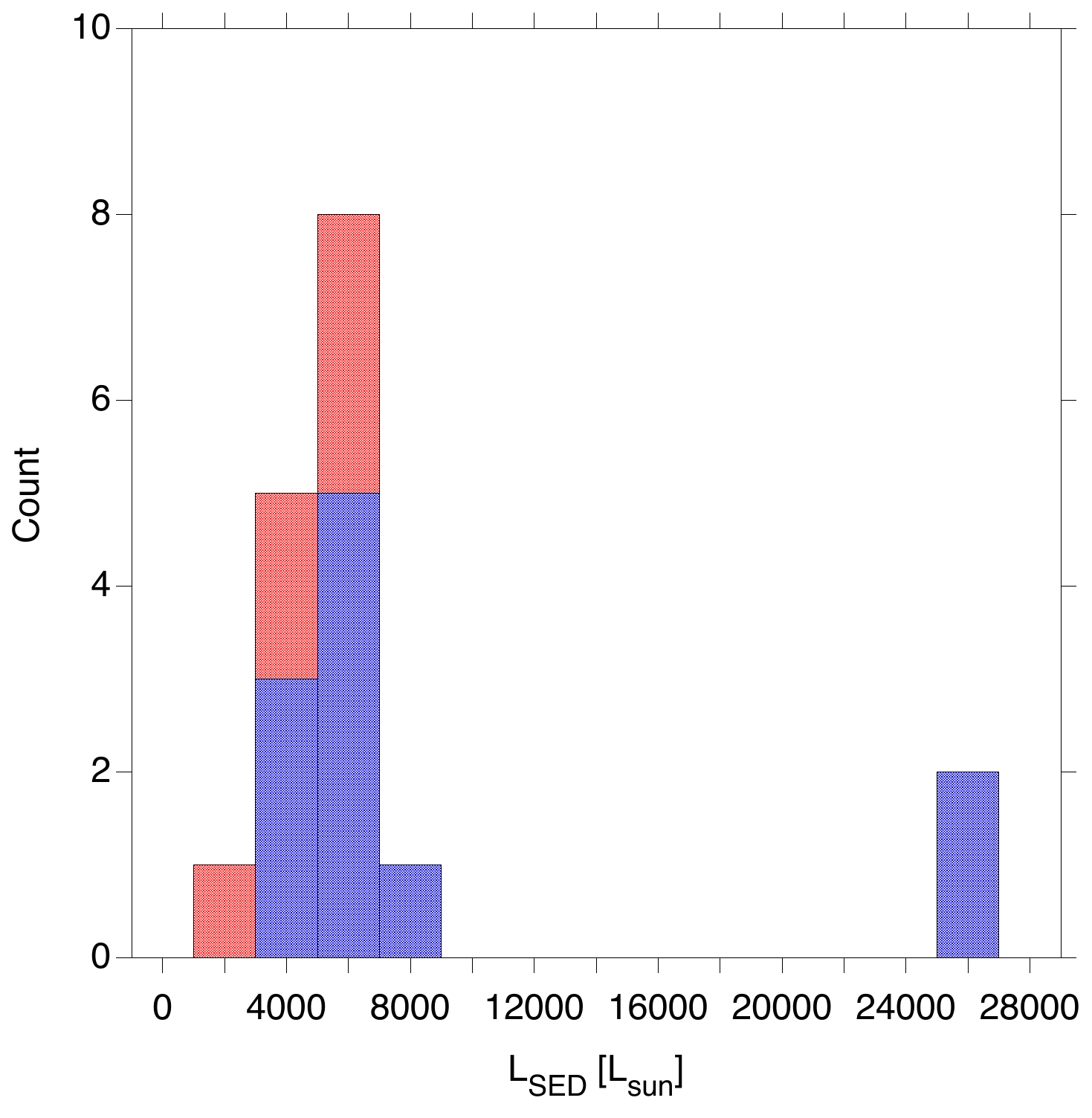}
    \caption{Luminosity distribution of the inner-GB sample (blue for SE objects, red for DE objects). 
    }
   \label{f:lum}
\end{figure}

Figure~\ref{f:lum} shows the luminosity distribution for our inner-GB sample. The median luminosity is 5600\,$L_\odot$ and the range covered is 2100 to 8700\,$L_\odot$ for the majority of the objects. There are two outliers at a higher luminosity, 25000\,$L_\odot$. In this sense, our objects appear in many respects similar to the sources in the GB discussed by \citet{vandvehabi90} and also by \citet{groeblom05}, \citet{jimeenge15}, and \citet{blometal18}, for instance. Using the results of \citet{kara14}, the median luminosity translates into an initial mass of $\approx$\,1.2\,$M_\odot$ if the star has solar metallicity and lies at the peak of the AGB [using the results of \citet{ventetal18} gives the same result]\footnote{The metallicity of the Bulge is a complicated issue. There exists sub-solar, solar,  and super-solar metallicity populations \citep[see e.g.][]{garcpereetal18}. However, the initial mass estimate is relatively independent of metallicity. Following \citet{kara14} we find that a metallicity of $Z$\,=\,0.007 results in $M_{\rm i}$\,$\approx$\,1\,$M_\odot$, while $Z$\,=\,0.03 results in $M_{\rm i}$\,$\approx$\,1.4\,$M_\odot$.}. This means that these objects will not become carbon stars, and their masses fall well below the limit for the hot-bottom-burning (HBB) process.  Using instead the often used, but older, results of \citet{vasswood93} and the same assumptions, the median luminosity translates into an initial mass of $\approx$\,1.6\,$M_\odot$, still well below the HBB mass limit. The initial mass estimates translates to an age in the range 4\,$-$\,7\,Gyr for these sources. For the two higher-luminosity stars, 25\,000\,$L_\odot$ indicates an initial mass of about 4.3\,$M_\odot$ if at solar metallicity and at the peak of the AGB \citep{kara14}. That is, these are much younger stars (about 0.3\,Gyr) that may have started HBB. Such stars were also found by \citet{groeblom05} towards the GC. Alternatively, they may be foreground stars.

We have compared our estimated luminosities with those obtained from the pulsational periods using different period-luminosity (PL) relations. \citet{whitetal91} derived a PL relation using a mixture of Galactic and Large Magellanic Cloud objects of O-type with periods in the range 120 to 2000$^{\rm d}$ ($\approx$\,40 objects; $M_{\rm bol}$\,=\,--2.55\,$\log P$\,+\,1.85 mag). For the eleven inner-GB objects with determined Mira-like variability, this relation gives luminosities that are, on average, two times higher than our results. \citet{groeetal20} used Magellanic Clouds objects of O-type with periods in excess of 1000$^{\rm d}$ (11 objects; $M_{\rm bol}$\,=\,--2.97\,$\log P$\,+\,2.59 mag). In comparison, this relation gives luminosities that are, on average, three times higher than our results. Clearly our objects lie below these PL relations. A similar conclusion was reached by \citet{woodetal98} and \citet{blometal98} for their samples of OH/IR stars. \citet{woodetal98} argued that this is an effect of lengthened periods due to extensive mass loss (and hence lower present stellar masses), possibly strengthened by metallicity effects. The only exceptions are the 25\,000\,$L_\odot$ objects. The luminosities obtained from the \citet{whitetal91} relation and their pulsational periods are two times lower than our results. This may be a strong indication that these are actually foreground stars.

%
\subsection{Mass-loss rates and gas kinematics}
\label{s:massloss}

\begin{figure}
  \includegraphics[width=7.3cm]{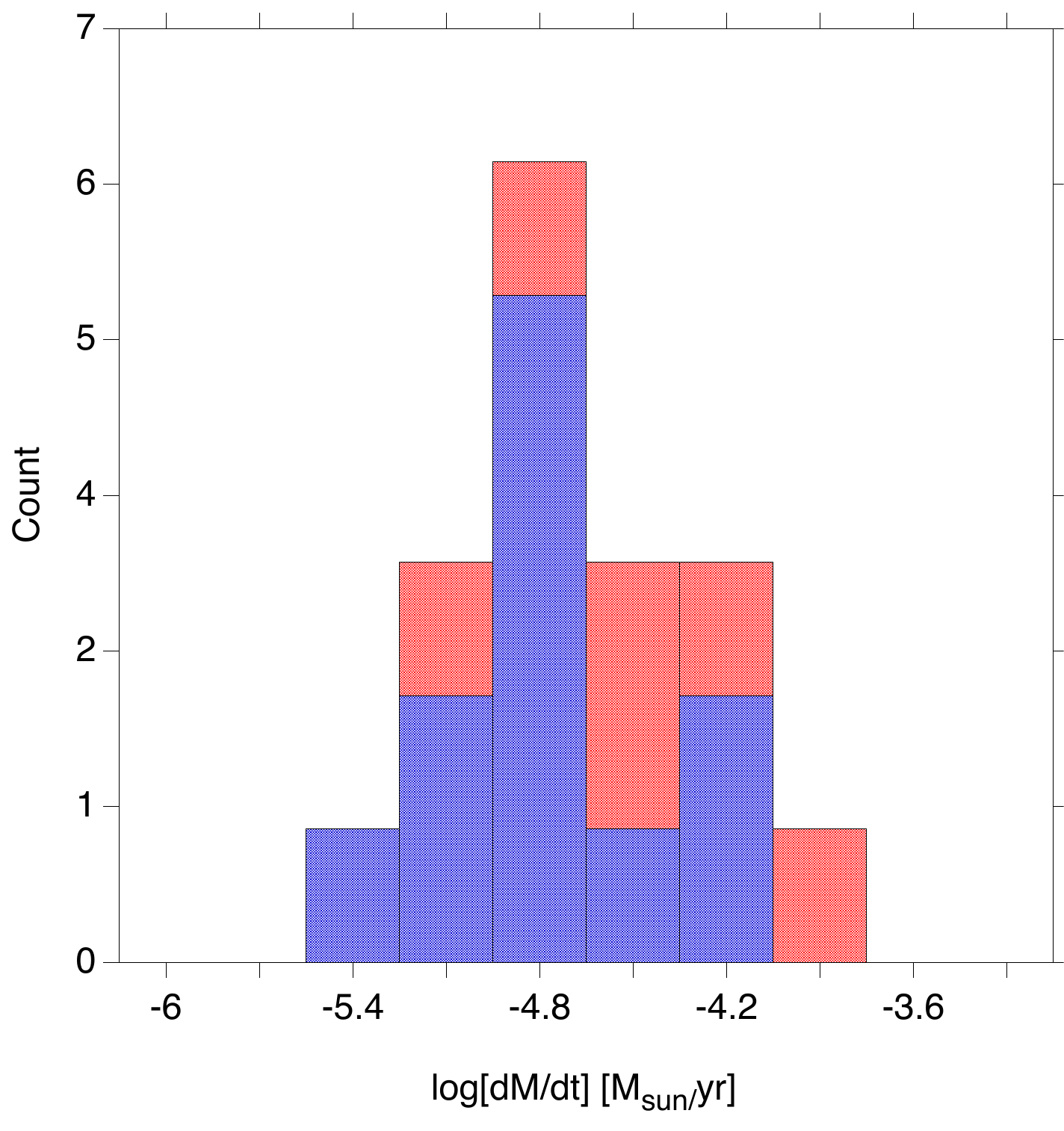}
    \caption{Gas mass-loss rate distribution of the inner-GB sample (blue for SE objects, red for DE objects).
    }
   \label{f:mass_loss}
\end{figure}

The gas-mass-loss rates and terminal gas-expansion velocities have been estimated using a radiative transfer code and the $^{12}$CO $J$\,=\,2--1 and 3--2 lines. The results are summarised in Table~\ref{t:summary_inner_GB}. The gas-mass-loss-rate distribution for the 17 stars in the inner-GB sample, see Fig~\ref{f:mass_loss}, has median of 2$\times$10$^{-5}$\,$M_\odot$\,yr$^{-1}$. This means that this is definitely a high-mass-loss-rate sample compared with, for instance, a solar neighbourhood sample of O-type AGB stars with a median mass-loss rate of 3$\times$10$^{-7}$\,$M_\odot$\,yr$^{-1}$ [the results of \citet{olofetal02} and \citet{gonzetal03a} as summarised by \citet{ramsetal09}]. In terms of mass-loss characteristics, the inner-GB sub-sample stars are similar to nearby Miras like \object{GX~Mon}, \object{IK~Tau}, and \object{WX~Psc}, whose masses are estimated to lie in the range 1.1\,$-$\,1.5\,$M_\odot$ \citep{denuetal17,danietal17}. \citet{jimeenge15} and \citet{blometal18} derived mass-loss rates for their samples of stars similar to ours by solving the momentum equations for gas and dust (coupled via friction) and obtained mass-loss rates that are a factor of a few higher than ours, but this way of estimating mass-loss rates is not without its problems \citep{ramsetal08}.

\begin{figure}
  \includegraphics[width=7.5cm]{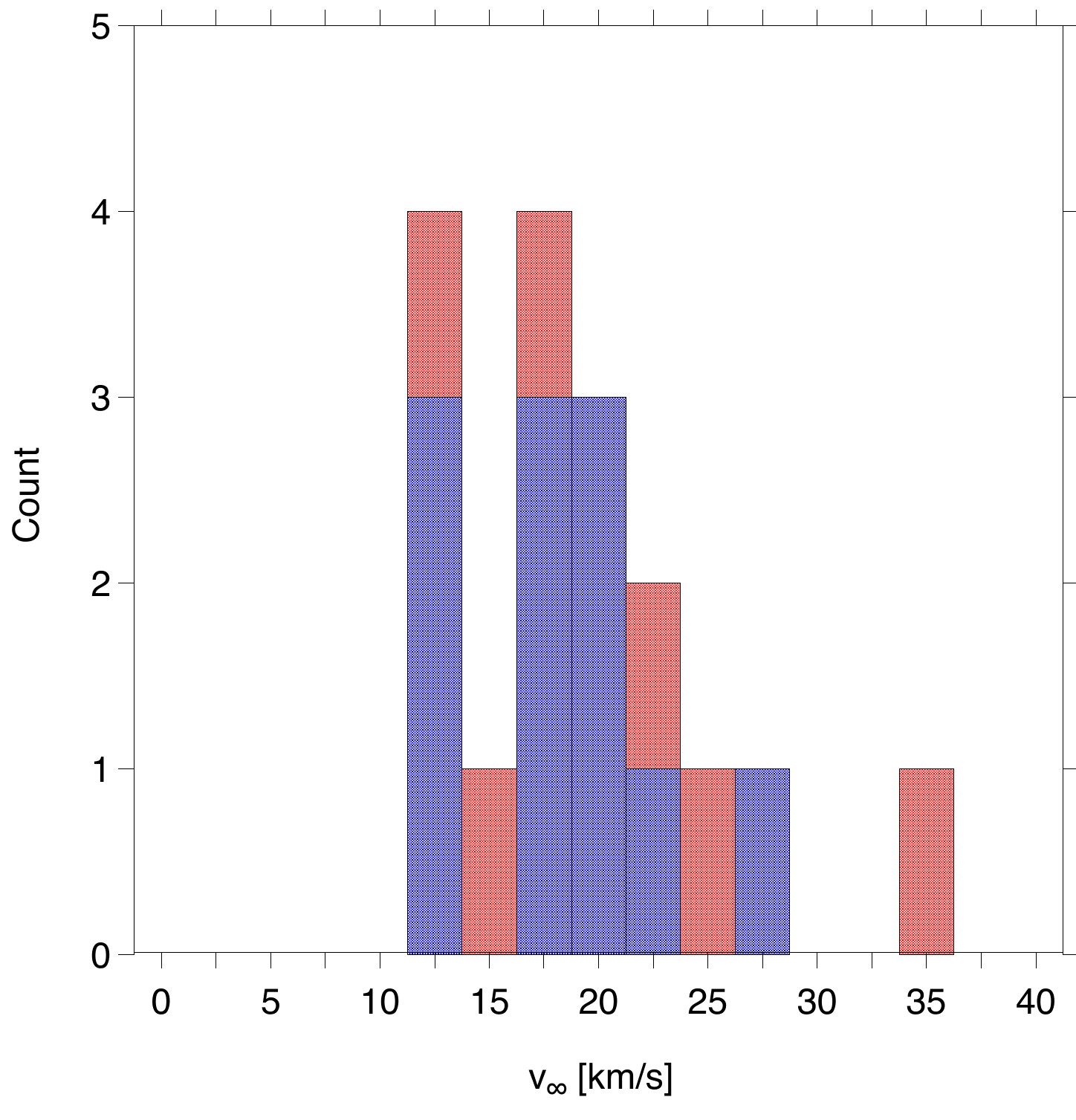}
    \caption{Terminal gas expansion velocity distribution of the inner-GB sample (blue for SE objects, red for DE objects).
    }
   \label{f:exp_vel}
\end{figure}

The terminal gas-expansion velocity distribution of our inner-GB sources, see Fig~\ref{f:exp_vel},  has a median of 18\,km\,s$^{-1}$.  This result can be compared with the peaks of the velocity distributions of OH/IR stars in the Galactic Plane (12\,km\,s$^{-1}$), the GB (14\,km\,s$^{-1}$), and the GC (19\,km\,s$^{-1}$) \citep{sjouetal98}, and the median of 7\,km\,s$^{-1}$ for a solar neighbourhood sample  \citep{ramsetal09}. It has been argued that higher-velocity sources are drawn from a younger population than the lower-velocity sources \citep{lindetal92b}, but it could also be an effect of our stars reaching the tip of the AGB and the end of their the mass-loss evolution, since, in general, mass-loss rate and expansion velocity are positively correlated \citep{hoefolof18}. We find no such correlation for our inner-GB sub-sample, but this is most likely due to the limited range measured for, and the relatively large uncertainties in the estimates of, both these quantities. Nevertheless, our objects fall in the correct regime of this correlation. Finally, observations of OH 1612\,MHz masers and CO line emission towards AGB stars in the Large Magellanic Cloud suggest that the gas expansion velocities are lower at lower metallicity \citep{goldetal17,groeetal16}. Hence, the high expansion velocities of our inner-GB sample sources suggest that they are not low-metallicity objects.

As discussed in Sect.~\ref{s:coradtrans} the $^{12}$CO $J$\,=\,3--2/2--1 line intensity ratio depends sensitively on the mass-loss rate, and to a lesser degree on the kinetic temperature. We find a significant discrepancy, by factors of 1.5--3, between the observational and model ratios (for the derived mass-loss rate) in eleven of our inner-GB sources, and under-estimates are as common as over-estimates. The most likely explanation is a mixture of an inadequate circumstellar model and an inadequate S/N of the observational data. Higher quality data on more $^{12}$CO line transitions are required in order to address this issue.

The angular sizes (in diameter) corresponding to the $R_{1/2}$ values, as calculated using Eq.(\ref{e:rph}), for our inner-GB sources lie in the range 1\farcs6--8\farcs4. Since the MRSs are about 2.5\arcsec\ and 1\arcsec\ for the $J$\,=\,\mbox{2--1} and \mbox{3--2} lines, respectively, there is a risk that we lose flux. We have made a crude estimate for a $\dot{M}_{\rm g}$\,=\,10$^{-4}$\,$M_\odot$\,yr$^{-1}$ and $\varv_\infty$\,=\,20\,km\,s$^{-1}$ model source at a distance of 8.2\,kpc. In this rather extreme case, the $R_{1/2}$ value corresponds to 8\farcs6 in diameter. When observed with beam sizes that equal the MRSs, about 60\,\% and 50\,\% of the total flux densities are lost in the $J$\,=\,\mbox{2--1} and \mbox{3--2} lines, respectively. For our median values of 2$\times$10$^{-5}$\,$M_\odot$\,yr$^{-1}$ and 18\,km\,s$^{-1}$ the loss is much less, about 14\,\% in both lines. This means, that, on average, the estimated mass-loss rates are not significantly affected by any missing flux, only at the extreme upper end, the underestimate may be about a factor of two. The effect becomes less if the $R_{1/2}$ values are smaller than predicted by Eq.(\ref{e:rph}), see discussion below. 

\begin{figure}
  \includegraphics[width=7.3cm]{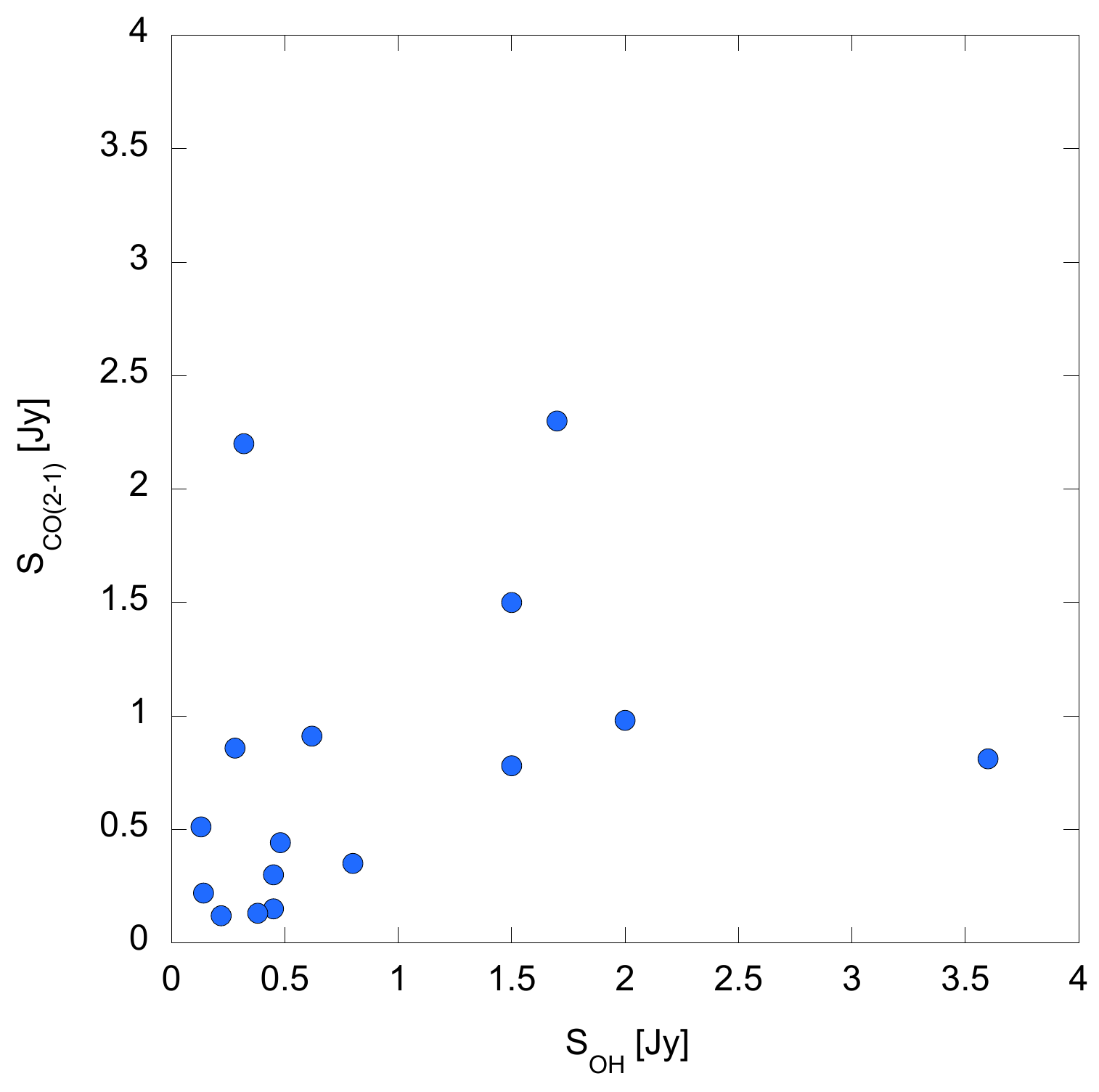}
    \caption{Flux densities of the $^{12}$CO($J$\,=\,\mbox{2--1}) and OH 1612\,MHz lines plotted versus each other for the inner-GB sample.
    }
   \label{f:sco_soh}
\end{figure}

Both the CO and OH 1612\,MHz line intensities are expected to depend on the mass-loss rate \citep{baudhabi83,ramsetal08}. We plot the $^{12}$CO $J$\,=\,\mbox{2--1} flux density versus that of the OH 1612\,MHz line for the inner-GB sample in Fig.~\ref{f:sco_soh}. As expected, there is a correlation, but it is not particularly tight. This is most likely due to the OH line being a significantly poorer mass-loss-rate estimator than the CO line. Nevertheless, it shows that the strength of the OH 1612\,MHz line can be used, as a first approximation, to estimate the expected strengths of the CO rotational lines.

%
%
%
\subsection{CO photodissociation and brightness distributions}
\label{s:phdiss}

We have used the results of \citet{sabeetal19} to calculate the photodissociation radii of $^{12}$CO and $^{13}$CO, $R_{1/2}$, for each source as outlined in Sect.~\ref{s:coradtrans}. These are based on the assumption of an interstellar UV radiation field with a strength applicable to the solar neighbourhood. It is possible, even very likely, that this radiation field is considerably stronger in the region of the GC. This will decrease the CO photodissociation radius, although the effect is relatively moderate for higher-mass-loss-rate objects. An increase by a factor of four in the strength of the UV radiation field leads to a decrease in $R_{1/2}$ for $^{12}$CO by only 30\,\% for a 2$\times$10$^{-5}$\,$M_\odot$\,yr$^{-1}$ and 15\,km\,s$^{-1}$ object, and even less for higher mass-loss rates, as shown by \citet{sabeetal19}. 

Nevertheless, this opens up the question of how much the uncertainty in the CO photodissociation radius will affect our estimated mass-loss rates. Decreasing $R_{1/2}$ by a factor of three for a model with our median values, 2$\times$10$^{-5}$\,$M_\odot$\,yr$^{-1}$ and 18\,km\,s$^{-1}$, results in flux densities lowered by 34\,\% and 6\,\% for the $^{12}$CO $J$\,=\,\mbox{2--1} and \mbox{3--2} lines, respectively. This corresponds to a mass-loss rate under-estimated by about 20\,\%, well within the uncertainty of this estimate. Consequently, the UV radiation field in the inner GB must be substantially higher than that in the solar neighbourhood to have any effect on our estimated mass-loss rates.

Our estimated source sizes are (deconvolved) FWHMs of Gaussian fits to the velocity-integrated brightness distributions (the $\theta_{\rm s}$ results are summarised in Tables~\ref{t:12co21}-\ref{t:13co32}), and the question is if they contain any useful physical information, for instance, on the photodissociation radii. We have used our radiative transfer model to test this for the $^{12}$CO(\mbox{2--1}) line, since this has the most extended brightness distribution and should consequently be the most-affected one of our observed lines. The adopted models all have $\varv_\infty$\,=\,18\,km\,s$^{-1}$, and we use three mass-loss rates, 4$\times$10$^{-6}$, 2$\times$10$^{-5}$, and 1$\times$10$^{-4}$\,$M_\odot$\,yr$^{-1}$ (this covers well the mass-loss-rate range of our inner-GB sub-sample). The distance is 8.2\,kpc and the beam size is 0\farcs19. The result is that for these conditions, the $\theta_{\rm s}$ values of the velocity-integrated model brightness distributions are essentially independent of the photodissociation radius. The latter has to be decreased by a factor of ten to give a noticeable result on $\theta_{\rm s}$, a decrease by 40\,\%. Thus, we can conclude that our estimated source sizes, in all likelihood, give no information on the photodissociation radius. Relations between Gaussian fits to brightness distributions and photodissociation radii of nearby, lower-mass-loss-rate sources are discussed in \citet{ramsetal20}.

\begin{figure}
  \includegraphics[width=5.6cm]{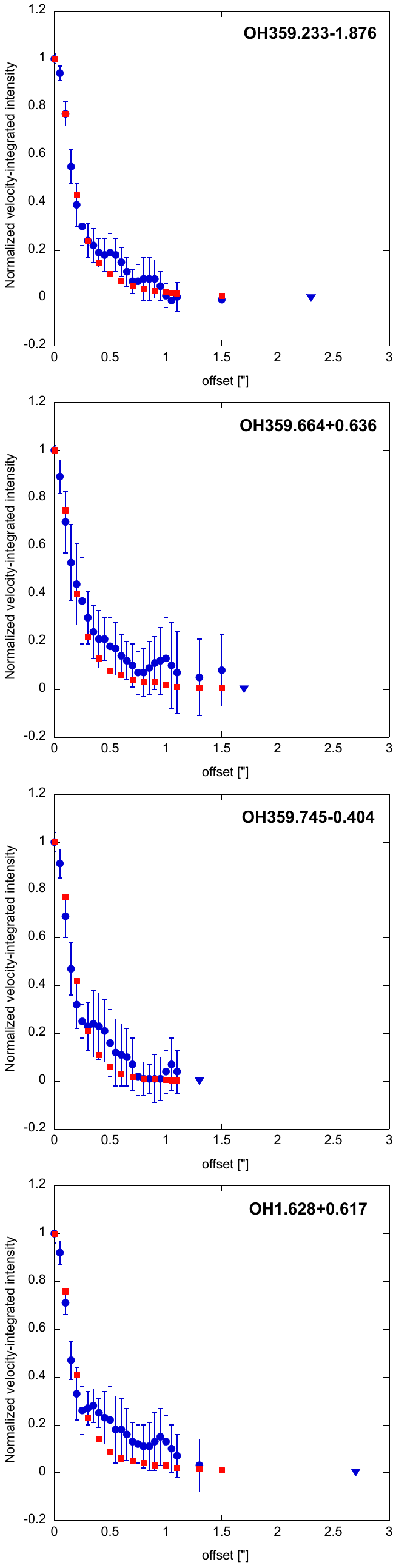}
    \caption{Azimuthally averaged observed brightness distributions of the $^{12}$CO(\mbox{2--1}) line towards four objects (blue filled circles), normalised intensity scales. The error bars are $\pm$\,1$\sigma$. The blue triangle gives the photodissociation radius in angular units, $R_{1/2}/D$. The red squares shows the brightness distribution of the best-fit radiative transfer model for each source obtained with a beam that equals the synthesised beam.
    }
   \label{f:bright_distr}
\end{figure}   

Finally, we test whether or not our radiative transfer modelling is able to reproduce the observed brightness distributions. In Fig.~\ref{f:bright_distr}, we compare the azimuthally averaged observed brightness distributions of the $^{12}$CO(\mbox{2--1}) line towards four of our sources (these are inner-GB objects having the largest $\theta_{\rm s}$ and the highest S/N data), with the brightness distributions of the best-fit radiative transfer models for each source (obtained with a Gaussian beam having a FWHM that equals that of the synthesised beam). As can be seen, the narrow and bright central peak of the observed brightness distributions are well reproduced in the models. On the contrary, the observed extended low-surface-brightness emission in the observational data is clearly under-estimated in the models. Within the adopted circumstellar model, the only way to selectively increase the model brightness in the external parts is to raise the minimum kinetic temperature, $T_{\rm min}$ (it should be noted here that the physical properties of the external parts of AGB CSEs are particularly poorly constrained). However, in order to still fit the observed velocity-integrated intensities, this requires a corresponding decrease in the mass-loss rate, to the extent that the under-estimate of the brightness in the external parts persists. Hence, the only way to get a better fit to the observed brightness distribution is to modify the circumstellar model, such as the geometry, or allow a deviation from an $r^{-2}$-law for the density distribution.

There are also suggestions of modulations in the brightness distributions that can be due to ring-, arc- or spiral-like structures in the density distribution. Clearly, a study of structure in the CSEs requires higher S/N data and larger MRSs than provided by our data set.

%
%
\subsection{Gas-to-dust mass ratios}

Our estimate of the gas-to-dust mass ratio, $R_{\rm gd}$, is obtained as $\dot{M_{\rm g}}$/$\dot{M_{\rm d}}$, and the results for the inner-GB objects are summarised in Table~\ref{t:summary_inner_GB}. It is clear that we must separate the discussions of the SE and DE objects. For the SE sources the resulting $R_{\rm gd}$ values range from about 70 to 780 with a median of 200. This is in the range found for other O-type AGB stars \citep[e.g.][and references therein]{ramsetal08}. On the contrary, for the DE sources the $R_{\rm gd}$ values are much lower, in the range 5\,$-$\,143. This is further discussed in Sect.~\ref{s:de_objects}.

%
%
\subsection{The circumstellar $^{12}$CO/$^{13}$CO abundance ratio}

\begin{figure}
  \includegraphics[width=7.3cm]{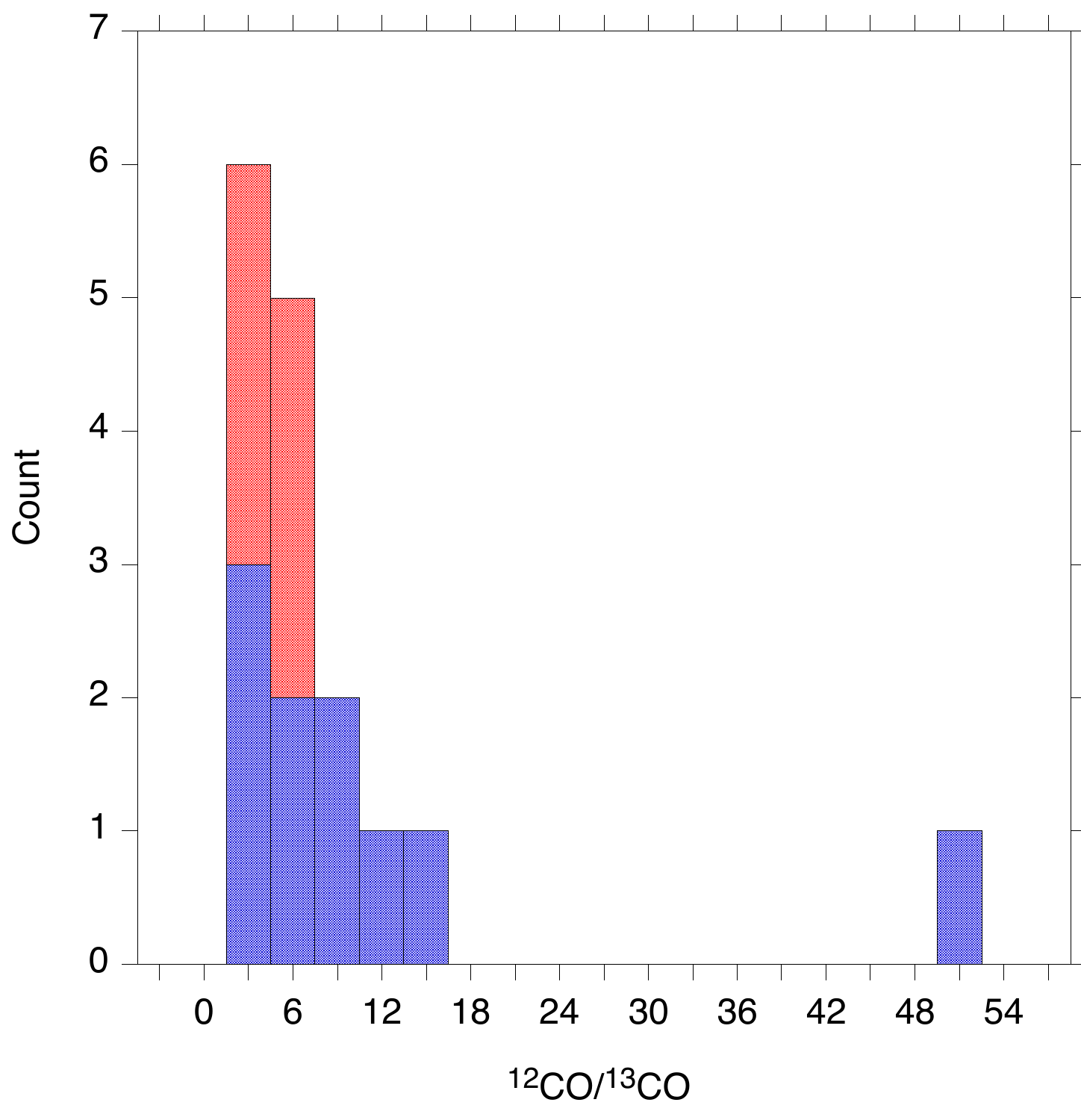}
    \caption{Circumstellar $^{12}$CO/$^{13}$CO abundance ratio distribution for the inner-GB sample (blue for SE objects, red for DE objects).
    }
   \label{f:isotopes}
\end{figure}

The circumstellar $^{12}$CO/$^{13}$CO abundance ratio has been estimated through a radiative transfer analysis, in order to correct for any optical depth effects, of a single $^{13}$CO line within the circumstellar model determined from the $^{12}$CO lines. As such, it is associated with some considerable uncertainty, at least a factor of four as discussed in Sect.~\ref{s:coradtrans}. The results are summarised in Table~\ref{t:summary_inner_GB} and presented in Fig.~\ref{f:isotopes}. For the inner-GB objects the $^{12}$CO/$^{13}$CO abundance ratios range from 1.6 to 16 (with an outlier at 50) with a median of 5. 

The circumstellar $^{12}$CO/$^{13}$CO abundance ratio is of particular interest if it is a good measure of the stellar $^{12}$C/$^{13}$C isotope ratio. There are good reasons to believe that this is the case. The two main processes that can alter the molecular ratio are isotope-selective chemistry and photodissociation. The processes determining the initial $^{12}$CO/$^{13}$CO abundance ratio are in all likelihood not isotope-selective. Further out in the CSE, the photodissociation is isotope-selective since it occurs in lines and the optical depth is lower for the $^{13}$CO lines than the corresponding $^{12}$CO lines. However, this can be counter-acted by the exchange reaction that favours the formation of $^{13}$CO at the expense of $^{12}$CO in low-temperature gas \citep{watsetal76}. These CSE processes have been studied in detail by \citet{sabeetal19,sabeetal20}, and the conclusion is that the circumstellar $^{12}$CO/$^{13}$CO abundance ratio in the external part of the CSE of a high-mass-loss-rate object is only slightly above the initial $^{12}$CO/$^{13}$CO value, which is determined by the stellar $^{12}$C/$^{13}$C isotope ratio. This means that, if anything, we overestimate the latter. \citet{sabeetal20} also tested the effect of a higher UV radiation field. It will lead to a further slight increase in the circumstellar $^{12}$CO/$^{13}$CO abundance ratio in the external parts of the CSE, but the effect is marginal for the mass-loss rates of our objects.

The very low stellar $^{12}$C/$^{13}$C isotope ratios inferred in this way are theoretically problematic, since the stars in our inner-GB sample are relatively low-mass stars (1.2\,$-$\,1.4\,$M_\odot$) that do not go through HBB. The latter is the only viable nucleo-synthetic way of producing very low $^{12}$C/$^{13}$C isotope ratios, close to the equilibrium value of the CNO-cycle $\approx$\,3.5, but it requires a stellar mass in excess of 4\,$M_\odot$ \citep{karaluga16}. In the case of lower-mass stars this isotope ratio is set by the first dredge-up during red giant branch (RGB) evolution, and it is difficult, in theory, to obtain isotope ratios below 20 \citep{karaluga16}, unless there is some efficient extra mixing such as thermohaline mixing, which is introduced through a parameterised approach [see \citet{karalatt14} for a discussion of different mixing processes]. 

Our result should be put into the context of results for other O-type red giants. \citet{ramsolof14} found circumstellar $^{12}$CO/$^{13}$CO abundance ratios that range from 6 to 66 with a median of 13 in their sample of solar neighbourhood, most likely lower-mass, mass-losing, O-type AGB stars. For the high-mass-loss-rate objects \object{GX~Mon}, \object{IK~Tau}, and \object{WX~Psc}, with estimated masses in the range 1.0 to 1.5\,$M_\odot$ (see above), the results are 11, 10, and 13, respectively. For their remaining ten sources with mass-loss rates in excess of 5$\times$10$^{-6}$\,$M_\odot$\,yr$^{-1}$, but without mass estimates, the median value is 15. The work of \citet{hinketal16}, based on stellar atmosphere synthetic spectra of O-type AGB stars (mainly Miras with weak, if any, mass loss), emphasises these results: they found a median stellar $^{12}$C/$^{13}$C of 17, and no source with a reliable ratio lower than 10 in their sample of 33 stars. Results for O-type giants on the RGB has also been obtained. \citet{gilr89} estimated stellar $^{12}$C/$^{13}$C isotopic ratios in the range 6\,$-$\,17 for stars in open clusters and masses in the range 1\,$-$\,2\,$M_\odot$. \citet{gilrbrow91} found an average $^{12}$C/$^{13}$C of 13 for stars at the tip of the RGB and at the Horizontal Branch in the open cluster M67. \citet{tsuj07} found stellar $^{12}$C/$^{13}$C isotopic ratios in the range 8\,$-$\,16 for a sample of eight RGB stars. Considering the uncertainties of our estimates, it is currently not possible to judge whether or not the stars in our sample have significantly lower inferred stellar $^{12}$C/$^{13}$C isotope ratios than those estimated by others for RGB and AGB stars. On the other hand, there is no obvious reason why our analysis should lead to a systematic under-estimate of the $^{12}$CO/$^{13}$CO abundance ratio by factors of two to three. Finally, we note here that \citet{delfetal97} also estimated very low circumstellar $^{12}$CO/$^{13}$CO abundance ratios for a small sample of OH/IR stars, and that \citet{justetal15} inferred high $^{17}$O/$^{18}$O isotope ratios for a sample of five OH/IR stars. The latter result strongly suggest HBB processing. The masses of the stars in these two studies are not known, but they, most likely, belong to a population of higher-mass AGB stars.

\begin{table*}[t]
\caption{Results for the inner-GB sources.}
\begin{tabular}{l c c c c r c c c}
\hline \hline
Source                                  &  SED  & $P$\,$^{(a)}$   & $L$\,$^{(b)}$    & $\dot{M}_{\rm g}$\,$^{(c)}$       & $R_{\rm gd}$\,$^{(d)}$  &  $\varv_{\rm sys}$\,$^{(e)}$  &  $\varv_\infty$\,$^{(e)}$  &  $^{12}$CO/$^{13}$CO\,$^{(f)}$  \\
                                        &  type & [days]          & [$L_\odot$]      & [$M_\odot$\,yr$^{-1}$]  &               &  [km\,s$^{-1}$]        &  [km\,s$^{-1}$]  \\
\hline 
1\,-\,\phantom{0}\object{OH358.083+0.137}                         &  DE   & np              & \phantom{2}2100  & 2$\times$10$^{-5}$      &   29           &  \phantom{1}--24        &  22                 &  \phantom{1}5 \\
4\,-\,\phantom{0}\object{OH358.505+0.330}                         &  DE   & np              & \phantom{2}6500  & 1$\times$10$^{-4}$      &  125           &  \phantom{1}--30        &  24                 &  \phantom{1}5 \\
5\,-\,\phantom{0}\object{OH359.140+1.137}                         &  DE   & 1000            & \phantom{2}6100  & 1$\times$10$^{-5}$      &    8           &  --136                  &  12                 &  \phantom{1}3 \\
7\,-\,\phantom{0}\object{OH359.220+0.163}                         &  DE   & np              & \phantom{2}3400  & 8$\times$10$^{-5}$      &   67           &  --137                  &  16                 &  \phantom{1}7 \\
8\,-\,\phantom{0}\object{OH359.233$-$1.876}                       &  DE   & np              & \phantom{2}3400  & 3$\times$10$^{-5}$      &   20           &  \phantom{1}--15        &  18                 &  \phantom{1}4 \\
9\,-\,\phantom{0}\object{OH359.467+1.029}                         &  SE   & \phantom{1}605  & \phantom{2}5600  & 2$\times$10$^{-5}$      &  200           &  \phantom{--}110        &  18                 &  13 \\
10\,-\,\object{OH359.543$-$1.775}                       &  DE   & mp              & \phantom{2}5400  & 2$\times$10$^{-5}$      &   12           &  \phantom{1--}90        &  36                 &  \phantom{1}2 \\
11\,-\,\object{OH359.664+0.636}                         &  SE   & \phantom{1}550  & \phantom{2}6100  & 2$\times$10$^{-5}$      &  250           &  --158                  &  13                 &  \phantom{1}5 \\
12\,-\,\object{OH359.745$-$0.404}                       &  SE   & \phantom{1}730  & \phantom{2}6000  & 1$\times$10$^{-5}$      &  100           &  \phantom{--}186        &  21                 &  \phantom{1}2 \\
13\,-\,\object{OH359.805+0.200}\,$^{(g)}$               &  SE   & \phantom{1}500  & \phantom{2}4300  & 5$\times$10$^{-6}$      &   71           &  \phantom{1}--51        &  21                 &  \phantom{1}4 \\
14\,-\,\object{OH359.826+0.153}                         &  SE   & \phantom{1}500  & \phantom{2}5600  & 6$\times$10$^{-6}$      &  171           &  \phantom{1}--26        &  17                 &  \phantom{1}7 \\
15\,-\,\object{OH359.902+0.061}                         &  SE   & \phantom{1}590  & \phantom{2}4300  & 1$\times$10$^{-5}$      &  250           &  --133                  &  19                 &  \phantom{1}4 \\
16\,-\,\object{OH0.173+0.211}\,$^{(g)}$                 &  SE   & \phantom{1}512  & \phantom{2}3300  & 1$\times$10$^{-5}$      &  333           &  \phantom{1--}52        &  18                 &  $\ldots$     \\
17\,-\,\object{OH0.221+0.168}                           &  SE   & \phantom{1}663  & \phantom{2}5300  & 1$\times$10$^{-5}$      &  200           &  \phantom{1--}64        &  23                 &            10 \\
19\,-\,\object{OH0.739+0.411}                           &  SE   & \phantom{1}570  & 25000            & 3$\times$10$^{-5}$      &  176           &  \phantom{1}--21        &  12                 &  \phantom{1}8 \\
21\,-\,\object{OH1.221+0.294}                           &  SE   & \phantom{1}715  & \phantom{2}8700  & 7$\times$10$^{-5}$      &  778           &  --261                  &  13                 &            50 \\
22\,-\,\object{OH1.628+0.617}                           &  SE   & \phantom{1}830  & 25000            & 5$\times$10$^{-5}$      &  250           &  \phantom{1}--86        &  27                 &            16 \\                        
\hline
\end{tabular}
\label{t:summary_inner_GB}
\tablefoot{$^{(a)}$ The adopted period from Table~\ref{t:summary_periods} (np = non-periodic, mp = non-Mira multi-periodic variablity). $^{(b)}$ Obtained by integrating the fit to the extinction-corrected SEDs. $^{(c)}$ The uncertainty is estimated to be a factor of five. $^{(d)}$ The gas-to-dust mass ratio estimated as $\dot{M}_{\rm g}/\dot{M}_{\rm d}$. The uncertainty is estimated to be a factor of seven. $^{(e)}$ Estimated as weighted averages of the results for the detected CO lines. The uncertainties are estimated to be $\pm$\,1.0 and $\pm$\,1.5\,km\,s$^{-1}$ for the systemic and expansion velocities, respectively. $^{(f)}$ The uncertainty is estimated to be a factor of four. $^{(g)}$ Only the $^{12}$CO($J$\,=\,2--1) line is used for the mass-loss-rate estimate.}
\end{table*}

%
%
%
\subsection{Objects with detached dust CSEs}
\label{s:de_objects}

About a third of our inner-GB objects belongs to a distinct class of sources where the SEDs indicate that the dust CSEs are detached from the central source, that is their mass loss has ceased some time ago, the DE objects, see Fig.~\ref{f:all_seds}. The fits to their SEDs also suggest that their central stars are warmer. These objects are non-periodic, or, in one case (\object{OH359.140+1.137}), show non-Mira-like variability of very low amplitude. These characteristics strongly indicate that the DE objects are the most evolved in our total sample, that is they lie close to the tip of the AGB, and in some cases may even have left the AGB. The two sources with resolved structure in the CO line brightness distributions are both of the DE type, and they are probably the objects that are most evolved. As a guideline the maximum estimated inner radius, 600\,au, corresponds to a time scale of $\approx$\,150\,yr for an expansion velocity of 20\,km\,s$^{-1}$. It may be that our way of selecting the sample has lead to a bias towards objects with DE characteristics, but it is difficult to quantify the magnitude of any bias, if present.

Unfortunately, our sample is too small to allow a detailed comparison of the circumstellar characteristics of the SE and DE sources. This can only be done by increasing the sample size and by observing emission from more $^{12}$CO and $^{13}$CO transitions. For illustration, we have separated the SE and DE objects in the figures presenting the luminosity, mass-loss rate, expansion velocity, and $^{12}$CO/$^{13}$CO distributions.

The estimated gas-to-dust-mass ratios warrants a discussion, however. For the DE sources we find a range in $R_{\rm gd}$ from 5 to 143, and a median of 36. These are low values, and a possible explanation is that the SED fits give erroneous results for $\dot{M_{\rm d}}$. Furthermore, the dust temperatures at the inner radii of the dust CSEs are remarkably similar for all the DE sources, 185$\pm$25\,K. The fact that the SEDs of all the DE objects peak at the same wavelength, $\approx$\,25\,$\mu$m, is in line with this. In the context of the adopted model, an isotropic dust density distribution with different inner radii, this is difficult to understand. This opens up the question of an alternative, non-isotropic, dust density distribution for the DE sources as, for instance, suggested by \citet{decietal19} for evolved OH/IR stars. This can be analysed along the lines of \citet{wiegetal20}, but it would lead too far to test this out in detail in this paper. The conclusion for now is that there is no obvious way to produce the amount of emission we see with significantly less dust, and hence increase the $R_{\rm gd}$. The main argument is that the optically thin part of the dust emission spectrum starting at about 30\,$\mu$m to longer wavelengths needs a certain amount of dust at relatively low temperatures. There could be warmer dust contributing at those wavelengths but that would have to be hidden by high optical depths at shorter wavelengths.

Finally, we note here that the estimated inner radii of the dust CSEs of the DE sources are small enough that they will not affect the intensities of the CO lines discussed in this paper. Our radiative transfer calculations show that the intensities of the CO lines are not affected by increasing the inner radius of the CO envelope to, for instance, 600\,au. For OH, the situation is different. Using the results of \citet{netzknap87} we estimate that the radius of the OH shell is $\approx$\,2000\,au for a mass-loss rate of 2$\times$10$^{-5}$\,$M_\odot$\,yr$^{-1}$ and an expansion velocity of 18\,km\,s$^{-1}$ (our median values). However, if the interstellar UV radiation field is stronger in the inner GB than in the solar neighbourhood, this would lead to smaller molecular envelopes. As discussed in Sect.~\ref{s:phdiss}, the CO photodissociation radii of the inner-GB sub-sample sources cannot be estimated from our data. In fact, they could be substantially smaller without affecting the observed source sizes and the estimated mass-loss rates. Since the H$_2$O and OH molecules are photodissociated in the continuum, and not in lines as CO [see \citet{vandi88} for a description of photodissociation in continuum and in lines], the size of the OH shell is more affected by an increased UV radiation field than is the CO envelope. An OH shell about three times smaller than that obtained for a solar neighbourhood UV radiation field, that is about 700\,au for our median values of mass-loss rate and expansion velocity, fits very well a scenario where sources with CSE inner radii in excess of $\approx$\,700\,au would not be detectable as OH/IR stars, and hence not end up in our sample.

%
%
%
\subsection{Suspected foreground objects}
\label{s:foreground}

For five of our sample objects the luminosities obtained, if placed at our adopted distance to the GC, lie at the maximum luminosity of an AGB star ($\approx$\,50\,000\,$L_\odot$) or considerably higher. These objects are therefore in all likelihood foreground objects, Table~\ref{t:summary_foreground}. They all belong to the SE class. Periods have been determined for two, two are non-periodic, and for the remaining star the photometry fluxes are saturated. All five objects are detected in 324\,GHz continuum. For these objects, the SEDs shown in Fig.~\ref{f:all_seds} are not extinction-corrected. It should be noted that neither of these sources, nor any other sources in our total sample, have a useful parallax in Gaia EDR3 \citep{gaiaetal20}.

\begin{table*}[t]
\caption{Results for the suspected foreground objects.}
\begin{tabular}{l c c c c c r c c c }
\hline \hline
Source                        & SED   & $P$\,$^{(a)}$  & $L$\,$^{(b)}$  & $D_{\rm L}$ & $\dot{M}_{\rm g}$\,$^{(c)}$       & $R_{\rm gd}$\,$^{(d)}$  &  $\varv_{\rm sys}$\,$^{(e)}$  &  $\varv_\infty$\,$^{(e)}$  &  $^{12}$CO/$^{13}$CO\,$^{(f)}$ \\
                              & type  & [days]         & [$L_\odot$]    & [kpc]       & [$M_\odot$\,yr$^{-1}$]  &               &  [km\,s$^{-1}$]        &  [km\,s$^{-1}$]  \\
\hline 
2\,-\,\phantom{0}\object{OH358.162+0.490}     & SE    &  1230          & 5600           &  1.2        & 1$\times$10$^{-5}$      & 100           &  \phantom{--0}3\phantom{:}       &  23\phantom{:}                 &  6\phantom{:}                  \\
3\,-\,\phantom{0}\object{OH358.235+0.115}     & SE    & \phantom{1}874 & 5600           &  0.7        & 3$\times$10$^{-7}$      &  10           &  $-$21\phantom{:}                 &  19\phantom{:}                 & 8:                    \\
6\,-\,\phantom{0}\object{OH359.149$-$0.043}   & SE    &  np            & 5600           &  2.8        & $\ldots$                & $\ldots$      &  \phantom{--}48:                  &  26:                           &  $\ldots$             \\
18\,-\,\object{OH0.548$-$0.059}               & SE    &  np            & 5600           &  2.8        & $\ldots$                & $\ldots$      &  --48:                            &  25:                           &  $\ldots$             \\
20\,-\,\object{OH1.095$-$0.832}               & SE    &  sat           & 5600           &  1.2        & 3$\times$10$^{-6}$      &  30           & \phantom{--}11\phantom{:}         &  22\phantom{:}                 &  3\phantom{:}              \\                        
\hline
\end{tabular}
\label{t:summary_foreground}
\tablefoot{$^{(a)}$ The adopted period from Table~\ref{t:summary_periods} (np = non-periodic, sat = saturated data). $^{(b)}$ Assumed to be the same as the median value for the inner-GB sample. $^{(c)}$ The uncertainty is estimated to be a factor of five. $^{(d)}$ The gas-to-dust mass ratio estimated as $\dot{M}_{\rm g}/\dot{M}_{\rm d}$. The uncertainty is estimated to be a factor of seven. $^{(e)}$ Estimated as weighted averages of the results for the detected CO lines. The uncertainties are estimated to be $\pm$\,1.0 and $\pm$\,1.5\,km\,s$^{-1}$ for the systemic and expansion velocities, respectively. A colon indicates an uncertain value. $^{(f)}$ The uncertainty is estimated to be a factor of four. A colon indicates an uncertain value.}
\end{table*}

Two objects, \object{OH358.162+0.490} and \object{OH1.095$-$0.832}, are the, by far, strongest emitters in the $^{12}$CO($J$\,=\,2--1) line in our total sample, Table~\ref{t:12co21}. The CO line profiles have high S/N, and have the characteristics expected of a thermal (i.e. not masing) line from an AGB CSE, Figs.~\ref{f:13co_spectra}, \ref{f:12co21_spectra}, and \ref{f:12co32_spectra}. The brightness distributions are centrally peaked and resolved. The latter is reflected in the $^{13}$CO($J$\,=\,3--2) line profiles which are double-peaked as expected for circumstellar, optically thin, resolved line emission. Their OH 1612\,MHz lines are the, by far, strongest in our total sample, and their line profiles are the characteristic double-peaked ones \citep{seveetal97a}. Their systemic and terminal gas expansion velocities are 3 and 11\,km\,s$^{-1}$ and 23 and 22\,km\,s$^{-1}$, respectively, as determined from our CO data. They are also the two strongest 324\,GHz continuum sources in our total sample. Both coincide with strong IRAC sources, Fig.~\ref{f:co_spitzer}. OH358.162+0.490 show long-period Mira-like variability, while for OH1.095$-$0.832 the photometry data are saturated. These objects show all the characteristics of being AGB stars.  

In the case of \object{OH358.235+0.115}, the CO line data have relatively low S/N, Fig.~\ref{f:all_images}, but the emissions are centrally peaked and coincide with a strong 5.8\,$\mu$m IRAC source, Fig.~\ref{f:co_spitzer}. The OH 1612\,MHz line shape is not the characteristic one, being triple-peaked rather than double-peaked and having no sharp outer edges \citep{seveetal97a}. The systemic and gas expansion velocities estimated from the CO and OH data differ, the results from the CO data are $-$21 and 19\,km\,s$^{-1}$, respectively. The object is relatively strong in 324\,GHz continuum. The variability is periodic, so this object is most likely an AGB star despite the strange OH line profile. 

\object{OH359.149$-$0.043} exhibits very strange brightness distributions in all three CO lines in the form of arc-like structures extending about 2\farcs5, Fig.~\ref{f:all_images}. Nevertheless, the emission lies in the vicinity of a strong 8\,$\mu$m IRAC source, Fig.~\ref{f:co_spitzer}. The CO line shapes appear reasonable from a circumstellar perspective (Figs.~\ref{f:13co_spectra}, \ref{f:12co21_spectra}, and \ref{f:12co32_spectra}), although the $^{13}$CO($J$\,=\,3--2) line is somewhat sharply peaked, but there is a problem with interstellar CO line emission that cannot be removed by eliminating short-baseline data without losing flux in the circumstellar CO lines. The OH 1612\,MHz line shape is double-peaked, but narrow (covering about 15\,km\,s$^{-1}$) and not apparently of the characteristic type \citep{seveetal97a}. The CO data suggest a considerably higher gas expansion velocity, 26\,km\,s$^{-1}$. A weak continuum source at 324\,GHz is detected at the position of the near-IR source. This is tentatively an AGB star, but the structures in the CO line brightness distributions suggest that its circumstellar characteristics are not of the normal type, so we refrain from modelling the CO line emission.

\object{OH0.548$-$0.059} is also an object with strange characteristics. The CO line brightness distributions are centrally peaked, Fig.~\ref{f:all_images}, just about resolved, and they coincide with a strong 5.8\,$\mu$m IRAC source, Fig.~\ref{f:co_spitzer}. The CO line profiles are more difficult to interpret. The $^{13}$CO($J$\,=\,3--2) line is broad (covering about 90\,km\,s$^{-1}$) and covers the same velocity range as the OH 1612\,MHz line (Fig.~\ref{f:13co_spectra}), although the OH data are of limited quality and the double-peak structure is not obvious \citep{lindetal92a}. Unfortunately, the $^{12}$CO line spectra are of very low quality (Figs.~\ref{f:12co21_spectra} and \ref{f:12co32_spectra}), mainly due to the problem with interstellar CO line emission along the line of sight, and they do not provide any insight into the circumstellar characteristics. The ALMA continuum fluxes lie high above the fit to the SED, Fig.~\ref{f:all_seds}. Also in this case we refrain from modelling the CO line emission.

The most reasonable assumption is that the three objects for which we attempt a CO line modelling are of the same character as our inner-GB objects, that is they have a luminosity of 5600\,$L_\odot$, but lie closer to us. We note that for the two objects with determined periods, this luminosity lies well below those obtained from PL relations (see Sect.~\ref{s:lum} for a discussion on this). Using a luminosity of 5600\,$L_\odot$ and the non-extinction-corrected SEDs, we derive the distances $D_{\rm L}$ (the same approach is used for the remaining two sources), see Table~\ref{t:summary_foreground}. They lie in the range 0.7 to 1.2\,kpc. This may indicate that the distances are under-estimated, since it is more likely that the sources are spread out over the distance to the GC. Their systemic velocities are within the $|\varv_{\rm sys}|$\,$\la$\,20\,km\,s$^{-1}$ range.

We have derived their circumstellar characteristics using the $D_{\rm L}$ distance. For \object{OH358.162+0.490} and \object{OH1.095$-$0.832} the gas mass-loss rates lie 2 and 7 times, respectively, below the median for the inner-GB sample. For \object{OH358.235+0.115} the gas mass-loss rate is as low as 3$\times$10$^{-7}$\,$M_\odot$\,yr$^{-1}$, much lower than expected for a proper OH/IR star, suggesting that either the circumstellar model is wrong for this object, or its distance is substantially larger. In fact, its $R_{\rm gd}$ is as low as 10 suggesting that something is wrong. For the two other objects the $R_{\rm gd}$ values are 30 and 100, also low compared to the inner-GB sample. We note that the $R_{\rm gd}$ estimates are, to a first approximation, distance-independent (the necessity to apply extinction correction will change this). The $^{12}$CO/$^{13}$CO ratios are 6 and 3 for \object{OH358.162+0.490} and \object{OH1.095$-$0.832}, respectively, in line with the results for the inner-GB objects. The corresponding result for \object{OH358.235+0.115}, 8, must be considered very uncertain.

Finally, the three objects that we consider to be firmly established AGB stars among the foreground objects are the only objects in our full sample in which we detect also the H$^{13}$CN($J$\,=\,\mbox{4--3}) line, Table~\ref{t:h13cn43} and Fig.\ref{f:h13cn43_spectra}. Despite being a C-bearing species, hydrogen cyanide (HCN) is ubiquitously detected also towards O-type AGB stars \citep{schoetal13}. However, its circumstellar abundance is down by more than two orders of magnitude compared to that estimated in C-type objects \citep{schoetal13}. Consequently, the HCN rotational lines in O-type objects are considerably weaker than the CO rotational lines, for instance, in the range 10--20\% for the HCN($J$\,=\,\mbox{4--3})/CO($J$\,=\,\mbox{3--2}) intensity ratio \citep{schoetal13,ramsolof14}. Therefore, our detections of the rarer isotopologue, H$^{13}$CN, were done at surprisingly high intensity levels, about 10--30\% of that of the $^{12}$CO($J$\,=\,\mbox{3--2}) lines.  Thus, this result further strengthen the conclusion that our sources are rich in $^{13}$C. We note here that, as opposed to the case of CO, HCN is photodissociated in the continuum and hence we do not expect any effects of circumstellar isotope-selective processes \citep{sabeetal20}.

%
%
%
%
\section{Conclusions}

Observing OH/IR stars in CO line radiation towards regions of high extinction is a difficult task due to the contamination of ubiquitous interstellar CO along the line of sight. At the distance of the GC this is further exacerbated by the weakness of the circumstellar line and continuum radiation. Nevertheless, through the use of the ALMA interferometer both of these problems have been, at least partly, circumvented, and with observations of two rotational lines of $^{12}$CO, one of $^{13}$CO, and mm-wave continuum emission, we can infer interesting characteristics of a sample of 22 OH/IR stars in directions within 2$^\circ$ of the GC, the inner GB. All objects are detected in at least one CO line, and eight of them are detected in 324\,GHz continuum. The CO line brightness distributions coincide with infrared point sources. All sources are resolved in the line emission. Only two of them show circumstellar structure (a cavity and a bipolar structure), one shows a very strange brightness distribution (in this context), while the rest have centrally peaked, circular symmetric emission (at 0\farcs15 resolution). Thirteen objects show Mira-like variability with well-determined periods, seven show non-periodic or very-low-amplitude variability, one show multi-period non-Mira-like variability, while the fluxes are saturated for the remaining object. 

Spectral energy distributions have been constructed from extinction-corrected photometry and luminosities have been estimated. This leads to a division (based on luminosity criteria) of our sample into 17 objects that are most likely located in the inner GB, and five objects that are most likely foreground sources. Further, the sub-sample of inner-GB objects is divided into two classes depending on the SED characteristics: eleven objects have standard dust CSEs (SE), while six have detached dust CSEs and indications of warmer central stars (DE). The SE sources all have long-period Mira-like variability, while the DE sources are predominantly non-periodic. A reasonable explanation for this is that the DE sources have evolved further than the SE sources. Most likely they have reached a stage where the mass-loss rate has decreased significantly, and some of them may have already left the AGB, evolving to the left in the Hertzsprung-Russell diagram towards the white dwarf stage. The two sources with resolved circumstellar structure in CO line emission, in both cases in the form of a cavity and a bipolar structure, belong to the DE class and are probably the most evolved. The two classes appear to be rather distinct, and this suggests that the transition from one stage to the other is fast, of the order $\la$\,100\,yr, if our evolutionary interpretation is correct. It is also apparent that, when data at longer wavelengths ($\ga$\,70\,$\mu$m) are available, the SED fits are often not very good, the flux densities are underestimated. This applies to ten objects, and it indicates more complex geometries of the dust CSEs, a varying mass-loss-rate history, a possible effects of grain size, and/or the use of inadequate dust optical properties.

For the inner-GB sub-sample, we derived luminosities that imply that the sources (except for two higher-luminosity sources) are lower-mass stars of median mass $\approx$\,1.2\,$-$\,1.6\,$M_\odot$, and a population age of $\approx$\,4\,$-$\,7\,Gyr (the uncertainty depends on the stellar evolution model being used). The two higher-luminosity objects, with estimated masses of $\approx$\,4.3\,$M_\odot$, belong to a much younger population ($<$\,1\,Gyr), unless they are foreground sources. For the sources with determined variability period, the luminosities are, at least, a factor of two lower than those estimated from commonly used PL relations. This may be interpreted as an effect of substantial mass loss, lowering the present stellar mass, possibly augmented by a higher metallicity. 

Despite the relatively low mass for the inner-GB sub-sample stars, the mass-loss rates are high, a median of 2$\times$10$^{-5}$\,$M_\odot$\,yr$^{-1}$, and the CSEs expand at high terminal velocities, a median of 18\,km\,s$^{-1}$. Consequently, these objects fall on the high-end tail of AGB stellar mass-loss-rate and gas expansion velocity distributions estimated for solar-neighbourhood O-type AGB stars \citep{ramsetal09}. The high expansion velocities suggest that our inner-GB sources are not low-metallicity objects.  Further, the inner-GB sources have very low estimated circumstellar $^{12}$CO/$^{13}$CO abundance ratios, a median of five, and by inference very low stellar $^{12}$C/$^{13}$C isotope ratios. This is theoretically very difficult to understand, since our sources are not expected to go through HBB. Such low ratios can possibly be set during the evolution on the Red Giant Branch if effective extra mixing is present. Unfortunately, our sample size is not large enough to allow a comparison between the luminosities, mass-loss rates, gas expansion velocities, and circumstellar $^{12}$CO/$^{13}$CO abundance ratios of the SE and DE sources. 

The estimated gas-to-dust ratios for the inner-GB SE sources have a median of 200 and quite some spread, in line with what has been reported in other studies of different samples of AGB stars. On the contrary, the estimated low gas-to-dust ratios for the DE sources indicate that the dust-mass-loss rate estimates are wrong, most likely due to the adopted geometry of their dust CSEs being wrong. A further indication of the latter is the uniformity of the dust temperature at the inner radius of the dust CSE, which is difficult to explain within the adopted isotropic dust CSE.

Our data do not allow an observational determination of the CO photodissociation radii of our inner-GB sources. In fact, these radii can be substantially smaller than calculated using Eq.\,(\ref{e:rph}), for instance, as a result of a stronger interstellar UV radiation field close to the GC (however, the CO photodissociation radius of a higher-mass-loss-rate object is only moderately sensitive to the strength of the UV field), without affecting the estimated source sizes and mass-loss rates. On the other hand the size of the OH shell is more susceptible to a change in the UV radiation field than is the CO envelope. An OH shell three times smaller than that estimated for a solar neighbourhood interstellar UV radiation field would nicely explain why we have no objects with inner dust-CSE radii larger than $\approx$\,600\,au. They would not show any OH\,1612\,MHz emission, and hence would not end up in our sample.

Our CO line radiative transfer modelling fails in two respects. First, the observed $^{12}$CO $J$\,=\,3$-$2/2$-$1 line intensity ratios (primarily determined by the mass-loss rate in the model) are not reproduced in several cases. This could be caused by the limited S/N of the observational data, but it may also indicate that the use of the same circumstellar model for all sources is a too crude approach. Second, the observed presence of extended, low-level, brightness emission in many cases is also not reproduced in the models. There are several possible reasons for this, such as a mass-loss rate varying with time or a non-isotropic CSE.

Three of the foreground sources are likely AGB stars, while the nature of the remaining two is uncertain. We present results for the AGB stars assuming they have a luminosity of 5600\,$L_\odot$, but note that for the two sources with determined periods, this value is significantly below those predicted by PL relations.

In summary we have reached the following conclusions:
\begin{itemize}
\item{For a sample of 22 OH/IR stars in the direction of the inner GB (within 2$^\circ$ of the GC), we detect all in at least one CO rotational line using ALMA. Eight of them are detected in 324\,GHz continuum.}
\item{Periods of Mira-like pulsational variability have been determined for 12 objects, and one period is taken from the literature. They fall in the range about 500 to 1200 days. Six objects show no periodic variability.} 
\item{Based on luminosity criteria, 17 of these objects are believed to be located in the inner GB. The conclusion points below, except the last one, refer to this sub-sample.}
\item{The median luminosity, 5600\,$L_\odot$, corresponds to an initial mass in the range 1.2\,$-$\,1.6\,$M_\odot$. Two stars have significantly higher luminosities, 25\,000\,$L_\odot$, suggesting that they are higher-mass stars, or lie in the foreground.}
\item{For the lower-mass stars the luminosities fall well below those predicted by established period-luminosity relations, most likely due to these stars having reached the end of the AGB evolution.}
\item{The objects are further divided into stars that are approaching the tip of the AGB (about two thirds of the objects) and stars that have recently left the AGB. The former stars all show Mira-like variability, while the latter stars show no, or very-low-amplitude, periodic variability. In addition, the latter stars appear to be warmer.} 
\item{The median gas-mass-loss rate, gas terminal expansion velocity, gas-to-dust mass ratio, and circumstellar $^{12}$CO/$^{13}$CO abundance ratio are estimated to be 2$\times$10$^{-5}$\,$M_\odot$\,yr$^{-1}$, 18\,km\,s$^{-1}$, 200 (excluding the sources that have recently left the AGB), and 5, respectively.}
\item{Both the median mass-loss rate and gas expansion velocity lie at the high end of values found for AGB stars. The latter can be taken as an indication that these stars are not low-metallicity objects.}
\item{The low circumstellar $^{12}$CO/$^{13}$CO abundance ratios, and hence inferred stellar $^{12}$C/$^{13}$C isotope ratios, are most likely produced through efficient dredge-up on the RGB.}
\item{Resolved structure in the CO line brightness distributions are seen for only two sources. These belong to the group of stars that are believed to have left the AGB.}
\item{It is unlikely that the UV radiation field in the inner GB is so high that it has any effect on our estimated mass-loss rates and $^{12}$CO/$^{13}$CO abundance ratios. On the contrary, the OH shell size is more sensitive to the strength of the UV radiation field, and this may explain why there are no sources in our sample with an inner radius of the dust CSE larger than 600\,au.}
\item{Our circumstellar CO line radiative transfer modelling fails in two respects. The observed $^{12}$CO $J$\,=\,3$-$2/2$-$1 line intensity ratios are not reproduced in several cases. The observed presence of extended, low-level, brightness emission in many cases is also not reproduced in the models.}
\item{Finally, the remaining five sources are likely foreground objects. Three are likely AGB stars, while the nature of two is uncertain.}
\end{itemize}

More firmly establishing the findings presented in this paper requires the observations of a larger sample of inner-GB OH/IR stars (in particular of the DE type) in more $^{12}$CO lines (to significantly improve the circumstellar model and the gas-mass-loss-rate estimates), in more $^{13}$CO lines (to significantly improve the estimates of the circumstellar $^{12}$CO/$^{13}$CO abundance ratio), and in continuum (to put better constraints on the long-wavelength behaviour of the SEDs). Given that our data on a sample of 22 stars were obtained with a short observing time, it is certainly feasible to achieve such an improvement within a reasonable amount of observing time. Such a sample of equidistant sources would be excellent for studying the evolution of the mass-loss characteristics of solar-type stars at the tip, and slightly beyond, the AGB.

%
%
%
%

\begin{acknowledgements}
This paper makes use of the following ALMA data: ADS/JAO.ALMA\#2016.1.00983.S. ALMA is a partnership of ESO (representing its member states), NSF (USA) and NINS (Japan), together with NRC (Canada) and NSC and ASIAA (Taiwan) and KASI (Republic of Korea), in cooperation with the Republic of Chile. The Joint ALMA Observatory is operated by ESO, AUI/NRAO and NAOJ. We acknowledge support from the Nordic ALMA Regional Centre (ARC) node based at Onsala Space Observatory. The Nordic ARC node is funded through Swedish Research Council grant No 2017-00648. This research has made use of the VizieR catalogue access tool, CDS, Strasbourg, France (the original description of the VizieR service was published in \citet{ochsetal00}), and the NASA/IPAC Infrared Science Archive (IRSA), which is funded by the US National Aeronautics and Space Administration (NASA) and operated by the California Institute of Technology. This publication makes use of data products from the Akari satellite developed and operated by the Japan Aerospace Exploration Agency (JAXA) in cooperation with institutes in Europe and Korea, the Infrared Astronomical Satellite (IRAS) developed and operated by the Netherlands Agency for Aerospace Programs (NIVR), NASA, and the UK Science and Engineering Research Council (SERC), the ESA missions the Infrared Space Observatory (ISO) and the Herschel Space Observatory, the Mid Course Space Experiment (MSX) operated by the  Ballistic Missile Defense Organization (BMDO), the Panoramic Survey Telescope and Rapid Response System (Pan-STARRS), a collaboration between the University of Hawaii, MIT Lincoln Laboratory, Maui High Performance Computing Center and Science Applications International Corporation, the Spitzer Space Telescope operated by the Jet Propulsion Laboratory, California Institute of Technology, under NASA contract 1407, the Two Micron All Sky Survey (2MASS), a joint project of the University of Massachusetts and the Infrared Processing and Analysis Center/California Institute of Technology funded by NASA and the National Science Foundation, the UKIDSS Galactic Plane Survey on the United Kingdom Infrared Telescope (UKIRT), the Visible and Infrared Survey Telescope for Astronomy (VISTA) operated by ESO, and the Wide-field Infrared Survey Explorer (WISE), a joint project of the University of California, Los Angeles, and the Jet Propulsion Laboratory/California Institute of Technology funded by NASA. We acknowledge the use of the ADS bibliographic services. We are grateful to the referee, Anita Richards, for constructive and clarifying comments. We thank Tom Henry for retrieving the WISE data used for the variability study of our sample. NR acknowledges support from the Royal Physiographic Society in Lund through the Stiftelse Walter Gyllenbergs fond and M{\"a}rta och Erik Holmbergs donation. SS acknowledges support from UNAM-PAPIIT Program IA104822. 
\end{acknowledgements}



%
%
%
%
%
\begin{appendix}

\section{Observational results}
\label{a:obs_res}

\subsection{Tables}
\label{a:tables}

Tables with the observational results for the $^{12}$CO($J$\,=\,\mbox{2--1} and \mbox{3--2}), $^{13}$CO($J$\,=\,\mbox{3--2}), and H$^{13}$CN($J$\,=\,\mbox{4--3}) lines, and the 222, 324, and 339\,GHz continuum data are presented here.

\begin{table*}[h!]
\caption{Results for the $^{12}$CO $J$\,=\,2--1 line.}
\begin{tabular}{lccccccl}
\hline \hline
Source                          &  $S$\,$^{(a)}$  &  $\int S{\rm d}\varv$  &  $\varv_{\rm c}$\,$^{(b)}$  &  $\varv_\infty$\,$^{(b)}$  &  $\beta$            &  $\theta_{\rm s}$\,$^{(c)}$  &  Remarks  \\
                                &  [Jy]           &  [Jy\,km\,s$^{-1}$]       &  [km\,s$^{-1}$]      &  [km\,s$^{-1}$]     &                     &  [\arcsec]       &  \\
\hline 
1\,-\,\phantom{0}\object{OH358.083+0.137}       &  0.35           & \phantom{00}8.7           &  \phantom{0}--24     &  19                 &  \phantom{--}0.5    & 0.38             & Low S/N\\
2\,-\,\phantom{0}\object{OH358.162+0.490}       &  6.0\phantom{0} & 250\phantom{.0}           &  \phantom{--00}3     &  20                 &  \phantom{--}0.0    & 1.46     \\
3\,-\,\phantom{0}\object{OH358.235+0.115}       &  0.45           & \phantom{0}14\phantom{.0} &  \phantom{0}--25     &  19                 &  \phantom{--}0.0    & 0.63             & \\
4\,-\,\phantom{0}\object{OH358.505+0.330}       &  1.5\phantom{0} & \phantom{0}31\phantom{.0} &  \phantom{0}--30     &  25                 &  --0.5              & 0.70             & $B$\,$>$\,25\,m data used \\
5\,-\,\phantom{0}\object{OH359.140+1.137}       &  0.78           & \phantom{0}12\phantom{.0} &  --137               &  11                 &  \phantom{--}0.5    & 0.44      \\
6\,-\,\phantom{0}\object{OH359.149$-$0.043}     &  0.31           & \phantom{0}11\phantom{.0} &  \phantom{--0}53     &  27                 &  \phantom{--}1.0    & $\ldots$         & $B$\,$>$\,50\,m data used; Brightness distr.?\\
7\,-\,\phantom{0}\object{OH359.220+0.163}       &  2.3\phantom{0} & \phantom{0}50\phantom{.0} &  --138               &  15                 &  \phantom{--}0.5    & 1.25             & Resolved structure\\
8\,-\,\phantom{0}\object{OH359.233$-$1.876}     &  0.98           & \phantom{0}23\phantom{.0} &  \phantom{0}--16     &  17                 &  \phantom{--}0.5    & 0.60      \\
9\,-\,\phantom{0}\object{OH359.467+1.029}       &  0.30           & \phantom{0}11\phantom{.0} &  \phantom{--}110     &  16                 &  \phantom{--}0.5    & 0.58        \\
10\,-\,\object{OH359.543$-$1.775}               &  0.15           & \phantom{00}7.3           &  \phantom{--0}95     &  32                 &  \phantom{--}0.5    & 0.31             & Low S/N \\
11\,-\,\object{OH359.664+0.636}                 &  0.91           & \phantom{0}16\phantom{.0} &  --158               &  13                 &  \phantom{--}0.5    & 0.84       \\
12\,-\,\object{OH359.745$-$0.404}               &  0.44           & \phantom{00}9.9           &  \phantom{--}184     &  22                 &  \phantom{--}0.5    & 0.69       \\
13\,-\,\object{OH359.805+0.200}                 &  0.22           & \phantom{00}2.9           &  \phantom{0}--48     &  15                 &  \phantom{--}0.3    & $\ldots$         & $B$\,$>$\,25\,m data used; Low S/N  \\
14\,-\,\object{OH359.826+0.153}                 &  0.12           & \phantom{00}2.5           &  \phantom{0}--26     &  16                 &  \phantom{--}0.3    & 0.51             & $B$\,$>$\,50\,m data used; Low S/N  \\
15\,-\,\object{OH359.902+0.061}                 &  0.51           & \phantom{00}8.4           &  --123               &  13                 &  \phantom{--}0.5    & 0.43             & $B$\,$>$\,25\,m data used; Low S/N \\
16\,-\,\object{OH0.173+0.211}                   &  0.31           & \phantom{00}7.5           &  \phantom{--0}52     &  18                 &  \phantom{--}1.0    & 0.52             & $B$\,$>$\,25\,m data used   \\
17\,-\,\object{OH0.221+0.168}                   &  0.13           & \phantom{00}4.9           &  \phantom{--0}60     &  26                 &  \phantom{--}1.0    & 0.44             & $B$\,$>$\,50\,m data used \\
18\,-\,\object{OH0.548$-$0.059}                 &  0.10           & \phantom{00}2.0           &  \phantom{0}--53     &  22                 &  --0.5              & 0.59             & $B$\,$>$\,50\,m data used; Low S/N \\
19\,-\,\object{OH0.739+0.411}                   &  0.81           & \phantom{0}13\phantom{.0} &  \phantom{0}--21     &  12                 &  \phantom{--}0.3    & 1.04             & $B$\,$>$\,25\,m data used; Line profile? \\
20\,-\,\object{OH1.095$-$0.832}                 &  2.1\phantom{0} & 104\phantom{.0}           &  \phantom{--0}10     &  23                 &  --0.3              & 0.91    \\
21\,-\,\object{OH1.221+0.294}                   &  2.2\phantom{0} & \phantom{0}44\phantom{.0} &  --261               &  14                 &  \phantom{--}0.5    & 0.57     \\
22\,-\,\object{OH1.628+0.617}                   &  0.86           & \phantom{0}34\phantom{.0} &  \phantom{0}--84     &  30                 &  \phantom{--}1.0    & 0.80             \\                        
\hline
\end{tabular}
\label{t:12co21}
\tablefoot{$^{(a)}$ See text for details on the uncertainties of the flux densities, but we estimate that for lines stronger than 1.0\,Jy the uncertainty is about 10\,\%, while for lines weaker than 0.4\,Jy the uncertainty increases to about 50\,\% for the weakest lines. $^{(b)}$ The uncertainties in the expansion and centre velocities are estimated to be about $\pm$2 and $\pm$1.5\,km\,s$^{-1}$, respectively. $^{(c)}$ Average of the deconvolved major and minor axes of a two-dimensional Gaussian fit to the brightness distribution.}
\end{table*}

\FloatBarrier

\begin{table*}[h!]
\caption{Results for the $^{12}$CO $J$\,=\,3--2 line.}
\begin{tabular}{lccccccl}
\hline \hline
Source                          &  $S$\,$^{(a)}$  &  $\int S{\rm d}\varv$  &  $\varv_{\rm c}$\,$^{(b)}$  &  $\varv_\infty$\,$^{(b)}$  &  $\beta$            &  $\theta_{\rm s}$\,$^{(c)}$  &  Remarks  \\
                                &  [Jy]           &  [Jy\,km\,s$^{-1}$]       &  [km\,s$^{-1}$]      &  [km\,s$^{-1}$]     &                     &  [\arcsec]       &  \\
\hline 
1\,-\,\phantom{0}\object{OH358.083+0.137}       &  0.68           & \phantom{0}20\phantom{.0} &  \phantom{0}--23     &  21                 &  \phantom{--}0.5    & 0.52             & Low S/N\\
2\,-\,\phantom{0}\object{OH358.162+0.490}       &  9.4\phantom{0} & 320\phantom{.0}           &  \phantom{--00}4     &  26                 &  \phantom{--}0.5    & 0.70     \\
3\,-\,\phantom{0}\object{OH358.235+0.115}       &  0.91           & \phantom{0}32\phantom{.0} &  \phantom{0}--21     &  19                 &  \phantom{--}0.0    & 0.46             & Low S/N \\
4\,-\,\phantom{0}\object{OH358.505+0.330}       &  2.2\phantom{0} & \phantom{0}57\phantom{.0} &  \phantom{0}--34     &  27                 &  --0.2              & $\ldots$        & Low S/N \\
5\,-\,\phantom{0}\object{OH359.140+1.137}       &  0.26           & \phantom{00}5.2           &  --134               &  13                 &  \phantom{--}0.0    & $\ldots$         & $B$\,$>$\,25\,m data used\\
6\,-\,\phantom{0}\object{OH359.149$-$0.043}     &  2.0\phantom{0} & \phantom{0}57\phantom{.0} &  \phantom{--0}47     &  27                 &  \phantom{--}1.0    & $\ldots$         & $B$\,$>$\,25\,m data used; Brightness distr.?\\
7\,-\,\phantom{0}\object{OH359.220+0.163}       &  2.1\phantom{0} & \phantom{0}40\phantom{.0} &  --139               &  13                 &  \phantom{--}0.5    & 1.26             & Brightness distr.? \\
8\,-\,\phantom{0}\object{OH359.233$-$1.876}     &  0.65           & \phantom{0}15\phantom{.0} &  \phantom{0}--11     &  21                 &  \phantom{--}0.5    & 0.31             & Low S/N   \\
9\,-\,\phantom{0}\object{OH359.467+1.029}       &  0.74           & \phantom{0}18\phantom{.0} &  \phantom{--}110     &  15                 &  \phantom{--}0.5    & 0.58             & Low S/N   \\
10\,-\,\object{OH359.543$-$1.775}               &  0.18           & \phantom{00}8.4           &  \phantom{--0}90     &  37                 &  \phantom{--}0.5    & 0.27             & Low S/N \\
11\,-\,\object{OH359.664+0.636}                 &  0.28           & \phantom{00}4.8           &  --158               &  14                 &  \phantom{--}0.5    & $\ldots$         & Low S/N       \\
12\,-\,\object{OH359.745$-$0.404}               &  0.25           & \phantom{00}6.4           &  \phantom{--}189     &  20                 &  \phantom{--}0.5    & $\ldots$          & $B$\,$>$\,25\,m data used; Low S/N        \\
13\,-\,\object{OH359.805+0.200}                 &  $<$0.3         & $\ldots$                  &  $\ldots$            &  $\ldots$           &  $\ldots$           & $\ldots$          \\
14\,-\,\object{OH359.826+0.153}                 &  0.25           & \phantom{00}6.7           &  \phantom{0}--26     &  19                 &  \phantom{--}0.0    & $\ldots$ \\
15\,-\,\object{OH359.902+0.061}                 &  0.29           & \phantom{00}6.5           &  --129               &  11                 &  \phantom{--}0.5    & 0.24             & Low S/N \\
16\,-\,\object{OH0.173+0.211}                   &  $<$0.6         & $\ldots$                  &  $\ldots$            &  $\ldots$           &  $\ldots$           & $\ldots$ \\
17\,-\,\object{OH0.221+0.168}                   &  0.85           & \phantom{0}13\phantom{.0} &  \phantom{--0}66     &  15                 &  \phantom{--}0.5    & 0.31             & $B$\,$>$\,25\,m data used; Low S/N \\
18\,-\,\object{OH0.548$-$0.059}                 &  0.45           & \phantom{00}8.6           &  \phantom{0}--47     &  23                 &  \phantom{--}0.0    & 0.35             & $B$\,$>$\,50\,m data used; Low S/N \\
19\,-\,\object{OH0.739+0.411}                   &  1.3\phantom{0} & \phantom{0}27\phantom{.0} &  \phantom{0}--21     &  15                 &  \phantom{--}0.5    & 0.33             & $B$\,$>$\,25\,m data used \\
20\,-\,\object{OH1.095$-$0.832}                 &  1.8\phantom{0} & \phantom{0}86\phantom{.0} &  \phantom{--0}11     &  22                 &  \phantom{--}0.0    & 0.83    \\
21\,-\,\object{OH1.221+0.294}                   &  3.0\phantom{0} & \phantom{0}43\phantom{.0} &  --261               &  10                 &  \phantom{--}0.5    & $\ldots$ \\
22\,-\,\object{OH1.628+0.617}                   &  0.45           & \phantom{0}15\phantom{.0} &  \phantom{0}--87     &  25                 &  \phantom{--}0.0    & $\ldots$       \\                        
\hline
\end{tabular}
\label{t:12co32}
\tablefoot{$^{(a)}$ The upper limits are 3$\sigma$. See text for details on the uncertainties of the flux densities, but we estimate that for lines stronger than 2.0\,Jy the uncertainty is about 10\,\%, while for lines weaker than 1.0\,Jy the uncertainty increases to about 50\,\% for the weakest lines. $^{(b)}$ The uncertainties in the expansion and centre velocities are estimated to be about $\pm$3 and $\pm$2\,km\,s$^{-1}$, respectively. $^{(c)}$ Average of the deconvolved major and minor axes of a two-dimensional Gaussian fit to the brightness distribution.}
\end{table*}

\FloatBarrier

\begin{table*}[h!]
\caption{Results for the $^{13}$CO $J$\,=\,3--2 line.}
\begin{tabular}{lccccccl}
\hline \hline
Source                          &  $S$\,$^{(a)}$  &  $\int S{\rm d}\varv$  &  $\varv_{\rm c}$\,$^{(b)}$  &  $\varv_\infty$\,$^{(b)}$  &  $\beta$            &  $\theta_{\rm s}$\,$^{(c)}$  &  Remarks  \\
                                &  [Jy]           &  [Jy\,km\,s$^{-1}$]       &  [km\,s$^{-1}$]      &  [km\,s$^{-1}$]     &                     &  [\arcsec]       &  \\
\hline 
1\,-\,\phantom{0}\object{OH358.083+0.137}                 &  0.18           & \phantom{0}6.4            &  \phantom{0}--26     &  26                 &  \phantom{--}0.5    & 0.37      \\
2\,-\,\phantom{0}\object{OH358.162+0.490}                 &  1.7\phantom{0} & 74\phantom{.0}            &  \phantom{--00}2     &  23                 &  \phantom{--}0.0    & 0.72     \\
3\,-\,\phantom{0}\object{OH358.235+0.115}                 &  0.04           & \phantom{0}2.2            &  \phantom{0}--17     &  18                 &  \phantom{--}0.0    & 0.16             & Low S/N\\
4\,-\,\phantom{0}\object{OH358.505+0.330}                 &  0.91           & 25\phantom{.0}            &  \phantom{0}--26     &  20                 &  \phantom{--}0.5    & 0.93             & Resolved structure\\
5\,-\,\phantom{0}\object{OH359.140+1.137}                 &  0.19           & \phantom{0}3.1            &  --135               &  12                 &  \phantom{--}0.5    & 0.27      \\
6\,-\,\phantom{0}\object{OH359.149$-$0.043}               &  0.28           & \phantom{0}3.1            &  \phantom{--0}46     &  25                 &  \phantom{--}1.0    & $\ldots$         & $B$\,$>$\,25\,m data used; Brightness distr.?\\
7\,-\,\phantom{0}\object{OH359.220+0.163}                 &  0.51           & 16\phantom{.0}            &  --135               &  19                 &  \phantom{--}0.3    & 0.72             & Resolved structure\\
8\,-\,\phantom{0}\object{OH359.233$-$1.876}               &  0.24           & \phantom{0}6.3            &  \phantom{0}--15     &  18                 &  \phantom{--}0.5    & 0.28      \\
9\,-\,\phantom{0}\object{OH359.467+1.029}                 &  0.10           & \phantom{0}2.8            &  \phantom{--}110     &  23                 &  \phantom{--}0.5    & 0.43        \\
10\,-\,\object{OH359.543$-$1.775}               &  0.13           & \phantom{0}6.7            &  \phantom{--0}87     &  40                 &  \phantom{--}0.5    & 0.24             & Line profile? \\
11\,-\,\object{OH359.664+0.636}                 &  0.10           & \phantom{0}1.9            &  --157               &  13                 &  \phantom{--}0.5    & 0.35       \\
12\,-\,\object{OH359.745$-$0.404}               &  0.13           & \phantom{0}3.9            &  \phantom{--}186     &  20                 &  \phantom{--}0.5    & 0.34       \\
13\,-\,\object{OH359.805+0.200}                 &  0.05           & \phantom{0}1.4            &  \phantom{0}--51     &  21                 &  \phantom{--}0.3    & 0.32             & Low S/N  \\
14\,-\,\object{OH359.826+0.153}                 &  0.05           & \phantom{0}1.5            &  \phantom{0}--25     &  16                 &  \phantom{--}0.0    & 0.35             & Low S/N  \\
15\,-\,\object{OH359.902+0.061}                 &  0.06           & \phantom{0}2.2            &  --133               &  19                 &  \phantom{--}0.0    & 0.37             & $B$\,$>$\,25\,m data used; Low S/N \\
16\,-\,\object{OH0.173+0.211}                   &  $<$0.08        & $\ldots$                  &  $\ldots$            &  $\ldots$           &  $\ldots$          & $\ldots$\\
17\,-\,\object{OH0.221+0.168}                   &  0.05           & \phantom{0}1.9            &  \phantom{--0}67     &  27                 &  \phantom{--}1.0    & 0.36             & $B$\,$>$\,25\,m data used; Low S/N \\
18\,-\,\object{OH0.548$-$0.059}                 &  0.15           & \phantom{0}6.0            &  \phantom{0}--45     &  31                 &  \phantom{--}0.5    & 0.37             & $B$\,$>$\,50\,m data used; Line profile? \\
19\,-\,\object{OH0.739+0.411}                   &  0.58           & \phantom{0}9.2            &  \phantom{0}--21     &  10                 &  \phantom{--}0.5    & 0.50             & Line profile? \\
20\,-\,\object{OH1.095$-$0.832}                 &  0.60           & 33\phantom{.0}            &  \phantom{--0}12     &  33                 &  --0.3              & 0.64    \\
21\,-\,\object{OH1.221+0.294}                   &  0.11           & \phantom{0}4.1            &  --258               &  17                 &  \phantom{--}0.0    & 0.31     \\
22\,-\,\object{OH1.628+0.617}                   &  0.03           & \phantom{0}2.1            &  \phantom{0}--87     &  25                 &  --0.3              & 0.22             & Low S/N\\                        
\hline
\end{tabular}
\label{t:13co32}
\tablefoot{$^{(a)}$ The upper limit is 3$\sigma$. See text for details on the uncertainties of the flux densities, but we estimate that for lines stronger than 0.2\,Jy the uncertainty is about 10\,\%, while for lines weaker than 0.1\,Jy the uncertainty increases to about 50\,\% for the weakest lines. $^{(b)}$ The uncertainties in the expansion and centre velocities are estimated to be about $\pm$1.5 and $\pm$1\,km\,s$^{-1}$, respectively. $^{(c)}$ Average of the deconvolved major and minor axes of a two-dimensional Gaussian fit to the brightness distribution.}
\end{table*}


\begin{table*}[h!]
\caption{Results for the H$^{13}$CN $J$\,=\,4--3 line.}
\begin{tabular}{lccccccl}
\hline \hline
Source                          &  $S$\,$^{(a)}$  &  $\int S{\rm d}\varv$  &  $\varv_{\rm c}$\,$^{(b)}$  &  $\varv_\infty$\,$^{(b)}$  &  $\beta$            &  $\theta_{\rm s}$\,$^{(c)}$  &  Remarks  \\
                                &  [Jy]           &  [Jy\,km\,s$^{-1}$]       &  [km\,s$^{-1}$]      &  [km\,s$^{-1}$]     &                     &  [\arcsec]       &  \\
\hline 
2\,-\,\phantom{0}\object{OH358.162+0.490}                 &  1.5\phantom{0} & 50\phantom{.0}            &  \phantom{--00}3     &  25                 &  \phantom{--}1.0    & 0.31     \\
3\,-\,\phantom{0}\object{OH358.235+0.115}                 &  0.29           & 12\phantom{.0}            &  \phantom{0}--15     &  21                 &  \phantom{--}1.0    & 0.28             & Low S/N\\
20\,-\,\object{OH1.095$-$0.832}       &  0.16           & \phantom{0}6.2            &  \phantom{--0}10    &  28                 &  \phantom{--}1.0    & $\ldots$         & Low S/N     \\
\hline
\end{tabular}
\label{t:h13cn43}
\tablefoot{$^{(a)}$ See text for details on the uncertainties of the flux densities, but we estimate that for lines stronger than 1.0\,Jy the uncertainty is about 10\,\%, while for lines weaker than 0.5\,Jy the uncertainty increases to about 50\,\%. $^{(b)}$ The uncertainties in the expansion and centre velocities are estimated to be about $\pm$3 and $\pm$2\,km\,s$^{-1}$, respectively. $^{(c)}$ Average of the deconvolved major and minor axes of a two-dimensional Gaussian fit to the brightness distribution.}
\end{table*}

\FloatBarrier

\begin{table}[h!]
\caption{Results of the ALMA continuum observations at 222, 324, and 339\,GHz.}
\begin{tabular}{lrrrc}
\hline \hline
Source                        &  \multicolumn{3}{c}{$S$\,$^{(a)}$}                         &  $\theta_{\rm s}$\,$^{(b)}$    \\
\cline{2-4}
                              &   222             &   324                &  339    \\      
                              &                   &   [mJy]              &                 &  [\arcsec]\\
\hline 
1\,-\,\phantom{0}\object{OH358.083+0.137}               &  $<$\,0.8         &   $<$\,0.4           &  $<$\,1.8       &  $\ldots$  \\
2\,-\,\phantom{0}\object{OH358.162+0.490}               &  9.4              &   11\phantom{.0}     &  15\phantom{.0} &  0.10   \\
3\,-\,\phantom{0}\object{OH358.235+0.115}               &  4.1              &   6.8                &  11\phantom{.0} &  0.06 \\
4\,-\,\phantom{0}\object{OH358.505+0.330}               &  $<$\,0.8         &   $<$\,0.4           &  $<$\,2.0       &  $\ldots$  \\
5\,-\,\phantom{0}\object{OH359.140+1.137}               &  $<$\,0.8         &   $<$\,0.4           &  $<$\,2.2       &  $\ldots$  \\
6\,-\,\phantom{0}\object{OH359.149$-$0.043}             &  $<$\,0.7         &   0.7                &  $<$\,2.3       &  $\ldots$  \\
7\,-\,\phantom{0}\object{OH359.220+0.163}               &  $<$\,0.7         &   $<$\,0.4           &  $<$\,1.8       &  $\ldots$  \\
8\,-\,\phantom{0}\object{OH359.233$-$1.876}             &  $<$\,0.8         &   3.6                &  $<$\,1.8       &  0.17   \\
9\,-\,\phantom{0}\object{OH359.467+1.029}               &  $<$\,0.8         &   $<$\,0.4           &  $<$\,2.3       &  $\ldots$  \\
10\,-\,\object{OH359.543$-$1.775}             &  $<$\,0.7         &   $<$\,0.4           &  $<$\,1.7       &  $\ldots$  \\
11\,-\,\object{OH359.664+0.636}               &  $<$\,0.7         &   $<$\,0.4           &  $<$\,2.3       &  $\ldots$   \\
12\,-\,\object{OH359.745$-$0.404}             &  $<$\,0.7         &   $<$\,0.4           &  $<$\,2.2       &  $\ldots$  \\
13\,-\,\object{OH359.805+0.200}               &  $<$\,0.8         &   $<$\,0.4           &  $<$\,1.7       &  $\ldots$  \\
14\,-\,\object{OH359.826+0.153}               &  $<$\,0.6         &   $<$\,0.4           &  $<$\,1.6       &  $\ldots$  \\
15\,-\,\object{OH359.902+0.061}               &  $<$\,0.7         &   $<$\,0.4           &  $<$\,2.1       &  $\ldots$  \\
16\,-\,\object{OH0.173+0.211}                 &  $<$\,1.0         &   $<$\,0.6           &  $<$\,3.3       &  $\ldots$  \\
17\,-\,\object{OH0.221+0.168}                 &  $<$\,0.7         &   $<$\,0.4           &  $<$\,1.8       &  $\ldots$  \\
18\,-\,\object{OH0.548$-$0.059}               &  7.8              &   12\phantom{.0}     &  16\phantom{.0} &  0.14  \\
19\,-\,\object{OH0.739+0.411}                 &  1.5              &   4.0                &  $<$\,1.8       &  0.12  \\
20\,-\,\object{OH1.095$-$0.832}               &  8.3              &   14\phantom{.0}     &  15\phantom{.0} &  0.13 \\
21\,-\,\object{OH1.221+0.294}                 &  $<$\,0.7         &   $<$\,0.4           &  $<$\,1.6       &  $\ldots$ \\
22\,-\,\object{OH1.628+0.617}                 &  $<$\,0.7         &   1.0                &  $<$1.7    &  $\ldots$ \\                        
\hline
\end{tabular}
\label{t:continuum}
\tablefoot{$^{(a)}$ Flux densities obtained through a fit of a two-dimensional Gaussian to the brightness distribution. The 1$\sigma$ uncertainties in the flux densities are estimated to be 0.6, 0.5, and 2.0\,mJy at 222, 324, and 339\,GHz, respectively. The upper limits are 3$\sigma$ values. $^{(b)}$ Average of the deconvolved major and minor axes of a two-dimensional Gaussian fit to the 324\,GHz brightness distribution.}
\end{table}
%
%
%

\subsection{Images}
\label{a:images}

Images with the observational results for the $^{12}$CO($J$\,=\,\mbox{2--1} and \mbox{3--2}) and $^{13}$CO($J$\,=\,\mbox{3--2}) lines, and the 222, 324, and 339\,GHz continuum data are presented here, as well as the positions of the CO brightness distributions plotted on Spitzer IRAC images.

\begin{figure*}[h!]
  \includegraphics[width=18cm]{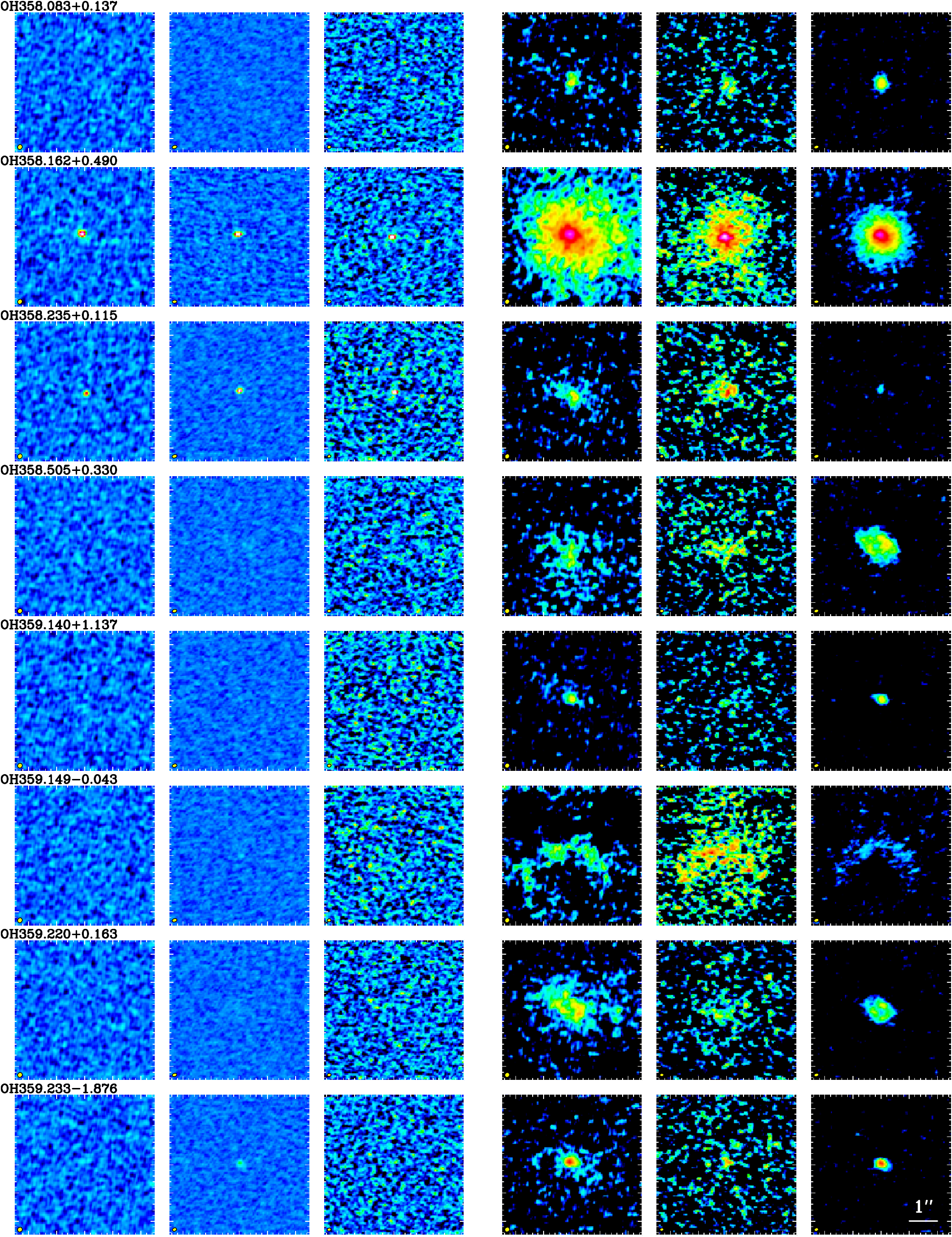} 
    \caption{Continuum images at 222, 324, and 339\,GHz in the 1st, 2nd, and 3rd column, respectively. The $^{12}$CO($J$\,=\,\mbox{2--1}), $^{12}$CO($J$\,=\,\mbox{3--2}), and $^{13}$CO($J$\,=\,\mbox{3--2}) brightness distributions integrated over the velocity range of the line in the 4th, 5th, and 6th column, respectively. Logarithmic scales are used in order to have the same scale for all objects. The image sizes are 5\arcsec $\times$5\arcsec . The synthesised beams are shown in the lower left corner of each panel.
    }
   \label{f:all_images}
\end{figure*}   

\begin{figure*}[h!]
  \includegraphics[width=18cm]{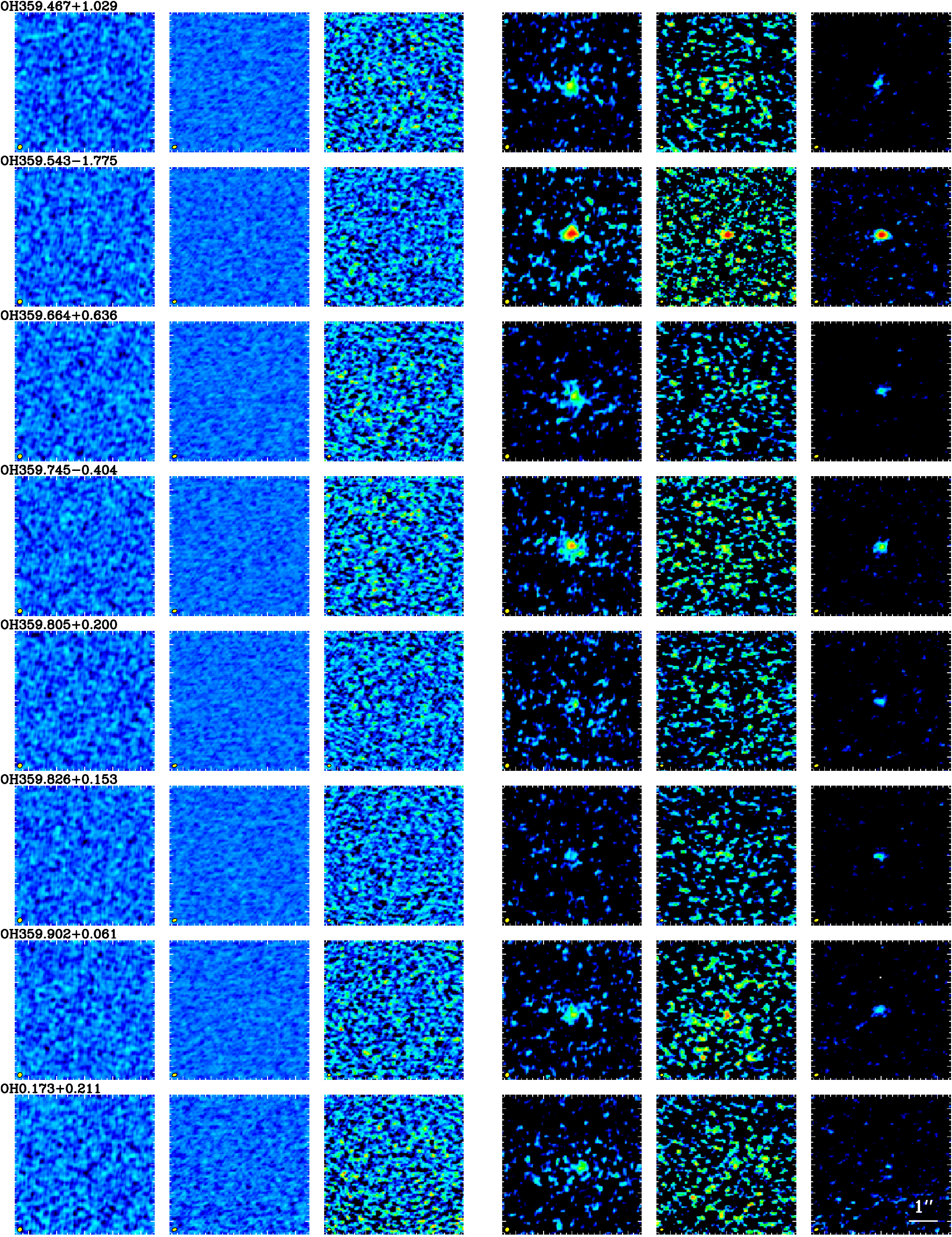} 
    \caption{Figure~\ref{f:all_images} continued.
    }
\end{figure*}   

\begin{figure*}[h!]
  \includegraphics[width=18cm]{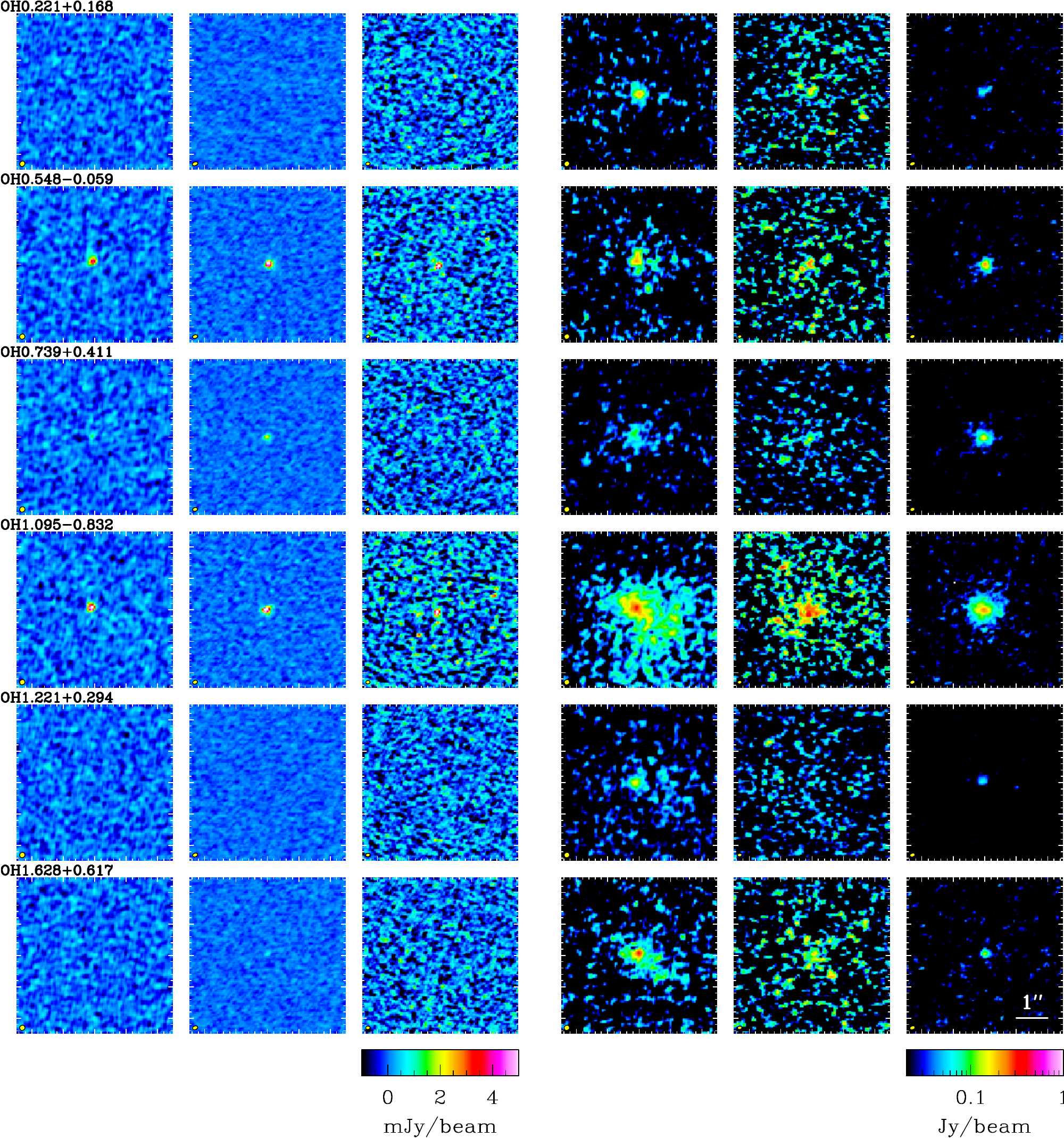} 
    \caption{Figure~\ref{f:all_images} continued.
    }
\end{figure*}   

\begin{figure*}[h!]
  \includegraphics[width=18cm]{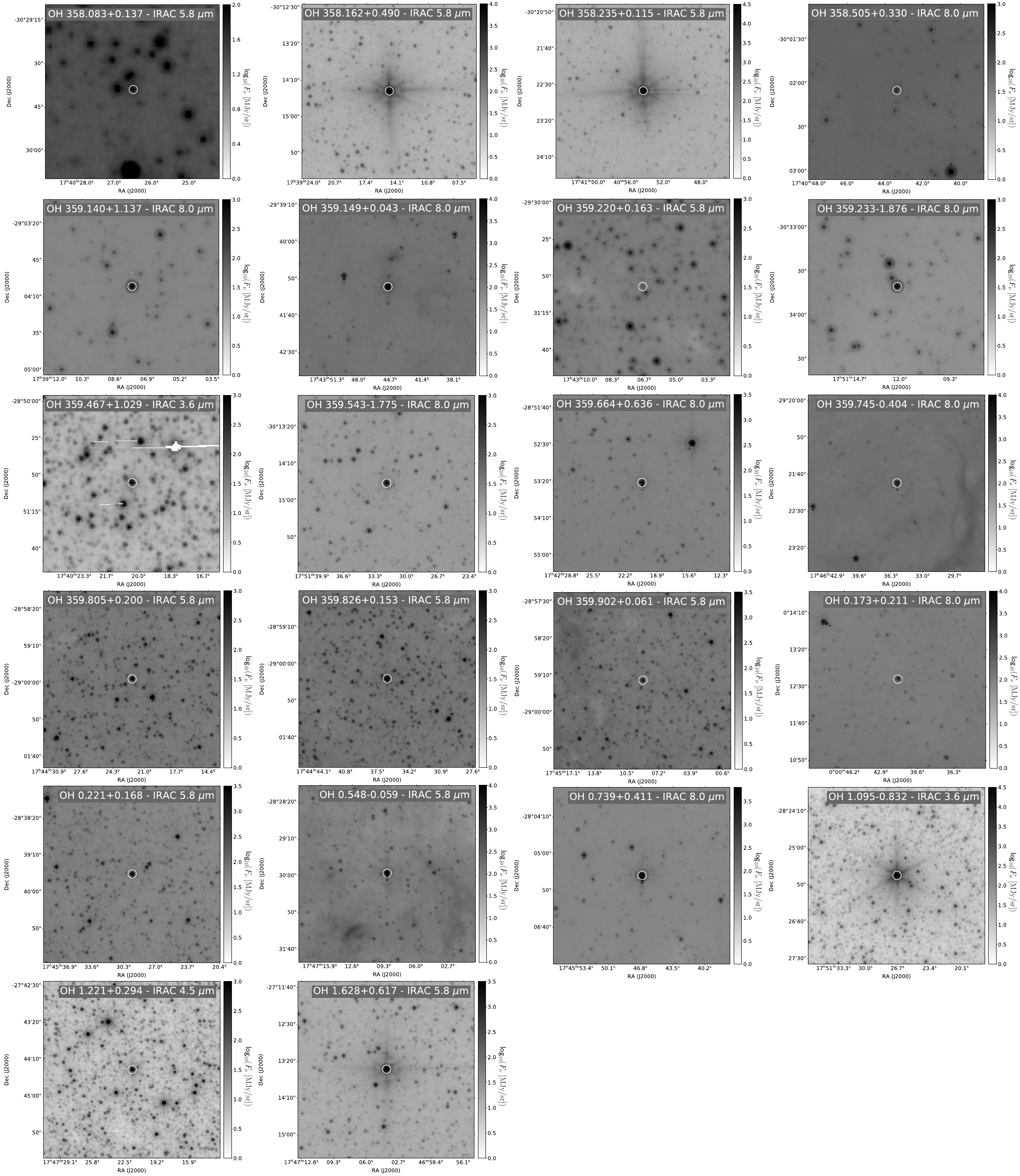}
    \caption{Central position of the $^{13}$CO($J$\,=\,3--2) brightness distribution (marked with a white circle) on a Spitzer IRAC 3.6, 4.5, 5.8, or 8.0\,$\mu$m image [for \object{OH0.173+0.211} the position of the $^{12}$CO($J$\,=\,2--1) brightness distribution is used, and for \object{OH359.149+0.043} the position of the 324\,GHz continuum peak is used]. The size of the CO line brightness distribution is significantly smaller than the white circle in all cases.
    }
   \label{f:co_spitzer}
\end{figure*}   

%
%
%
%
\FloatBarrier
\subsection{Spectra}
\label{a:spectra}

Spectra of the $^{12}$CO($J$\,=\,\mbox{2--1} and \mbox{3--2}) and H$^{13}$CN($J$\,=\,\mbox{4--3}) lines are presented here.

\begin{figure*}[h!]
  \includegraphics[width=18cm]{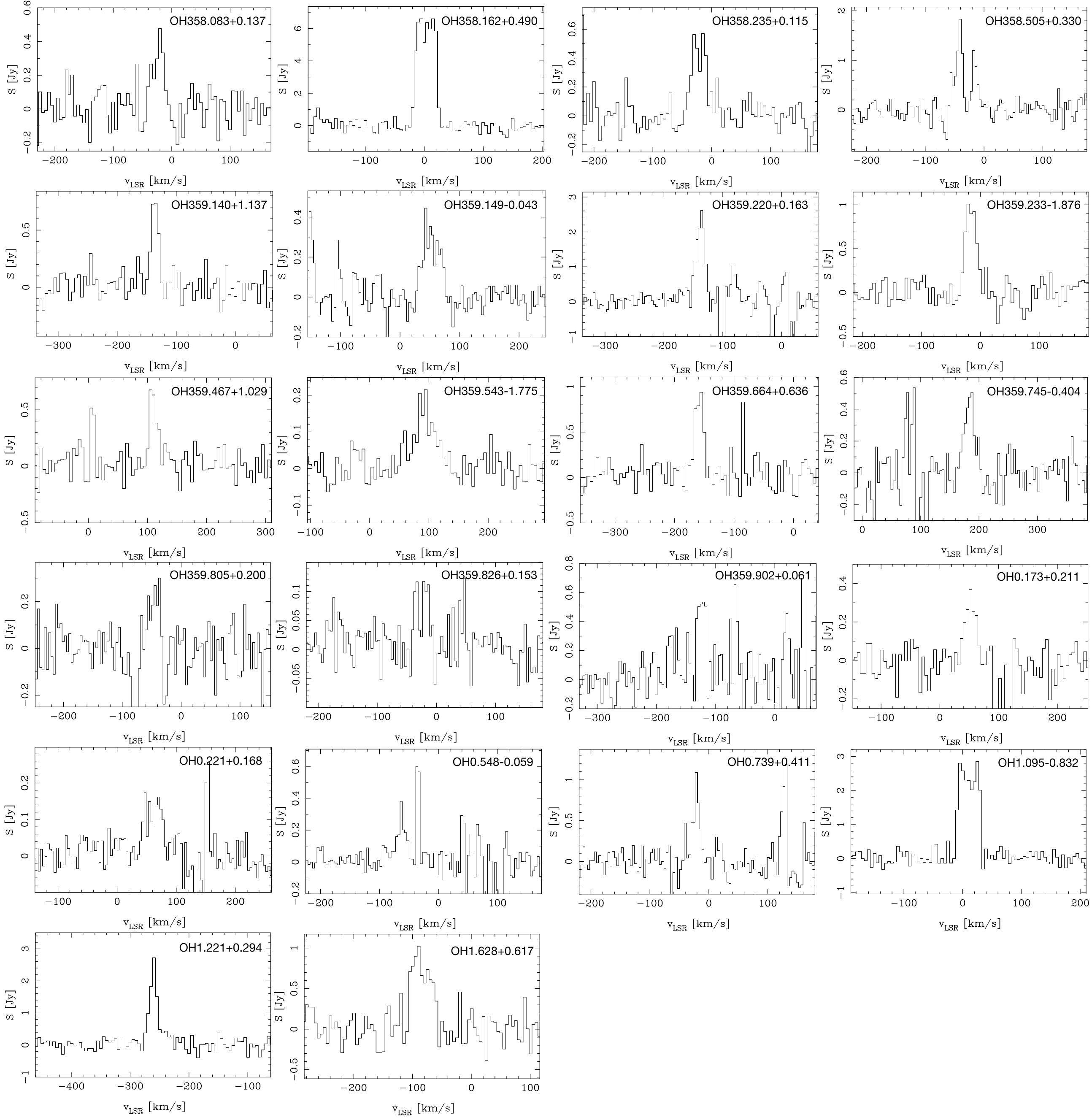} 
    \caption{$^{12}$CO($J$\,=\,\mbox{2--1}) spectra at a velocity resolution of 5\,km\,s$^{-1}$.
    }
   \label{f:12co21_spectra}
\end{figure*}   

\begin{figure*}
  \includegraphics[width=18cm]{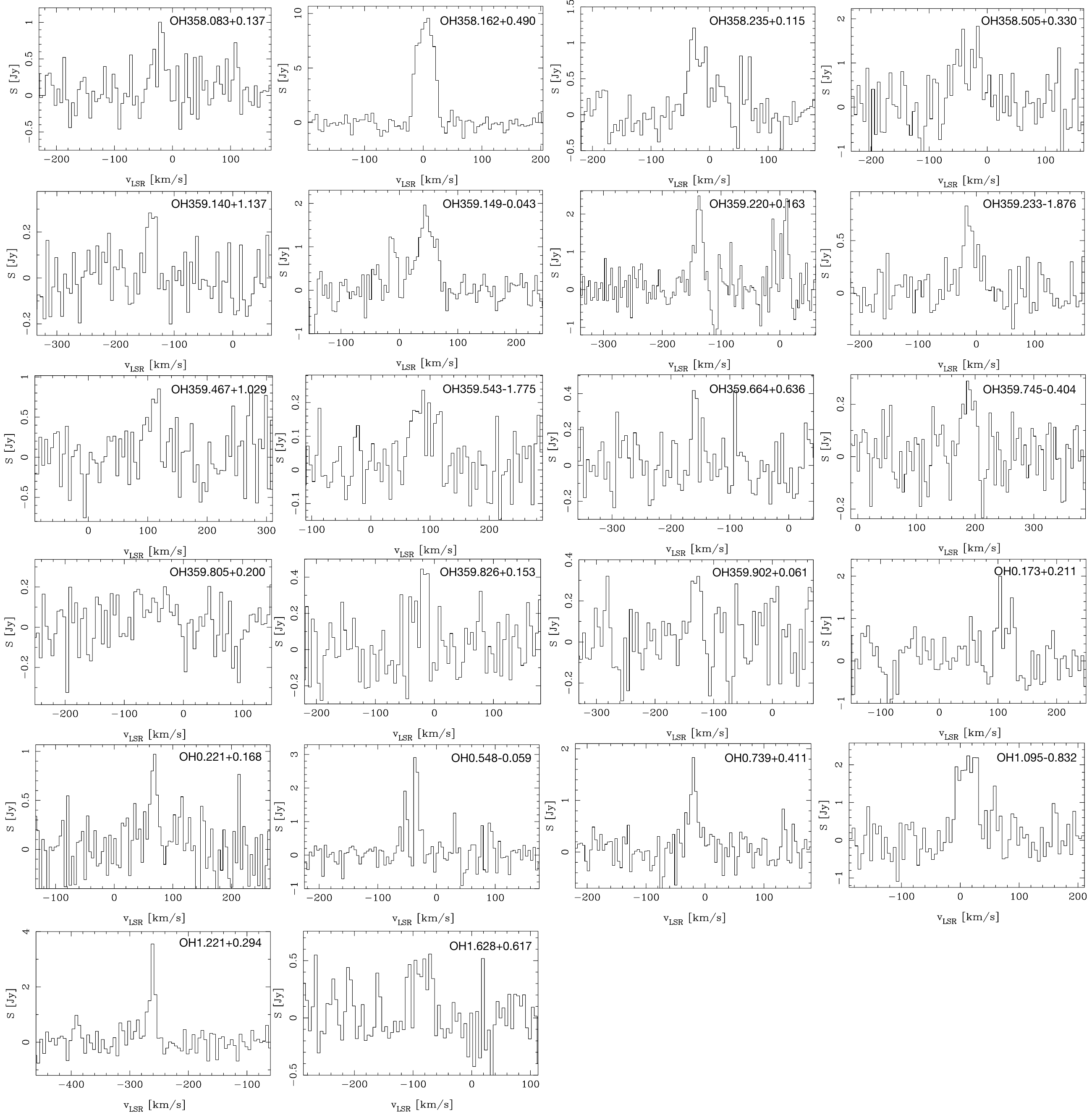}
    \caption{$^{12}$CO($J$\,=\,\mbox{3--2}) spectra at a velocity resolution of 5\,km\,s$^{-1}$. \object{OH359.805+0.200} and \object{OH0.173+0.211} were not detected in this line.
    }
   \label{f:12co32_spectra}
\end{figure*}   

\begin{figure*}
  \includegraphics[width=13.5cm]{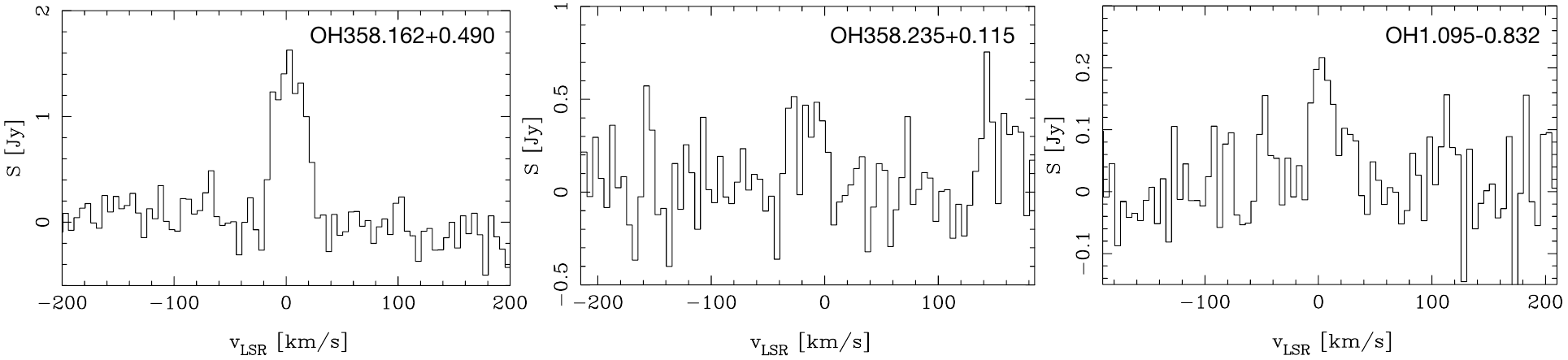} 
    \caption{H$^{13}$CN($J$\,=\,\mbox{4--3}) spectra at a velocity resolution of 5\,km\,s$^{-1}$.
    }
   \label{f:h13cn43_spectra}
\end{figure*}   

\end{appendix}

\end{document}